\documentclass[longauth]{aa}  

\usepackage{graphicx}
\usepackage{txfonts}
\usepackage{comment}
\usepackage{amssymb}
\usepackage[normalem]{ulem}
\usepackage{xcolor}
\usepackage{ragged2e}
\usepackage{amsmath}
\usepackage{float} 
\usepackage{placeins} 
\usepackage{hyperref} 
\usepackage{multicol}
\usepackage{wrapfig}

\begin{document} 

   \title{The birth of Be star disks  \\
   I.  From localized ejection to circularization
   }

   \titlerunning{Asymmetric mass ejection in Be stars}

   \author{J. Labadie-Bartz
          \inst{\ref{instParis}}
          \and
          A. C. Carciofi\inst{\ref{instBrazil}}
          \and
          A. C. Rubio\inst{\ref{instBrazil},~\ref{instMPA}} 
          \and
          D. Baade\inst{\ref{instESOGermany}}
          \and
          R. Siverd\inst{\ref{instRob}} 
          \and
          C. Arcos\inst{\ref{instChile}} 
          \and
          A. L. Figueiredo\inst{\ref{instBrazil}}
          \and
          Y. Naz{\'e}\inst{\ref{instBelgium1},\ref{instBelgium2}} 
          \and
          C. Neiner\inst{\ref{instParis}}
          \and
          T. Rivinius\inst{\ref{instESOChile}}
          \and
          N. D. Richardson\inst{\ref{instERAU}} 
          \and
          S. Nova\inst{\ref{instERAU}}
          \and
          M. L. Pinho\inst{\ref{instBrazil}}
          \and
          S. Bhattacharyya\inst{\ref{instESOChile},\ref{instSuman}}
          \and
          R. Leadbeater\inst{\ref{instRobin},\ref{instBeSS}} 
          \and
          J. Guarro Fló\inst{\ref{instBeSS},\ref{instJGF}} 
          \and
          V. Lecocq\inst{\ref{instBeSS}}
          \and
          G. Piehler\inst{\ref{instGeorg}} 
          \and
          J. Kozok\inst{\ref{instBeSS}}
          \and
          U. Sollecchia\inst{\ref{instBeSS}}
          \and
          E. Bryssinck\inst{\ref{instBryssinck}}
          \and
          C. Buil\inst{\ref{instBeSS}}
          \and
          J. Martin\inst{\ref{instJack}} 
          \and
          V. Desnoux\inst{\ref{instBeSS}}
          \and
          B. Heathcote\inst{\ref{instHeathcote}}
          \and
          P. Cacella\inst{\ref{instPaulo}}
          \and
          G. Bertrand\inst{\ref{instBertrand}} 
          \and
          J.J. Broussat\inst{\ref{instBeSS}}
          \and
          A. Ventura\inst{\ref{instBeSS}}
          \and
          R. Diz\inst{\ref{instBeSS}}
          \and
          A. Blais\inst{\ref{instBeSS}} 
          \and
          P. Somogyi\inst{\ref{instBeSS}}
          \and
          O. Thizy\inst{\ref{instThizy}} 
          \and
          O. Garde\inst{\ref{inst2SPOT}}
          \and
          S. Charbonnel\inst{\ref{inst2SPOT}}
          \and
          P. Le D\^u\inst{\ref{inst2SPOT}}
          \and
          L. Mulato\inst{\ref{inst2SPOT}}
          \and
          T. Petit\inst{\ref{inst2SPOT}}
          }

   \institute{
   LIRA, Observatoire de Paris, Universit\'e PSL, CNRS, Sorbonne Universit\'e, Universit\'e Paris Cit\'e, CY Cergy Paris Universit\'e, 92190 Meudon, France  \\
              \email{Jonathan.Bartz@obspm.fr} \label{instParis}
         \and
             Instituto de Astronomia, Geof{\'i}sica e Ci{\^e}ncias Atmosf{\'e}ricas, Universidade de S{\~a}o Paulo, Rua do Mat{\~a}o 1226, Cidade Universit{\'a}ria, B-05508-900 S{\~a}o Paulo, SP, Brazil \label{instBrazil}
        \and
        Max-Planck-Institut f\"ur Astrophysik, Karl-Schwarzschild-Str. 1, 85748 Garching b. M\"unchen, Germany \label{instMPA}
        \and
        European Organisation for Astronomical Research in the Southern Hemisphere (ESO), Karl-Schwarzschild-Str.\ 2, 85748 Garching b.\ M\"unchen, Germany \label{instESOGermany}
        \and
        Institute for Astronomy, University of Hawaii, 2680 Woodlawn, Honolulu, HI 96822, USA \label{instRob}
        \and
        Instituto de F\'isica y Astronom\'ia, Facultad de Ciencias, Universidad de Valpara\'iso,  Av. Gran Bretana 1111, Valpara\'iso, Chile \label{instChile}
        \and
        F.R.S.-FNRS Senior Research Associate \label{instBelgium1}
        \and
        Groupe d’Astrophysique des Hautes Energies, STAR, Universit\'e de Li\`ege, Quartier Agora (B5c, Institut d’Astrophysique et de G\'eophysique), All\'ee du 6 Ao\^ut 19c, 4000 Sart Tilman, Li\`ege, Belgium \label{instBelgium2}
        \and
        European Organisation for Astronomical Research in the Southern Hemisphere (ESO), Casilla 19001, Santiago 19, Chile \label{instESOChile}
        \and
        Department on Physics and Astronomy, Embry-Riddle Aeronautical University, 3700 Willow Creek Rd, Prescott, AZ 86301, USA\label{instERAU}
        \and
        CHRIST (Deemed to be University), Bangalore, India \label{instSuman}
        \and
        Three Hills Observatory, The Birches CA7 1JF, UK \label{instRobin}
        \and
        The Be Star Spectra (BeSS) database \label{instBeSS}
        \and
        Piera and SMM remote Observatories \label{instJGF}
        \and
        Selztal Observatory, Bechtolsheimer Weg 26, 55278 Friesenheim, Germany \label{instGeorg}
        \and
        BRIXIIS observatory (MPC:B96), Kruibeke, Belgium \label{instBryssinck}
        \and
        Huggins Spectroscopic Observatory UK\label{instJack}
        \and
        Glenpiper Observatory, 165 Sievers Lane, Glenhope, 3444, Victoria, Australia \label{instHeathcote}
        \and
        DogsHeaven Observatory X87, Brasilia, Brazil \label{instPaulo}
        \and
        La Montagne Observatory, 1 B Rue Jacques Prevert, 44620 La Montagne, France \label{instBertrand}
        \and
        Observatoire Belle Etoile, Revel, 38420, France \label{instThizy}
        \and
        2SPOT, 45, Chemin du Lac, 38690 Ch\^abons, France \label{inst2SPOT}
             }

   \date{Received ; accepted}

 
  \abstract
   {Classical Be stars are well known to eject mass to build up a disk, but the details governing the initial distribution and subsequent evolution of this matter into a disk are in general poorly constrained through observations.}
   {By combining high-cadence time-series spectroscopy with contemporaneous space photometry from the Transiting Exoplanet Survey Satellite (TESS), we have sampled about 30 mass ejection events in 13 Be stars. Our goal is to constrain the geometrical and kinematic properties of the ejecta as early as possible, facilitating the investigation into the material's initial conditions and evolution, and understanding its interactions with preexisting material.
   }
   {
   The photometric variability is analyzed together with measurements of the at-times rapidly changing emission features in order to identify the onset of outburst events and obtain information about the geometry of the ejecta and how it changes over time. Short-lived line asymmetries display oscillation cycles ({\v{S}}tefl frequencies), which are compared to photometric and stable spectroscopic frequencies.
   }
   {All Be stars observed with sufficiently high cadence during an outburst are found to exhibit rapid oscillations of line asymmetry with a single frequency in the days following the start of the event. For a given star this circumstellar frequency may differ only slightly from event to event even when the outbursts they are associated with have different properties. These circumstellar frequencies are typically between 0.5 to 2 d$^{-1}$, and are generally near photometric frequencies. They are slightly below prominent (generally stable) spectroscopic frequencies seen in photospheric absorption lines. The emission asymmetry cycles break down after roughly 5 -- 10 cycles, with the emission line profile converging toward approximate symmetry shortly thereafter. In photometry, several frequencies typically emerge at relatively high amplitude at some point during the mass ejection process.}
   {In all observed cases, freshly ejected material was initially constrained within a narrow azimuthal range, indicating it was launched from a localized region on the stellar surface. The material orbits the star with a frequency consistent with the near-surface Keplerian orbital frequency. This material circularizes into a disk configuration after several orbital timescales. This is true whether or not there was a preexisting disk at the time of the observed outburst. We find no evidence for precursor phases prior to the ejection of mass in our sample. The several photometric frequencies that emerge during outburst are at least partially stellar in origin. }
   
   \keywords{Stars: emission-line, Be --
                Stars: oscillations --
                Stars: winds, outflows --
                Stars: circumstellar matter --
                Techniques: spectroscopic --
                Techniques: photometric}

   \maketitle
%

\section{Introduction}
\label{sec:introduction}

Mass loss is ubiquitous in stars across the Hertzsprung-Russell diagram, but it is especially strong in massive stars for which it clearly shapes their stellar evolution. A prominent example is the radiative pressure on a line ensemble driving mass loss from massive stars \citep[i.e., winds;][]{1975ApJ...195..157C, 1988ApJ...335..914O, 2008A&ARv..16..209P, 2023Galax..11...68C}, both in the main sequence as well as different post-main sequence stages. Another important case is the mass loss that occurs in evolved stages of low- and intermediate-mass stars, such as red giant branch (RGB), asymptotic giant branch (AGB), and red supergiant (RSG) stars. Even though these are completely different objects, mass loss in this case seems to be the result of a small surface gravity combined with internal processes -- such as convective overshooting, pulsation -- and radiative pressure on grain and molecules \citep[e.g.,][]{2018A&ARv..26....1H}. In general, the matter lost by hot stars expands rapidly, and for cooler stars the circumstellar matter is contained in extended structures such as nebulae.

Another example where a low surface gravity is conducive to mass loss is Classical Be stars. These stars are unique in that the majority of their lost circumstellar matter is assembled in a gaseous Keplerian disk, with early-type Be stars also having a radiatively driven wind \citep{Rivinius2013}. These disks are known to be formed from outflowing material launched from the stellar surface, giving rise to emission lines, excess continuum flux, and polarization \citep[][and references therein]{Rivinius2013}. The central stars in such systems rotate near critically \citep{zorec2016} and are now understood to be nonradial pulsators as a rule \citep{2003A&A...411..229R,Walker2005,2016A&A...593A.106R,Baade2017,Semaan2018,JLB2022}. While rapid rotation lowers the effective gravity near the equator, one or more mechanisms are required to impart surface material with sufficient angular momentum to achieve orbital velocities and form an outflowing disk \citep{2006ASPC..355..219O}. The mass ejection mechanism is integral to the Be phenomenon. While nonradial pulsation seems to be a key aspect of the mass ejection mechanism, the currently proposed theoretical frameworks, for example, \citet{2020A&A...644A...9N}, have not been widely demonstrated on a large sample of classical Be stars yet. 

The total mass and its distribution in Be star disks (which dictate its observables) can vary on timescales from hours to decades, and a given star may lose its disk entirely, only to rebuild it at some later time. Stellar mass ejection is crucial for sustaining a disk as it otherwise dissipates via viscous forces \citep{1991MNRAS.250..432L, 2008ApJ...684.1374C, Carciofi2011} and radiative ablation \citep{2018MNRAS.479.4633K}. Such mass ejection often occurs in discrete episodes referred to as outbursts \citep[e.g.,][]{1998A&A...333..125R}. The amount of time during which the star is actively ejecting mass in an outburst event can be as brief as one to a few days \citep{JLB2022}. Longer outbursts (lasting weeks, months, or years) are also commonly observed \citep{1998A&A...335..565H,Rimulo2018,2018AJ....155...53L,Bernhard2018,Ghoreyshi2018}, although it is not always clear whether the disk is being built up by continuous stellar mass loss or from many frequent events \citep{2021MNRAS.502..242L}.

The long-term (i.e., months to years) behavior of Be disks is generally well described by the Viscous Decretion Disk (VDD) model \citep{Haubois2012}, and its application to long time-series observations tracking the formation and dissipation of the disk has provided important information about the mass and angular momentum injection rates, as well as the mass and angular momentum flux out of the system. These findings offer crucial clues regarding the efficiency of viscous transport processes \citep{Rimulo2018, Ghoreyshi2018,2021ApJ...912...76M}. Current models have so far explored a quite simplistic geometry for the mass ejection, assuming that material is ejected in a uniform ring around the star \citep[except for][who modeled a scenario where material is ballistically launched from a single spot on the stellar equator]{1997ASPC..121..494K}, with the exact amount of angular momentum to be in Keplerian rotation at that ring. From there, viscous forces enable the outward flow of disk material. In other words, current VDD implementations assume that there is already a symmetric disk or ring in place around the star, but it does not describe the initial formation of such a structure. There is no existing theory that describes the ``star-disk interface,'' whereby some initial density and velocity distribution of just-ejected material finally forms an axi-symmetric disk structure (or settles into a preexisting disk).

On average, emission-line profiles of Be disks are symmetric, implying that the disks are axi-symmetric. Disks may become perturbed, developing two-armed ($m=2$) density waves due to tidal forces from an orbiting binary companion \citep[e.g.,][]{2018MNRAS.473.3039P, 2020MNRAS.497.3525C}, and/or one-armed ($m=1$) density waves that do not require a companion \citep{1997A&A...318..548O}. The timescales of these density perturbations are long -- for $m=2$ waves they are equal to the binary orbital period (typically months) and $m=1$ waves have cycle lengths between about one to ten years. Despite the on-average symmetry of developed disks, there is no requirement that they are initially formed with a symmetric structure. Viscous shear and orbital phase mixing will inevitably transform an initially asymmetric matter distribution into a ring or disk-like structure. Once symmetry is achieved, the memory of the gas is lost, and it is impossible to trace back its history. 

The goal of this paper is to observationally probe the star-disk interface. Doing so requires a high observing cadence to sufficiently sample the relevant timescales involved that are primarily dictated by four factors. The first two are the stellar rotation and the Keplerian orbital periods which are similar to each other and are typically between $\sim$0.5 to $\sim$2 days (the ratio between rotation and Keplerian periods can range from slightly below to slightly above unity, depending on the fraction of critical rotation and the orbital distance). 
The third is the duration of the disk build-up and dissipation phases that together are typically several days and longer (this work focuses on shorter events). The fourth is the circularization timescale, that is, the time required for an initially asymmetric inner disk to become roughly symmetric, which is typically on the order of one week \citep[e.g.,][]{2011A&A...533A..75L}. Observations should then cover at least the first $\sim$10 days after the onset of an outburst event with multiple observations per day. From these data, it is then possible to determine the initial properties and evolution of ejected material.

Spectroscopy is a convenient tool for this objective since circumstellar material gives rise to line emission and within the line profile is encoded information about the material's geometric distribution and kinematic properties. For instance, an axi-symmetric ring or disk will result in a symmetric (typically double-peaked) line profile with equal intensity blue- and red-shifted peaks (referred to as the V and R peaks, respectively) as the observed kinematics will be symmetric around the systemic velocity of the system. On the other hand, an asymmetric mass distribution will result in blue/red emission line asymmetry ($V/R \neq 1$). Furthermore, as material orbits around the star this asymmetry will shift in radial velocity depending on its orbital phase as viewed from Earth. 

The critical observational signature is thus determining whether or not line emission originating close to the star (where material is expected to be hot and with relatively high velocity) is symmetric in the earliest stages of an outburst, or if it is initially asymmetric and travels across the line profile as material orbits. The variations of emission asymmetry are typically referred to as ``rapid V/R cycles'' \citep[e.g.,][]{1998A&A...333..125R}, because the intensities of the V and R peaks oscillate roughly in antiphase (see also Sect.~\ref{sec:EWs}). The term ``{\v{S}}tefl frequencies'' was coined to describe these V/R cycles \citep{2016A&A...588A..56B} following the pioneering work of \citet{1998ASPC..135..348S}, with an implicit interpretation -- that these variations are circumstellar.  {\v{S}}tefl frequencies may also manifest photometrically.

Observationally studying the initial phases of these events is challenging because of the unpredictable\footnote{While outbursts in most Be stars cannot at present be predicted, there are some exceptions where sufficient knowledge of the involved pulsation frequencies allows the timing of outbursts to be predicted such as in $\mu$~Cen, 25~Ori, and V442~And \citep{1998ASPC..135..343R, 2018pas8.conf...69B, 2021MNRAS.508.2002R}.}
nature of outbursts and the requirement for sampling the short-lived and rapid V/R variations. Nevertheless, such high-cadence spectroscopic datasets have been acquired and analyzed for a few Be stars during the early stages of an outburst, paving the way for the present study. Seemingly all of them are consistent with a scenario of ejecta being initially asymmetric and then forming or merging into an axi-symmetric disk (see next two paragraphs). That is, all reported cases initially showed rapid V/R cycles for some days before symmetry was realized. 

The Be star $\mu$ Cen was noted to have rapid V/R oscillations during the rising phase of outbursts, evolving to equal-intensity V and R peaks after reaching maximum emission strength \citep{1993A&A...274..356H}. With a high-cadence observing campaign (usually many spectra per night), and including archival spectra, \citet{1998A&A...333..125R} sampled multiple outburst events of $\mu$ Cen with sufficient cadence to track the evolution of emission features on sub-day timescales. During the early stages of these events (the first few days), cyclic variability was detected in emission features (especially helium) with cycle lengths of approximately 0.6 days. This timescale is similar to, but statistically distinct from, the dominant period of the stellar line profile variations (with a period of circa 0.5 d) caused by nonradial pulsation \citep{1998A&A...336..177R, 2001A&A...369.1058R}. The most relevant conclusion of \citet{1998A&A...333..125R} for this work is that the short-lived cyclic V/R oscillations present during the early phases of outbursts ``are consistent with an ejected cloud of gas which orbits the star a few times at a small radius until it is dispersed or merges with the disk or falls back to the star or some combination of these processes.'' In other words, the distribution of circumstellar material was initially localized (in azimuth), causing V/R asymmetry.

Other well-documented cases include $\upsilon$ Cyg, with V/R cycles of 0.67 d \citep[1.5 d$^{-1}$\xspace,][]{2005A&A...437..257N}, $\omega$ CMa  with V/R cycles of 1.49 d \citep[0.67 d$^{-1}$\xspace,][]{2003A&A...411..167S}, $\omega$ Ori with V/R cycles of 2.2 d (0.45 d$^{-1}$) and a main nonradial pulsation period of 0.971 d \citep{2002A&A...388..899N}, and potentially EW Lac with a signal at about 0.8 d (1.25 d$^{-1}$) that seems consistent with such V/R cycles \citep{2000A&A...362.1020F}. These are qualitatively similar to $\mu$ Cen, with V/R cycles that are near or slower than the dominant pulsation frequency (the latter being inferred from photospheric line profile variations). Extensive and very high quality observations of $\lambda$ Eri, comprising 840 spectra collected on 184 nights spanning the years 1984 -- 1988 \citep{1989ApJS...71..357S} showed many complex variations in both photospheric and circumstellar lines, including the types of V/R variations discussed above, with the author noting that ``Our picture is that emission originates from discrete blobs of ejected material.'' However, while the rapid V/R variations were obvious, no specific period was noted. For $\eta$ Cen, spectroscopy revealed a circumstellar {\v{S}}tefl frequency at 1.56 d$^{-1}$\xspace and two pulsational frequencies at 1.73 d$^{-1}$\xspace and 1.77 d$^{-1}$\xspace \citep{2003A&A...411..229R}. About 20 years later, photometry of $\eta$ Cen from the
BRITE (BRIght Target Explorer) Constellation of nano-satellites \citep{2014PASP..126..573W, 2016PASP..128l5001P} detected the same {\v{S}}tefl frequency as well as both pulsational frequencies from \citet{2003A&A...411..229R}, plus several other signals \citep{2016A&A...588A..56B}. $\alpha$ Eri is probably another case with a photometrically detected {\v{S}}tefl frequency (at 0.725 \,d$^{-1}$\xspace) slightly below its dominant pulsation frequency at 0.775 \,d$^{-1}$\xspace \citep{2011MNRAS.411..162G, 2016A&A...588A..56B}.

In this paper we present 33 mass-loss events in Be stars, the majority of which were observed at high cadence by space photometry and simultaneous spectroscopy. Section~\ref{sec:sample} describes the sample and the spectroscopic and photometric data acquired, with Sect.~\ref{sec:methods} introducing the measurements and methods used. The outcome for three representative stars are given in Sect.~\ref{sec:results}. In Sect.~\ref{sec:discussion} the results are discussed, especially with respect to the emission asymmetry oscillations. Conclusions are given in Sect.~\ref{sec:conclusions}. 
Appendix~\ref{sec:appendix_tbls} provides relevant tables, Appendix~\ref{sec:appendix_b} gives details about the remainder of the sample, Appendix~\ref{sec:pole_vs_edge} highlights mass ejection events in pole-on stars,  Appendix~\ref{sec:muCen} revisits old and new data for $\mu$~Cen, $\eta$ Cen, and $\omega$ CMa, and Appendix~\ref{sec:Xrays} discusses new and old X-ray observations of V767~Cen.

\section{Sample and data}
\label{sec:sample}


\subsection{Sample selection and observing strategy}

The Transiting Exoplanet Survey Satellite (TESS) space photometry mission (see Sect.~\ref{sec:TESS_data}) has been conducting a nearly all-sky survey since 2018. In doing so, it is observing virtually all Galactic Be stars (and systems in the Magellanic Clouds) brighter than $V\approx 14$\,mag. In order to make the most of the TESS photometry, we have been conducting an observing campaign to obtain time-series spectroscopy contemporaneous with TESS. A subset of our targets were observed with approximately daily or better cadence to achieve dense coverage during outburst events, should they occur during the TESS observing window. These targets were chosen based on a number of factors, including brightness, historical activity (i.e., having displayed frequent outbursts recently), and the status of the disk in the days and weeks leading up to the TESS observing window. In particular, targets both with and without strong preexisting disks were selected. All of the targets selected for high-cadence spectroscopy showed evidence of one or more mass ejection episode with only two exceptions (25~Cyg and QR~Vul did not experience an outburst during the high cadence monitoring in June-July 2022). Again, only targets with a reasonably high likelihood of exhibiting mass ejection were selected, so no inference can be made about the occurrence rates of outbursts in Be stars in general. Whenever possible, the observing cadence was increased upon detecting the first signs of an outburst (see Sect.~\ref{sec:flicker_obs}). Table~\ref{tbl:sample} lists the 13 targets presented in this work, plus six systems from the existing literature.

The TESS survey delivers critical information for our sample before, during, and after outburst events. This includes information about the (changing) frequency content of the star, and the changing broadband flux from the build up and dissipation of circumstellar material (Sect.~\ref{sec:VDD}). On the other hand, spectroscopy gives valuable insights into the kinematics of the ejecta. The combination of photometry and spectroscopy has already been demonstrated to provide strongly synergistic constraints to the nature of the circumstellar material in Be stars \citep[e.g.,][]{2015A&A...584A..85K}. The majority of our high-cadence spectroscopic datasets were obtained at the same time as the TESS observations of a given star. For convenience we primarily use the TESS Julian Date, defined as TJD = BJD - 2457000.

Our targets are mostly early-type Be stars (B3 and earlier), except for $\iota$ Lyr (B6IVe), introducing an obvious bias into our sample. This is a consequence of our selection criteria, since early-type Be stars tend to have much more frequent outbursts and typically demonstrate variability in their disks on shorter timescales compared to mid- and late-type Be stars \citep[e.g.,][]{1998A&A...335..565H,Bernhard2018}. Early-type Be stars also host disks with higher densities \citep{2017MNRAS.464.3071V}, increasing the visibility of the observational signatures associated with mass loss.

\subsection{TESS photometry}
\label{sec:TESS_data}

The NASA TESS \citep{Ricker2015} is a photometric mission performing wide-field photometry over nearly the entire sky. The 4 identical cameras of TESS cover a combined field of view of 24$^{\circ}$ $\times$ 96$^{\circ}$. TESS science operations began in August 2018, and the satellite has been surveying the sky in sectors of 27.4 days long, which continues to this day. Some regions of the sky are observed in multiple (consecutive) sectors. TESS records red optical light with a wide bandpass spanning roughly 600 -- 1000 nm, centered on the traditional Cousins $I$ band. For optimal targets, the noise floor is approximately 60 ppm h$^{-1}$.

Light curves of our targets were extracted from the TESS full frame images (FFIs) as in \citet{JLB2022}. The aperture size and shape were selected to include the pixels illuminated by the target star (including saturated columns and the ``spillover'' pixels) and avoid neighboring sources. The FFI cadence was 30 minutes during TESS cycles 1 and 2 (the first two years of the mission, sectors 1 -- 26), 10 minutes during cycles 3 and 4 (sectors 27 -- 55), and 200 seconds during cycles 5 and 6 (sectors 56 -- 83). While 2-minute cadence data are available for some of our targets, the FFI cadence is more than sufficient to sample the relevant timescales. 

With the large pixel scale of TESS (23\arcsec{}), flux from neighboring stars can contribute to the light curve extracted for the star within some given aperture (often referred to as blending). Based on a pixel-level variability analysis and considering all relevant nearby Gaia sources \citep[as in, e.g.,][]{2023A&A...676A..55L}, all detected signals can reliably be attributed to the target Be stars -- that is, blending has no impact. Besides the following two exceptions, none of our targets have neighboring stars within or near the adopted aperture that are less than 5 magnitudes fainter (in the Gaia G filter). 28 Cyg has a neighbor at a distance of 82 arcseconds that is $\sim$4.5 magnitudes fainter (HD 227992, B9) but does not contribute any detectable signals to the extracted light curve. V357~Lac has a neighbor 117 arcsec away and $\sim$4 magnitudes fainter (TYC 3619-1400-1, G2V), which also does not contribute any detectable signals.

\subsection{Spectroscopic data}
\label{sec:spec_data}

The spectroscopy used in this work is a combination of both professional and amateur efforts. The facilities are described in the following subsections. Employing multi-longitude observing sites is important for more fully sampling the rapid variations that occur early during outbursts, for providing insurance against bad weather to avoid prolonged gaps, and for confirming spectroscopic signals recorded on different instruments near the same time. A log of the spectroscopic observations and date ranges is given in Tab.~\ref{tbl:obs_log}.

\subsubsection{NRES}
\label{sec:spec_data_NRES}

Echelle spectra (${R\sim53000}$) were obtained with the Network of Robotic Echelle Spectrographs (NRES) attached to the 1-m telescopes of the Las Cumbres Observatory (LCO) Global Telescope network \citep{2013PASP..125.1031B}, including at the Wise observatory, the South African Astronomical Observatory, the Cerro Tololo Interamerican Observatory, and McDonald Observatory. The NRES spectra provide the main database upon which this paper is founded. NRES data can be retrieved from the LCO science archive\footnote{\url{https://archive.lco.global/}}.

\subsubsection{DAO}
\label{sec:spec_data_DAO}

Observations were obtained from the Dominion Astrophysical Observatory (DAO) 1.2-m telescope and McKeller spectrograph (with resolving power  ${R\sim17600}$), which covers H$\alpha$ and \ion{He}{I}\,$\lambda$6678 in the chosen observing mode \citep{2014RMxAC..45...69M}. DAO spectra can be downloaded from the DAO science archive\footnote{\url{https://www.cadc-ccda.hia-iha.nrc-cnrc.gc.ca/en/dao/}}.

\subsubsection{CHIRON}
\label{sec:spec_data_CHIRON}

For one of our targets, $\lambda$~Pav, spectra were acquired from the 1.5-m telescope located at the Cerro Tololo Inter-American Observatory (CTIO), Chile, using the CTIO High Resolution spectrometer (CHIRON) instrument \citep{2013PASP..125.1336T}, operated by the Small and Moderate Aperture Research Telescope System (SMARTS) Consortium. The instrument configuration was in ``slicer'' mode, providing a resolving power of ${R\sim80000}$. CHIRON spectra can be downloaded from the NOIRLab Astro data archive\footnote{\url{https://astroarchive.noirlab.edu/portal/search/}}.

\subsubsection{BeSS}
\label{sec:spec_data_BeSS}
The Be Star Spectra (BeSS) database\footnote{\url{http://basebe.obspm.fr}} \citep{BeSS} hosts a large collection of primarily amateur spectroscopy for Be stars. In addition, several observers specifically targeted stars in this sample with a high cadence during the TESS epochs to support the present work. The BeSS data serve two key purposes: enhancing coverage during the TESS observing period and extending the observed time baseline before and after the TESS visit. With hundreds of observers and instrumental setups, the data are inhomogeneous. The large majority of entries in BeSS encompass H$\alpha$, but numerous entries also include the nearby \ion{He}{I}\,$\lambda$6678 line, and some echelle spectra cover a broad array of useful lines. For this work we considered only spectra with resolving power ${R\sim10000}$ or higher. The BeSS database was also important in choosing which systems to monitor with high cadence through analysis of their historical behavior.

\section{Measurements and methods}
\label{sec:methods}

Past investigations of Be disks, especially those examining the temporal variability of observables, offer valuable insights into the observations presented herein and are thus summarized briefly in Sect.~\ref{sec:VDD}. Sections~~\ref{sec:flicker_obs} to \ref{sec:EWs} describe the measured quantities explored in this paper and the procedures to acquire them.

\subsection{Lessons learned from VDD studies}
\label{sec:VDD}

Hydrodynamic models such as the ones presented by \citet{Haubois2012} show that Be disks grow and dissipate in an inside-out fashion. When a diskless Be star becomes active, the inner disk rapidly fills up, while the outer regions follow suit at a much lower rate. When mass loss stops, the now unsupported disk dissipates, with the inner part returning to the star and the outer part being lost to the interstellar medium. From an observational standpoint, this manifests as a net brightening (indicative of growth) followed by a net dimming back to baseline (reflecting dissipation). The amplitude of the photometric variations traces the overall disk emitting area. The opposite happens when the disk is seen edge-on, as the disk growth results in a net dimming as it partially enshrouds the star. New outbursts may occur while a disk is dissipating, resulting in complex radial structure \citep{2001A&A...379..257R}.

Spectroscopically, the manifestation is more diverse. Clearly, the increase in disk mass results in stronger overall line emission, for example from H$\alpha$ and other Balmer lines, but the line profile displays a variability that depends crucially on the disk radial and azimuthal density distribution. When the disk is young and matter is confined to a very small volume close to the star, the V and R peaks should be at their maximum separation, roughly $2\times \upsilon_{\rm orb} \sin i$, where $\upsilon_{\rm orb}$ is the Keplerian orbital speed at the base of the disk and $i$ is the inclination angle. A decrease in peak separation reflects that material has moved outward, and a minimum peak separation value is associated with the largest line emission area. The dissipation of the disk can be tracked as the peak separation increases back to $2\times v_{\rm orb} \sin i$ \citep[e.g.,][]{2021ApJ...912...76M}. The above is strictly valid only for azimuthally symmetric disks. If mass ejection is highly (azimuthally) localized, the exact behavior of the V/R ratio and peak separation will depend on the matter distribution and the observing angle. 

Emission line wings respond relatively quickly to the ejection of mass by becoming wider and stronger, as they trace the high-velocity gas close to the star. For the same reason, the line wings decrease in strength quickly as the inner disk dissipates, resulting in a sharper transition between the continuum and emission peaks during dissipation.

\subsection{Observational signatures of mass ejection} \label{sec:flicker_obs}

For low and intermediate inclination angles, the net brightening in broadband photometry is the typical hallmark of an outburst (Sect.~\ref{sec:VDD}). Massive disks can increase the brightness by up to $\sim$0.5 magnitudes \citep[in the V or R band, e.g.,][]{2006ApJ...652.1617C,2017AJ....153..252L}. However, such large changes are usually only seen over longer timescales, as a massive disk might take months or many years to develop. On timescales of days to weeks, typical continuum variations of outburst events are at about the $\sim$5\% to 20\% level. These short-lived (days to weeks) and low-amplitude mass ejection events will be refereed to as ``flickers'' from now on.

Since there is a lag of weeks to months between a TESS observation and the availability of the photometry, we relied on spectroscopy for initially detecting flickers and subsequently increasing the observing cadence. This required immediate reduction and analysis of the spectra in order to maximally sample the earliest stages of these events. The most reliable indication of a new flicker is the sudden appearance of high velocity emission at similar locations in several emission lines. This may or may not coincide with overall increased emission strength (numerically lower equivalent width).

Once the full spectroscopic and photometric dataset is available for a given event, isolated flickers can generally be described as follows. At approximately the same time as the emergence of high-velocity emission, the brightness begins to increase. For the first several days, the new emission moves across the line profiles rapidly. The brightness increases to its peak value over a few days or longer (this is highly event dependent), and then returns toward the baseline flux on a timescale about twice as long as the brightening phase. The evolution of emission strength typically lags behind the photometry, with emission strength peaking some time after the photometric maximum. Emission strength then decays slower than the photometric flux. For some amount of time during the brightening event, enhanced rapid variations in photometry are usually evident. 

However, the observed photometric and spectroscopic behavior of a flicker depends on several factors, such as the strength (and radial density profile) of any preexisting disk, and the rate at which mass ejections occur. 
Observational signals become especially complex when flickers occur in rapid succession, such that it can be difficult to delineate discrete events. 
The inclination angle also plays an important role, and can dictate whether brightness and emission are correlated or anticorrelated, as discussed in \citet{1983HvaOB...7...55H, 2000ASPC..214...13H}.
For very high inclination angles ($i \gtrsim 80^{\circ}$; shell Be stars), the injection of material into the disk will cause a decrease in brightness \citep{Haubois2012}. Somewhere between intermediate and high inclination angles ($i \sim 70^{\circ}$), the growth of a disk may not have any net effect on the brightness, as the continuum flux added by the disk is canceled out by the blocking of stellar continuum flux by disk material \citep{Haubois2012}. One of the stars in the sample, $\iota$~Lyr, probably falls into this category as we observe mass ejection but without an obvious net change in brightness (Sect.~\ref{sec:iot_Lyr}). 

Nevertheless, the sudden emergence of rapidly varying high-velocity emission is a reliable marker of new mass ejection for all stars in our sample, regardless of other potential complications. Noncyclic net changes in brightness is a reliable photometric indicator of mass ejection.

\subsection{Photometric flicker timescales}
\label{sec:flicker}

For flickers that are both well-defined in photometry and are well sampled with spectroscopy, the start, peak, and end of each event were visually estimated from the TESS observations. The relevant quantities are the build-up time (the difference in time between the start of the brightening and peak brightness), the dissipation time (the time between peak brightness and the return to the base brightness), and the photometric amplitude (the difference in TESS flux between the peak and base brightness). The build-up duration should provide an upper limit to the amount of time the star spends actively ejecting mass. During the dissipation phase, the inner disk is decreasing in density as it is no longer being fed by the star. In several cases, the build-up and/or dissipation time cannot be measured due to either incomplete coverage of the event in the TESS data or the start of a new flicker prior to the return to baseline brightness.

\subsection{Line profile measurements}
\label{sec:EWs}

The most immediately useful line profile quantities for the purpose of this work are measures of the emission strength and emission asymmetry. Emission strength is simply measured by calculating the equivalent width (EW) over an appropriate range in velocity. 
By definition, EW is positive when a line is in absorption and negative when in emission.

The conventional way by which asymmetry in double-peaked emission lines is measured is by taking the ratio of the intensity of the V and R peaks, either relative to the continuum or using the integrated flux (the V/R ratio). However, this method is only valid when there are two well defined peaks, which is not always the case.

A more robust way to measure line asymmetries is by taking the ratio of the EW computed from the blue-shifted and red-shifted halves of an emission line profile, after adjusting for the systemic velocity. This is the primary method adopted for measuring line asymmetries in this work, according to the equation

\begin{equation}
\frac{\mathrm{EW}_{\rm V}}{\mathrm{EW}_{\rm R}}   = \frac{ \int_{\nu_1}^{c_1} (C+1-F_\nu) d\nu}{\int_{c_2}^{\nu_2} (C+1-F_\nu) d\nu}\,.
\label{eq:ewratio}
\end{equation}

\begin{figure}[!h]
   \centering
   \includegraphics[width=0.9\hsize]{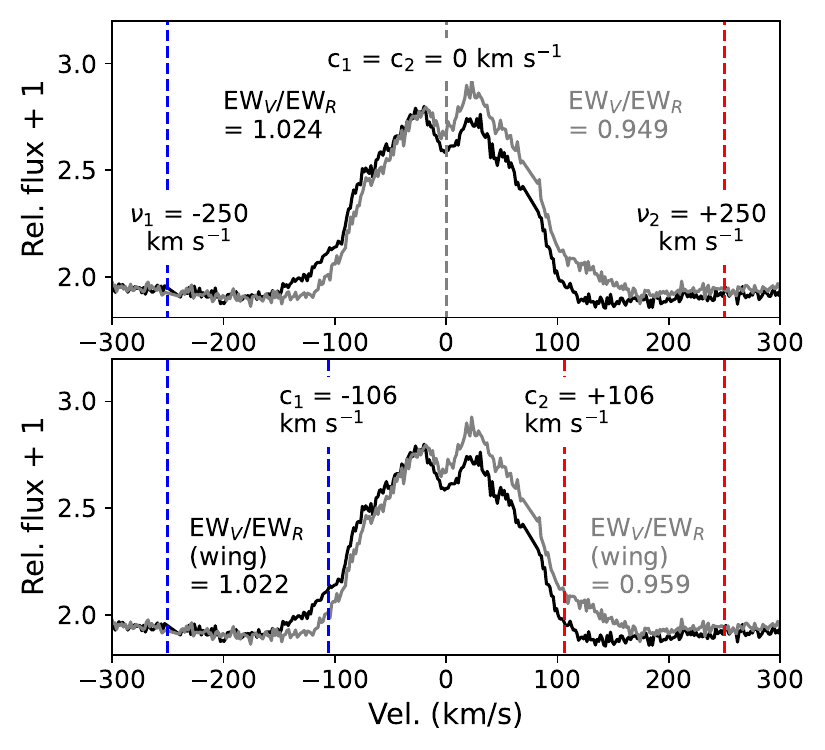}
      \caption{Two H$\alpha$ line profiles taken about 15 hours apart for one of our targets, V767~Cen, in black and gray (with corresponding EW$_{\rm V}$/EW$_{\rm R}$\xspace values printed) to illustrate the procedure for measuring EW$_{\rm V}$/EW$_{\rm R}$\xspace from Eq.~\ref{eq:ewratio}. A constant, $C=1$ is first added to the data. The outer integration limits are set at $\mp$250 km s$^{-1}$ for $\nu_1$ and $\nu_2$. In the top panel, the inner integration limits, $c_1$ and $c_2$, are set at 0 km s$^{-1}$ so that the full line profile is used. In the bottom panel, $c_1$ and $c_2$ are set at $\mp$${v\sin i}$ (106 km s$^{-1}$) so that only the wings are used in the calculation. 
              }
         \label{fig:illustrate_EWvr}
   \end{figure}

$\nu_1$ and $\nu_2$ are the outer integration limits for the calculation, which are chosen on a case-by-case basis for a given star based on the width of the emission features. $c_1$ and $c_2$ are the inner integration limits, which are both set to the line center (0 km s$^{-1}$ in velocity units) when calculating EW$_{\rm V}$/EW$_{\rm R}$\xspace for the full line profile, or are set to $\mp {v\sin i}$ when computing EW$_{\rm V}$/EW$_{\rm R}$\xspace for just the line wings, which generally trace higher-velocity material closer to the star (see Sect.~\ref{sec:distinguishing}). $F_\nu$ is the continuum normalized flux. The constant C is introduced because (parts of) a line can transition from net absorption to net emission, thus changing sign. In all measurements of EW$_{\rm V}$/EW$_{\rm R}$\xspace in this work, we set $C=1$ to guarantee that no changes in sign or division by near-zero values occurred. This is purely for mathematical convenience. 
Fig.~\ref{fig:illustrate_EWvr} illustrates the procedure for measuring EW$_{\rm V}$/EW$_{\rm R}$\xspace using the full H$\alpha$ emission line and only the wings according to Eq.~\ref{eq:ewratio}.

In some cases, there are sufficient observations of a system in a disk-less state, and then the photospheric line profile can be subtracted from each observation, yielding emission spectra. These can be more precise trackers of disk dynamics. However, for consistency, all reported measurements of EW and EW$_{\rm V}$/EW$_{\rm R}$\xspace were made without subtracting the photospheric line profiles (which are not available for all cases).

\subsection{Distinguishing between stellar and circumstellar signals}
\label{sec:distinguishing}

A complicating factor in measurements of the EW$_{\rm V}$/EW$_{\rm R}$\xspace (and V/R) asymmetry is that all Be stars are nonradial pulsators, and line profile variations (LPVs) due to pulsation will also manifest in line asymmetries irrespective of any (changing) emission. However, pulsational LPVs generally introduce weaker asymmetries compared to variations across the line profile due to emission. A major difference is that the dominant pulsational LPVs for these stars are apparently stable and coherent over the entire observing baseline, while the variations in emission are not.

For most of our sample, the spectroscopic data were sufficient in observing baseline, quantity, and quality to determine one or more pulsational frequencies via time-series analysis of the line profiles of photospheric absorption lines (e.g., \ion{He}{I}\,$\lambda$4388, \ion{C}{II}\,$\lambda$4267, \ion{Si}{III}\,$\lambda$4553, etc.). Although emission can appear in these lines in certain systems (e.g.,  with exceptionally hot and/or dense disks), one or more photospheric lines were generally emission-free such that they were suitable for detecting pulsational signatures. It should be noted that the presence of emission in a line does not necessarily prohibit detecting pulsation \citep[e.g.,][]{2020MNRAS.494..958N}. More in-depth analysis of the pulsational properties of the sample will be performed in subsequent works. However, the main point relevant to the message of this paper is that, for the majority of the sample, we identified one to a few pulsational frequencies. In all cases, these pulsational frequencies remained coherent throughout the observing baseline, their variations were confined to a velocity range slightly higher than, or within, literature values of ${v \sin i}$ \citep[mostly taken from][]{zorec2016}, and there are corresponding photometric frequencies in TESS (with one exception -- $\kappa$ CMa). 

Conversely, the EW$_{\rm V}$/EW$_{\rm R}$\xspace cycles measured from emission lines only exist during parts of the observing window, namely during outbursts. The asymmetry variations in a given emission line generally present much higher amplitude than the pulsational variations. Measurements of the EW$_{\rm V}$/EW$_{\rm R}$\xspace ratios of just the high-velocity H$\alpha$ wings (i.e., restricting the velocity range to only include the part of the line that is outside ${v \sin i}$, which should not be significantly contaminated by pulsation) reveal the same EW$_{\rm V}$/EW$_{\rm R}$\xspace cycles as measurements of EW$_{\rm V}$/EW$_{\rm R}$\xspace across the full line profile. An example of distinguishing between pulsational and circumstellar variability is provided for V767~Cen in Sect.~\ref{sec:V767_Cen}. A similar analysis was done for all systems in the sample having sufficient echelle spectroscopy, with essentially the same results.

\subsection{Determining the period of rapid EW$_{\rm V}$/EW$_{\rm R}$\xspace cycles} \label{sec:EWr_fit}

In order to determine the period or approximate cycle lengths of the emission asymmetry oscillations, the following function was fit to the EW$_{\rm V}$/EW$_{\rm R}$\xspace or V/R measurements:

\begin{equation} \label{eq:1}
y(t) = A + (B + C  t) \sin \left( 2 \pi  \frac{t - F}{D + E t} \right ) \,.
\end{equation}

\noindent This is simply a sinusoidal function where the amplitude and period are allowed to vary linearly with time. $A$ is a constant offset, $B$ is the amplitude, $C$ the rate of change in amplitude over time, $D$ is the period, $E$ the rate of change in the period, and $F$ is a phase term. Fits were performed by letting $E$ be a free (but reasonably constrained) parameter, or by setting $E = 0$ (i.e., not allowing the period to change over time). This is preferred over a standard frequency analysis (e.g., a Fourier periodogram), since changes in the amplitude and frequency of EW$_{\rm V}$/EW$_{\rm R}$\xspace can lead to broader and/or multiple frequency peaks. After applying this to all measured events, we found no improvement in the quality of the fit by letting the period vary, and so all plotted fits of Eq.~\ref{eq:1} and the determined periods have $E$ fixed at 0. To facilitate comparisons with other quantities, we report the frequencies (the inverse of the period) determined with Eq.~\ref{eq:1} with $E=0$ in Tab.~\ref{tbl:measurements}.

Errors on the frequency are estimated considering both the time baseline and the analytic uncertainty from Eq.~\ref{eq:1}, added in quadrature. The time baseline error term is taken to be one tenth of the inverse of the timespan between the first and last measurement used in the fit, and is usually the larger of the error terms. 

\subsection{Photometric frequencies}
\label{sec:phot}

TESS light curves are sensitive to changes in both the star and the circumstellar environment. As is typical for early-type Be stars, all systems in the sample show many signals in their TESS frequency spectrum. These signals tend to be clustered in ``frequency groups,'' where the typical configuration for a given star includes a group of low frequencies (usually $\lesssim 0.2$\,d$^{-1}$\xspace; hereafter ``$g0$''), a group usually between $\sim 0.5$ and 3\,d$^{-1}$\xspace (hereafter ``$g1$''), and another group ($g2$) located at approximately twice the frequency of $g1$ \citep[e.g.,][]{JLB2022}. There may be additional groups at higher frequencies (but usually still near-integer multiples of $g1$).  The total power of groups can vary over time (both in an absolute sense and relative to other frequency groups in the same star). Furthermore, with the short observing baseline of TESS, many frequencies are likely unresolved and blended, further compromising the accuracy of measured frequencies. 

Major changes to one or more frequency groups are usually seen during or close to an outburst. The typical signature is an increase in power mostly in the low-frequency side of one or more frequency groups. This enhancement begins near the start of the flicker and can persist for several days or longer (but the exact pattern depends on the star and can differ from event to event).

Frequency spectra were calculated from the TESS light curves in the usual fashion, using Period04 \citep{Lenz2005}. In most cases longer-term trends were removed to increase the visibility of higher-frequency signals. Different sections of the light curves can be analyzed to measure changes in the frequency spectrum over time.

\section{Results} \label{sec:results}

\begin{figure*}[!h]
   \centering
   \includegraphics[width=0.33\hsize]{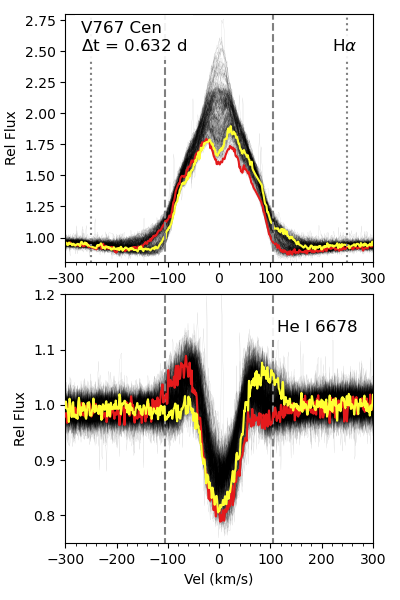}
   \includegraphics[width=0.33\hsize]{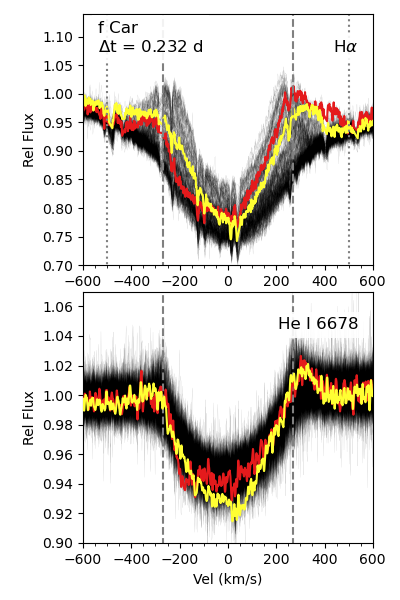}
   \includegraphics[width=0.33\hsize]{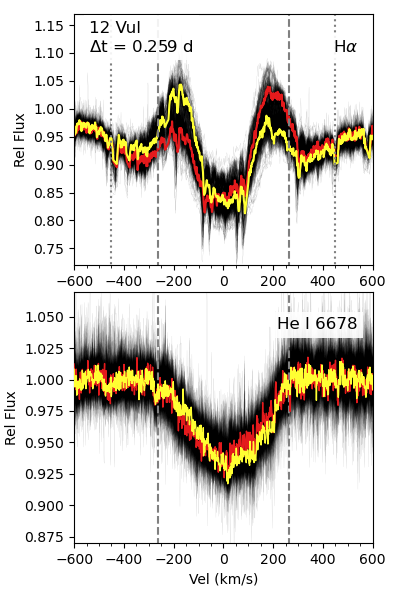}
      \caption{Line profiles of H$\alpha$ and \ion{He}{I}\,$\lambda$6678 for all observations (black) of V767~Cen, f~Car, and 12~Vul. For each star, the red and yellow curves are taken near to the start of a flicker, being separated in time by the amount printed in the top panels (the red curve is the earlier of the two). The dashed vertical lines are the literature ${v\sin i}$ values, and the dotted lines are the integration limits for EW and EW$_{\rm V}$/EW$_{\rm R}$\xspace for H$\alpha$. The $x$ axis for V767~Cen differs from the other two to better see the features of this low inclination system. 
              }
         \label{fig:example_LPFs}
   \end{figure*}

To illustrate the observed behavior, three examples -- V767~Cen, f~Car, and 12~Vul -- are analyzed and discussed in the following subsections. A similar analysis for the remaining 10 stars is presented in Appendix~\ref{sec:appendix_b}. The H$\alpha$ and \ion{He}{I}\,$\lambda$6678 lines for these three stars during the observational baselines are shown in Fig.~\ref{fig:example_LPFs}. Two epochs for each star are emphasized -- these observations were taken very near to the start of a flicker, and are separated by approximately half of the V/R (or EW$_{\rm V}$/EW$_{\rm R}$\xspace) oscillation period (or slightly less than this for f~Car) in order to demonstrate the typical rapid variations seen in the early stages of a flicker.

Purely for convenience, we selected these three stars to represent three qualitatively defined categories of behavior present in the larger sample. Some systems have a preexisting disk, and also have well-defined photometric and spectroscopic flicker signatures -- this is represented by V767~Cen. Some systems have no preexisting disk, but they do have flickers that are clearly delineated by their signals in both photometry and spectroscopy -- f~Car is in this class. Finally, in some cases, the photometric and/or spectroscopic variations are complex and it is difficult to pinpoint any discrete events, yet it is still evident that mass ejection activity occurs over some range in time (due to the presence of rapid changes in emission strength and asymmetry oscillations, secular changes in the continuum brightness, and enhancements in photometric frequency groups) -- 12~Vul is one such example. Nevertheless, all stars in our sample exhibit rapid V/R oscillations with a characteristic frequency during outbursts.

\subsection{V767~Cen (HD 120991; B2Ve)} \label{sec:V767_Cen}

Historically, the H$\alpha$ emission of V767 Cen is usually fairly strong, mostly single-peaked, and subject to variations on short and long timescales. Radial velocity (RV) motion indicative of binarity was not detected by \citet{2022MNRAS.510.2286N}, but only four RVs were measured and the low inclination angle also would make the RV shifts relatively small should there be a binary companion. V767~Cen is a $\gamma$ Cas analog, with hard X-rays \citep{2018A&A...619A.148N,2022MNRAS.512.1648N}. \citet{2003A&A...411..229R} detected photospheric line profile variations indicative of pulsation, but with insufficient sampling to recover a period. \citet{zorec2016} found ${v\sin i}$ = 106 $\pm$ 11 km s$^{-1}$ and $i = 25^{\circ} \pm 7^{\circ}$ for V767~Cen. In TESS photometry, \citet{2020MNRAS.498.3171N} noted long-term trends, the presence of frequency groups with variable amplitudes, and low-frequency red noise.

\begin{figure*}[h]
   \centering
   \includegraphics[width=0.95\hsize]{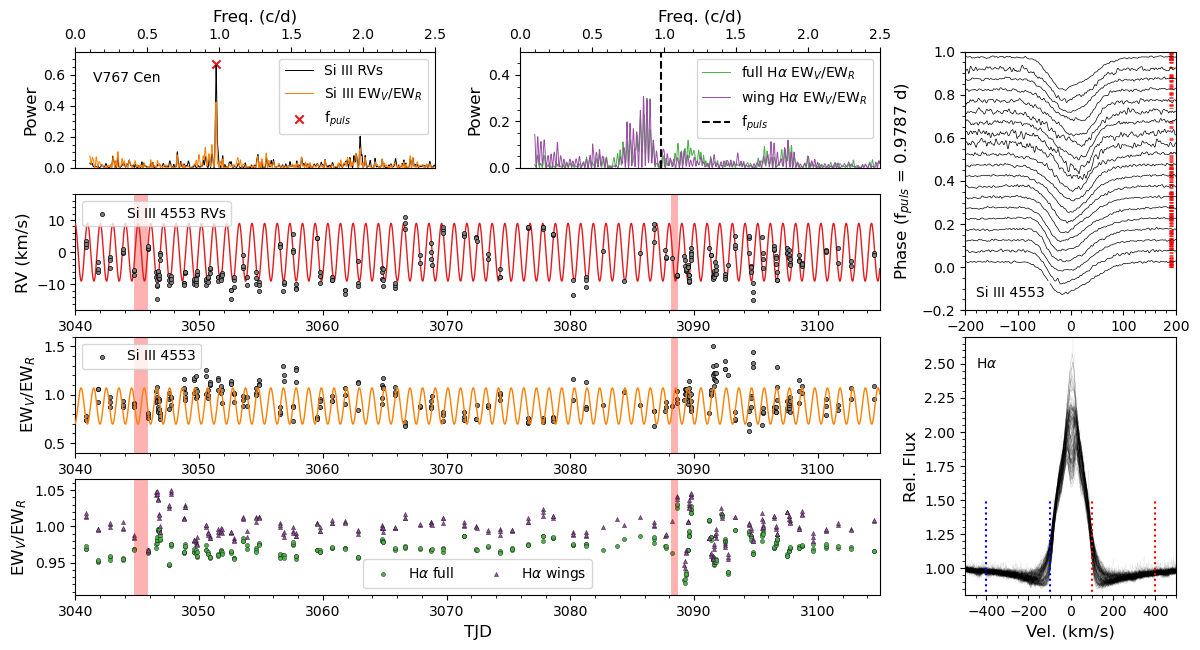}
      \caption{Line profiles and measurements of spectroscopic quantities for V767~Cen, illustrating the different character of pulsational variability versus variations in emission features. Top-left panel: Power spectrum determined from measurements of the \ion{Si}{III}\,$\lambda$4553 line, with black computed from RV measurements, and orange from the EW$_{\rm V}$/EW$_{\rm R}$\xspace ratio. The strongest peak, marked with a red ``x,'' is taken to be the dominant pulsation frequency. Top-middle panel:  Power spectrum computed from the H$\alpha$ EW$_{\rm V}$/EW$_{\rm R}$\xspace measurements using the full line profile (green) and the high-velocity wings (outside ${v\sin i}$; purple). Top-right panel:  \ion{Si}{III}\,$\lambda$4553 line phased to the dominant pulsation period (with 20 bins in phase). Bottom-right panel: H$\alpha$ line profiles where the inner set of vertical dotted lines mark $\pm{v\sin i}$, and the outer set marks the limits for the EW calculations. Three remaining left panels, from top to bottom:  \ion{Si}{III}\,$\lambda$4553 RVs, \ion{Si}{III}\,$\lambda$4553 EW$_{\rm V}$/EW$_{\rm R}$\xspace measurements, and  H$\alpha$ EW$_{\rm V}$/EW$_{\rm R}$\xspace measurements. Both the panels for \ion{Si}{III}\,$\lambda$4553 include the best-fit sinusoid at the frequency marked in the top-left panel. The vertical shaded regions in the bottom three left panels indicate the approximate start of the two clearest mass ejection events. 
              }
         \label{fig:V767_Cen_puls}
   \end{figure*}

Figure~\ref{fig:V767_Cen_puls} presents an analysis showing how photospheric versus circumstellar signals are distinguishable (Sect.~\ref{sec:distinguishing}). A skewed Gaussian (a Gaussian plus a linear term) was fit to each \ion{Si}{III}\,$\lambda$4553 observation, and the RV of the minimum of the absorption profile was determined. A frequency analysis of this RV time series shows a strong peak at a frequency of $f_{\rm puls}$ = 0.97866 \,d$^{-1}$\xspace (top left panel of Fig.~\ref{fig:V767_Cen_puls}), which we consider the dominant pulsation signal. The line profiles phased to this period (top right panel in Fig.~\ref{fig:V767_Cen_puls}) are typical of Be pulsators at low inclination angles \citep{2003A&A...411..229R, 2021MNRAS.508.2002R}, and a sinusoid at this period fits the measured RVs well (2nd panel on the left in Fig.~\ref{fig:V767_Cen_puls}). The EW$_{\rm V}$/EW$_{\rm R}$\xspace values of \ion{Si}{III}\,$\lambda$4553 are shown below the RVs in Fig.~\ref{fig:V767_Cen_puls}, with a frequency analysis finding the same frequency as with the RVs but with more scatter. 

The bottom panel of Fig.~\ref{fig:V767_Cen_puls} displays the EW$_{\rm V}$/EW$_{\rm R}$\xspace values for H$\alpha$ computed from the full line profile (out to $\pm$400 km s$^{-1}$) and from the wings (between $\pm{v\sin i}$ = 106\,km\,s$^{-1}$ and $\pm$400\,km\,s$^{-1}$). A frequency analysis of these H$\alpha$ quantities does not exhibit any peak at $f_{\rm puls}$ (top middle panel in Fig.~\ref{fig:V767_Cen_puls}), demonstrating that pulsation does not noticeably contribute to the variations we measure in H$\alpha$ emission. 
Comparing the EW$_{\rm V}$/EW$_{\rm R}$\xspace oscillations in the H$\alpha$ wings versus the whole line profile shows consistent periods but slightly different phases.

   \begin{figure*}[h!]
   \centering
   \includegraphics[width=0.94\hsize]{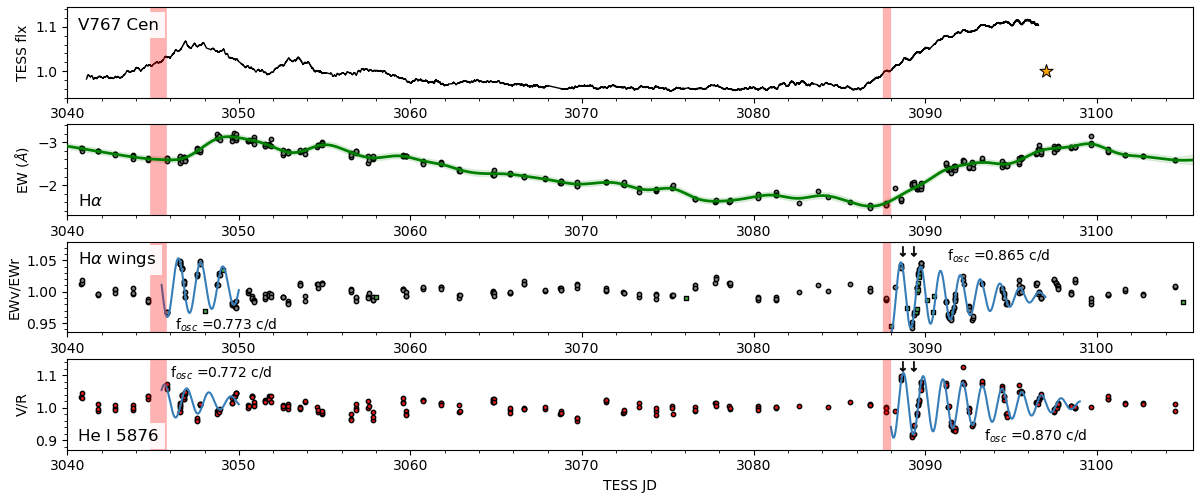}
      \caption{Observations of the Be star V767~Cen, showing two consecutive sectors of TESS photometry (first panel), H$\alpha$ EW with a Gaussian process regression (green curve) fit to the data (measured between $\pm$400 km s$^{-1}$, second panel),  H$\alpha$ EW$_{\rm V}$/EW$_{\rm R}$\xspace measurements using the emission wings (black circles for professional spectra, green squares for amateur spectra, third panel),
      and the V/R ratio of the two emission peaks of \ion{He}{I}\,$\lambda$5876 (4th panel). The fit of Eq.~\ref{eq:1} during the EW$_{\rm V}$/EW$_{\rm R}$\xspace and V/R oscillations are shown in the bottom two panels, with the corresponding frequency printed. The vertical red bars mark the approximate beginning of two flickers based on the first sign of emission asymmetry. The two arrows near TJD 3089 in the bottom two panels are the two epochs emphasized in Fig.~\ref{fig:example_LPFs}. The yellow star in the top panel marks the epoch of the X-ray observation (Appendix~\ref{sec:Xrays}). 
              }
         \label{fig:V767_Cen}
   \end{figure*}

Figure~\ref{fig:V767_Cen} features the two consecutive sectors (64 and 65) of TESS photometry and measurements of H$\alpha$ (EW and EW$_{\rm V}$/EW$_{\rm R}$\xspace) and \ion{He}{I}\,$\lambda$5876 (V/R). A Gaussian process regression was fit to the EW measurements to improve the visibility of their variations. The first indication of mass ejection activity begins near TJD 3045 and appears to be three small flickers in quick succession with decreasing photometric amplitudes. After $\sim$25 days of apparent quiescence (from TJD 3060 -- 3086), during which the H$\alpha$ emission strength steadily decreased, a new flicker started. In photometry, this event began at TJD 3086 and the brightness rose steadily until reaching a maximum just at the end of the TESS light curve. 
\ion{He}{I}\,$\lambda$6678 and \ion{He}{I}\,$\lambda$5876 showed essentially the same behavior, but the emission peaks were better defined in \ion{He}{I}\,$\lambda$5876. 

The higher frequency photometric variations in V767~Cen are much lower in amplitude compared to the flicker signatures. The photometric start of the flickers is therefore easier to pinpoint. Interestingly, the typical spectroscopic markers of a flicker seem to begin $\sim$1 day after the brightness starts to increase. The initial emission asymmetry and its oscillations are obvious as the emission grows during the beginning of the events. The oscillation frequencies during the two early outburst phases fit with Eq.~\ref{eq:1} are 0.773 $\pm$ 0.025 \,d$^{-1}$\xspace and 0.865 $\pm$ 0.012 \,d$^{-1}$\xspace for H$\alpha$ EW$_{\rm V}$/EW$_{\rm R}$\xspace (determined from the emission wings), and 0.772 $\pm$ 0.033 \,d$^{-1}$\xspace and 0.870 $\pm$ 0.010 \,d$^{-1}$\xspace for the V/R measurements of \ion{He}{I}\,$\lambda$5876 (as illustrated in Fig.~\ref{fig:V767_Cen}). The emission asymmetry frequencies are the same (within uncertainties) for H$\alpha$ and \ion{He}{I}\,$\lambda$5876 for a given event, but differ by about three sigma between the two events, and are about 15\% to 25\% slower than the dominant spectroscopic pulsation frequency ($f_{\rm puls}$ = 0.9787 \,d$^{-1}$\xspace).

For the first flicker, the asymmetry oscillations last for about three cycles until the approximate start of the next small flicker whereafter a continued fit of Eq.~\ref{eq:1} can no longer describe the data. For the flicker beginning near TESS JD 3088, the emission asymmetry oscillations are relatively coherent for six or seven cycles in H$\alpha$, and eight or nine cycles in \ion{He}{I}\,$\lambda$5876. All associated signals for this latter flicker are higher amplitude and longer lasting compared to earlier in the dataset. The two emphasized epochs in Fig.~\ref{fig:example_LPFs} correspond to the pair of downward arrows in Fig.~\ref{fig:V767_Cen} (near TJD 3089), at near opposite points in the first asymmetry cycle of the second flicker. An X-ray observation was taken with the Swift X-ray telescope near the end of the TESS observations (TJD 3097) at peak brightness -- these data are discussed in Appendix~\ref{sec:Xrays}.

\subsection{f~Car (HD 75311; B3Vne)} \label{sec:f_Car}

   \begin{figure*}
   \centering
   \includegraphics[width=1.0\hsize]{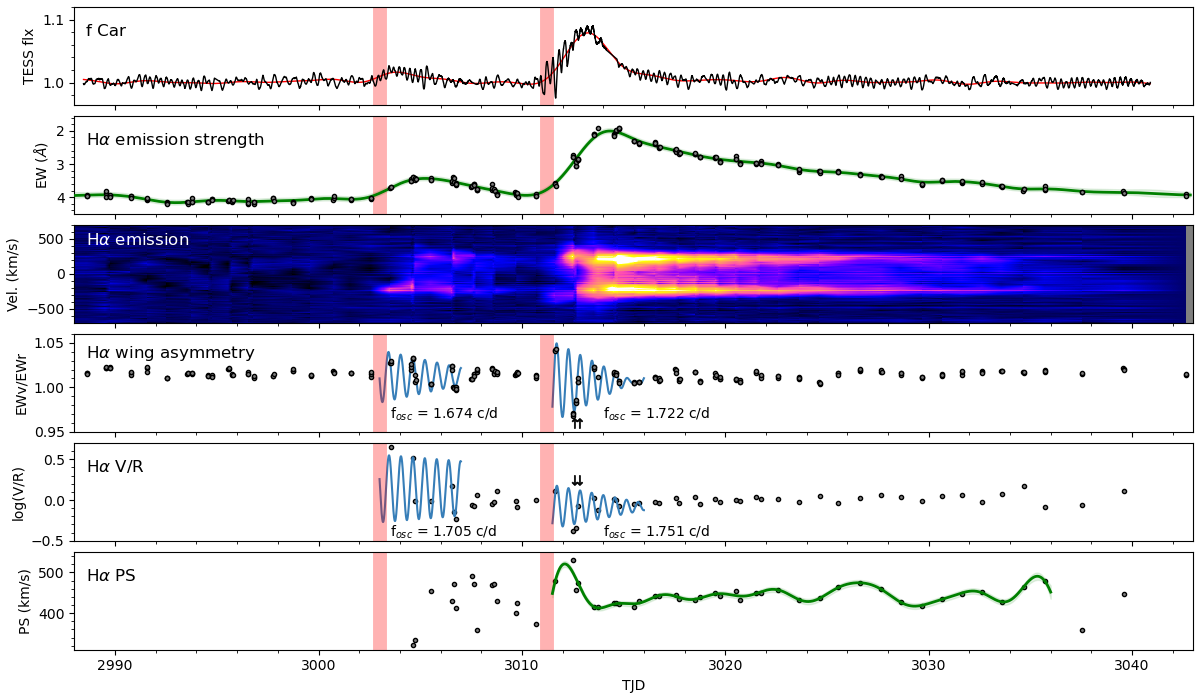}
      \caption{
      Observations of the Be star f~Car. Panels from top to bottom: TESS light curve (black), H$\alpha$ EW, dynamical spectrum of H$\alpha$ emission (after subtracting the photospheric profile), EW$_{\rm V}$/EW$_{\rm R}$\xspace of H$\alpha$, V/R and then peak separation (PS) of H$\alpha$ emission (with the photospheric profile subtracted). The start of two outburst events are indicated by vertical shaded rectangles. The red curve in the top panel traces the low-frequency signals (below 0.5 d$^{-1}$). The solid green curve in second and sixth panels are Gaussian Process Regression fits to (sections of) the measurements. The fit of Eq.~\ref{eq:1} is plotted for the first few days of the two flickers for EW$_{\rm V}$/EW$_{\rm R}$\xspace and V/R, with the corresponding frequencies indicated. The two downward arrows near TJD 3013 are the two epochs emphasized in Fig.~\ref{fig:example_LPFs}. 
              }
         \label{fig:f_Car}
   \end{figure*}

f Car (B3Vne) is a little studied Be star. Its ${v\sin i}$ is estimated to be $\approx$250 km s$^{-1}$ \citep{2022MNRAS.511.4404S} or 268 $\pm$ 18 km s$^{-1}$ and with an inclination angle of $i = 66^{\circ} \pm 16^{\circ}$ \citep{zorec2016}. This system is a useful example since it initially had no detectable disk until flickers occurred about two weeks into the TESS observing baseline, and so the growth of the new disk happened in a pristine environment.

Measurements from TESS and from contemporaneous spectroscopy of f~Car over a $\sim$50 day period are shown in Fig.~\ref{fig:f_Car}. TESS monitored f~Car for two consecutive sectors (62, 63), and the spectroscopic observations started about one week before the TESS observations, and ended a few days after. Two flickers occurred during this timespan which had similar behavior, but differed in their amplitude and duration. The V/R and EW$_{\rm V}$/EW$_{\rm R}$\xspace oscillations lasted for about 5 -- 6 cycles in H$\alpha$, with frequencies of about 1.7 \,d$^{-1}$\xspace. The TESS brightness and H$\alpha$ emission began to increase at approximately the same time, but H$\alpha$ evolved relatively slowly, especially the dissipation phase which was significantly longer than in photometry. The main difference is that for the first flicker, all new emission signatures essentially vanished $\sim$8 days after the start of the event, while for the second emission persisted for nearly 30 days. For the first flicker, the H$\alpha$ asymmetry oscillations had frequencies of 1.674 $\pm$ 0.036\,d$^{-1}$\xspace (EW$_{\rm V}$/EW$_{\rm R}$\xspace) and 1.705 $\pm$ 0.081\,d$^{-1}$\xspace (V/R), and for the second flicker the frequencies were 1.722 $\pm$ 0.032\,d$^{-1}$\xspace (EW$_{\rm V}$/EW$_{\rm R}$\xspace) and 1.751 $\pm$ 0.050\,d$^{-1}$\xspace (V/R).

\begin{figure*}
   \centering
   \includegraphics[width=1.0\hsize]{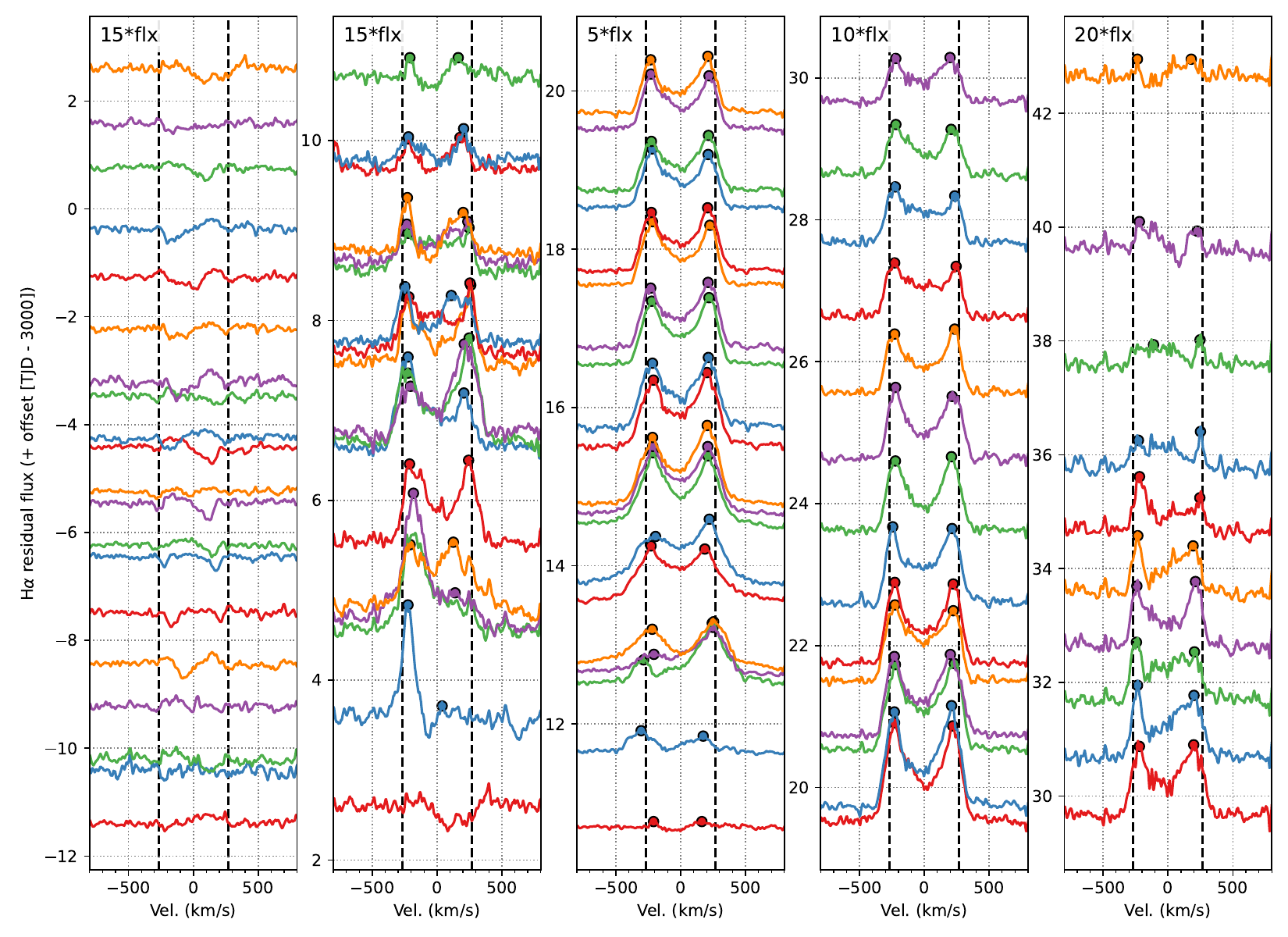}
      \caption{Emission H$\alpha$ line profiles for f~Car (after subtracting the average observed photospheric line profile) over the full time baseline shown in Fig.~\ref{fig:f_Car}. In each panel, time increases upward with the $y$-axis label corresponding to the TJD dates minus 3000 in Fig.~\ref{fig:f_Car}. In each panel, the flux is scaled by the amount shown in the upper left corner, to make features more visible. The $v \sin i = 268\rm\,km\,s^{-1}$ from \citet{zorec2016} is indicated by dashed black lines. The first panel is prior to the first flicker, the second panel shows the first smaller flicker, the third shows the first $\sim$10 days of the next flicker, and the last two panels show the dissipation phase. Solid circles are plotted at the emission peaks, which were used to determine V/R and PS when emission was present. 
              }
         \label{fig:f_Car_LPFs}
   \end{figure*}

Since f~Car had no emission for the first $\sim$2 weeks of the high cadence spectroscopic monitoring, the purely photospheric H$\alpha$ spectrum was determined from an average of the spectra taken before the first flicker. This photospheric spectrum was subtracted from all H$\alpha$ observations to provide the pure emission profiles. These are shown in Fig.~\ref{fig:f_Car_LPFs} for H$\alpha$, with solid circles indicating the intensity and radial velocity of the pure emission peaks (when applicable). This allowed for the measurements of H$\alpha$ V/R and peak separation (PS) values plotted in the bottom two panels of Fig.~\ref{fig:f_Car}. However, only at some epochs are there two well-defined peaks. The low amplitude variations within $\pm{v\sin i}$ prior to the first flicker (first column in Fig.~\ref{fig:f_Car_LPFs}) are consistent with pulsation.

The earliest emission profile sampling the first event (the second curve from the bottom in the second panel in Fig.~\ref{fig:f_Car_LPFs}) shows narrow emission only on the blue-shifted side of the line -- an obvious marker of the initial asymmetry of the ejecta. By the time of the next spectrum ($\sim$1 day later), there was emission on both sides of the line but with clearly unequal intensity, with the asymmetry oscillating in the following days. Concerning the H$\alpha$ PS, there does not seem to be any coherent pattern during the first event, likely due to its weakness and short duration.

For the second event, the earliest spectrum (the second curve from the bottom in the third panel in Fig.~\ref{fig:f_Car_LPFs}) has emission that appears nearly symmetric, but blue-shifted by $\sim$100 km~s$^{-1}$ (and so EW$_{\rm V}$/EW$_{\rm R}$\xspace $> 1$). The higher-cadence spectra taken in the following days show quickly varying asymmetry. The PS was initially high, but quickly dropped to a minimum about three days after the first appearance of emission.  Within one or two days after maximum H$\alpha$ EW, the emission is essentially symmetric. As the dissipation phase progresses, the slope between the continuum level and the emission peaks becomes steeper and the PS generally increases, both indicating inside-out dissipation (Sect.~\ref{sec:VDD}). Interestingly, the emission morphology after TJD 3030 (nearly 20 days after the start of the event, right-most panel in Fig.~\ref{fig:f_Car_LPFs}), becomes asymmetric and varies in its morphology despite the apparent lack of any new mass ejection. This is not obvious from the EW$_{\rm V}$/EW$_{\rm R}$\xspace or V/R values, but is clear from the line profiles. At this late stage in the dissipation phase, the emission is small and it is possible that pulsation is the cause of any apparent underlying variations (as in the first column of Fig.~\ref{fig:f_Car_LPFs} prior to the flickers).

\subsection{12~Vul (HD 187811; B2.5Ve)} \label{sec:12_Vul}

12 Vul (B2.5Ve) was found to have an inclination angle of $i = 60^{\circ} \pm 5^{\circ}$ \citep[using H$\alpha$ emission;][]{2023ApJ...948...34S} and $i = 52^{\circ} \pm 13^{\circ}$ with a ${v\sin i}$ = 264 $\pm$ 25\,km\,s$^{-1}$ \citep[using gravity darkened absorption lines;][]{zorec2016}. 12 Vul has experienced several emission-free phases (e.g,. between 2014 and 2019), and has had ``recurrent short-lived outbursts'' as sampled by Hipparcos photometry \citep{1998A&A...335..565H}. This is in agreement with the three available sectors of TESS data, showing short events with about 5 -- 15 d timescales. Molecular $^{12}$CO was detected in emission in a single observation of 12~Vul \citep{2021A&A...647A.164C}, which has never been reported in any other classical Be star, and is a feature associated with the B[e] class of stars. There does not seem to be any direct evidence for binarity in the literature, although the detection of $^{12}$CO emission could originate in a cool evolved companion. However, no companion was detected in three infrared interferometric observations taken at the Center for High Angular Resolution Astronomy (CHARA) array \citep{2024ApJ...962...70K}.

\begin{figure*}
   \centering
   \includegraphics[width=1.0\hsize]{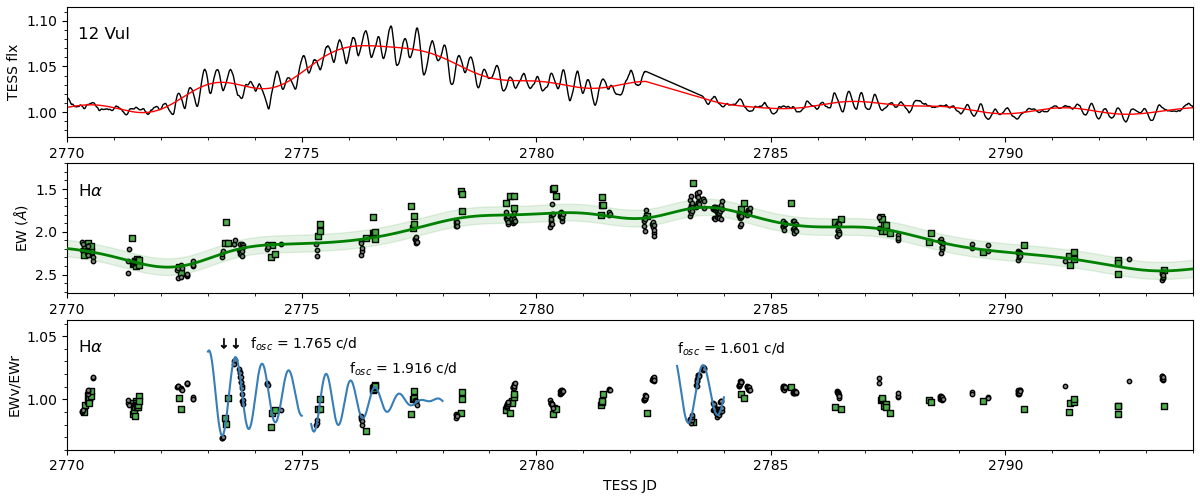}
      \caption{Observations of the Be star 12 Vul, similar to Fig.~\ref{fig:V767_Cen}. The two downward arrows near TJD 2773 are the two epochs emphasized in Fig.~\ref{fig:example_LPFs}. Green squares are measurements from amateur observations, and gray circles from professional data.
              }
         \label{fig:12_Vul}
   \end{figure*}

Figure~\ref{fig:12_Vul} shows the single-sector (54) TESS light curve and H$\alpha$ EW and EW$_{\rm V}$/EW$_{\rm R}$\xspace spanning about 25 days. Tens of spectra were taken in the $\sim$30 days prior to the TESS observations, showing weak and variable emission (most likely due to at least a few mass ejection events). Compared to f~Car and V767~Cen, individual events are less clearly delineated, but variations in photometry, emission strength, and emission asymmetry are still evident. The EW was measured inside of $\pm$450 km s$^{-1}$. Three sections of the EW$_{\rm V}$/EW$_{\rm R}$\xspace measurements are fit with Eq.~\ref{eq:1} resulting in frequencies of 1.765 $\pm$ 0.083 \,d$^{-1}$\xspace, 1.916 $\pm$ 0.069 \,d$^{-1}$\xspace, and 1.60 $\pm$ 0.15 \,d$^{-1}$\xspace. These epochs coincided with an increase in emission strength and/or in brightness (although the third timespan is poorly covered by TESS due to a gap in the middle of the sector). There are additional variations in EW$_{\rm V}$/EW$_{\rm R}$\xspace at other times, but extending the timespan covered by Eq.~\ref{eq:1} resulted in worse fits (i.e., the variation outside of the fit intervals is not phase coherent). 

Overall, the variations in 12~Vul seem qualitatively the same as in f~Car and V767~Cen, at least in terms of rapid emission asymmetry cycles appearing at the same time as increased emission and brightness. The main difference is that the notion of discrete events is more difficult to apply to 12~Vul, perhaps due to more quasi-continuous (but still variable) mass loss, and/or multiple  events closely spaced in time.

\subsection{Remainder of the sample}

The remaining 10 stars in our sample all display similar behavior as the above three cases of V767~Cen, f~Car, and 12~Vul. In particular, new emission is always initially asymmetric, and the asymmetry oscillates for several cycles. Each system has a characteristic frequency for its EW$_{\rm V}$/EW$_{\rm R}$\xspace oscillations, which can differ slightly from event to event. For example, for V442~And (Appendix~\ref{sec:V442_And}), nine outbursts are sufficiently sampled by H$\alpha$ spectroscopy, and the EW$_{\rm V}$/EW$_{\rm R}$\xspace oscillation frequencies for these are all between 0.332 \,d$^{-1}$\xspace and 0.362 \,d$^{-1}$\xspace (about a three sigma difference for the two most different frequencies, but it is possible our errors are underestimated). These systems are further discussed in Appendix~\ref{sec:appendix_b}. Information about the emission asymmetry cycles for the sample is provided in Tab~\ref{tbl:measurements}.

\subsection{Estimating the characteristic frequencies of each star}
\label{sec:characteristic_freq}

In Tab.~\ref{tbl:measurements}, the listed frequencies for each star include that measured from EW$_{\rm V}$/EW$_{\rm R}$\xspace as well as values derived from photospheric line profile variations. These can be compared to characteristic frequencies of the star, as defined below. The first of these is the orbital frequency at the stellar equator, defined as
\begin{equation}
    f_{\rm orb} = 
    \frac{1}{2\pi}
    \left( 
    \frac{GM}{R_{\rm e}^3}
    \right)^{1/2}
    \,,
\label{eq:orb_freq}
\end{equation}
where $M$ and $R_{\rm e}$ are the mass and equatorial radius of the star, respectively. This expression gives the frequency at which material orbiting just above the star’s surface would rotate if it were in a stable circular orbit.

Another important frequency is the critical rotation frequency, which corresponds to the stellar rotation rate at which centrifugal forces at the equator would balance gravitational forces. Under the Roche approximation for a star undergoing solid-body rotation, it is given by
\begin{equation}
    f_{\rm crit} = 
    \frac{1}{2\pi}
    \left[ 
    \frac{GM}{\left( 1.5R_{\rm p} \right)^3}
    \right]^{1/2}
    \,,
\label{eq:crit_freq}
\end{equation}
where $R_{\rm p}$  is the polar radius, and the factor of  $1.5 R_{\rm p}$  accounts for the fact that, in this approximation, the equatorial radius is 1.5 times the polar radius. Finally, a third frequency of interest is the rotation frequency, which can be written in simple terms as
\begin{equation}
    f_{\rm rot} = \frac{
    V_{\rm rot}
    }{
    2\pi R_{\rm e}
    }\,,
\label{eq:rot_freq}
\end{equation}
with $ V_{\rm rot}$ being the linear speed at the equator.  For a critically rotating star, $f_{\rm rot} = f_{\rm crit} = f_{\rm orb}$, but for subcritical rotation these quantities differ ($f_{\rm rot} < f_{\rm crit} < f_{\rm orb}$).

To calculate the frequencies discussed above, three quantities are required: $M$, $R_{\rm e}$, and $V_{\rm rot}$. Determining the fundamental parameters of Be stars is notoriously difficult due to several challenges. One major complication is the Stoeckley effect \citep{1968MNRAS.140..141S, 2004MNRAS.350..189T}, whereby gravity darkening changes the photospheric line profile making it difficult to measure the rotation rate. Additionally, the presence of a surrounding circumstellar disk introduces further complexity by partially obscuring the star, an effect known as veiling \citep[as discussed in, e.g.,][]{2023A&A...678A..47B}. 

\citet{zorec2016} studied the fundamental parameters of 233 Be stars from spectroscopic and photometric data. From our sample of 13 stars, all except 4 have been analyzed by \citeauthor{zorec2016}. Among the stars from the literature (second part of Tab.~\ref{tbl:sample}), all were investigated in that study. Consequently, we have fundamental parameters for a total of 15 stars in our sample. To obtain the three quantities listed above from the data provided in Table~4 of \citeauthor{zorec2016}, the following procedure was adopted.
\begin{enumerate}
    \item From the $\eta$ parameter defined in Eq.~1 of \citeauthor{zorec2016}, we determine the stellar flattening as $R_{\rm e}/R_{\rm p}= 1 + 0.5\eta$.
    \item The polar radius is estimated from the apparent effective surface gravity of the observed stellar hemisphere and the stellar mass as
    $(G\,M / g_{\rm pnrc})^{1/2}$, where the ``pnrc'' subscript denotes the ``parent nonrotating counterpart parameters'' as described in \citeauthor{zorec2016}.
    \item From the inclination angle and the $(v\sin\,i)_{\rm pnrc}$, corrected by the Stoeckley effect, we obtain $V_{\rm rot}$.
\end{enumerate}

In Tab.~\ref{tbl:zorec} the fundamental frequencies defined above are listed for the subset of Tab.~\ref{tbl:sample} for which data is available in \citet{zorec2016}. Additional stellar parameters (masses, equatorial and polar radii) from \citet{zorec2016} are also listed.

\subsection{Comparison of spectroscopic and photometric frequencies}
\label{sec:spec_vs_phot_freqs}

\begin{figure*}
   \centering
   \includegraphics[width=1\hsize]{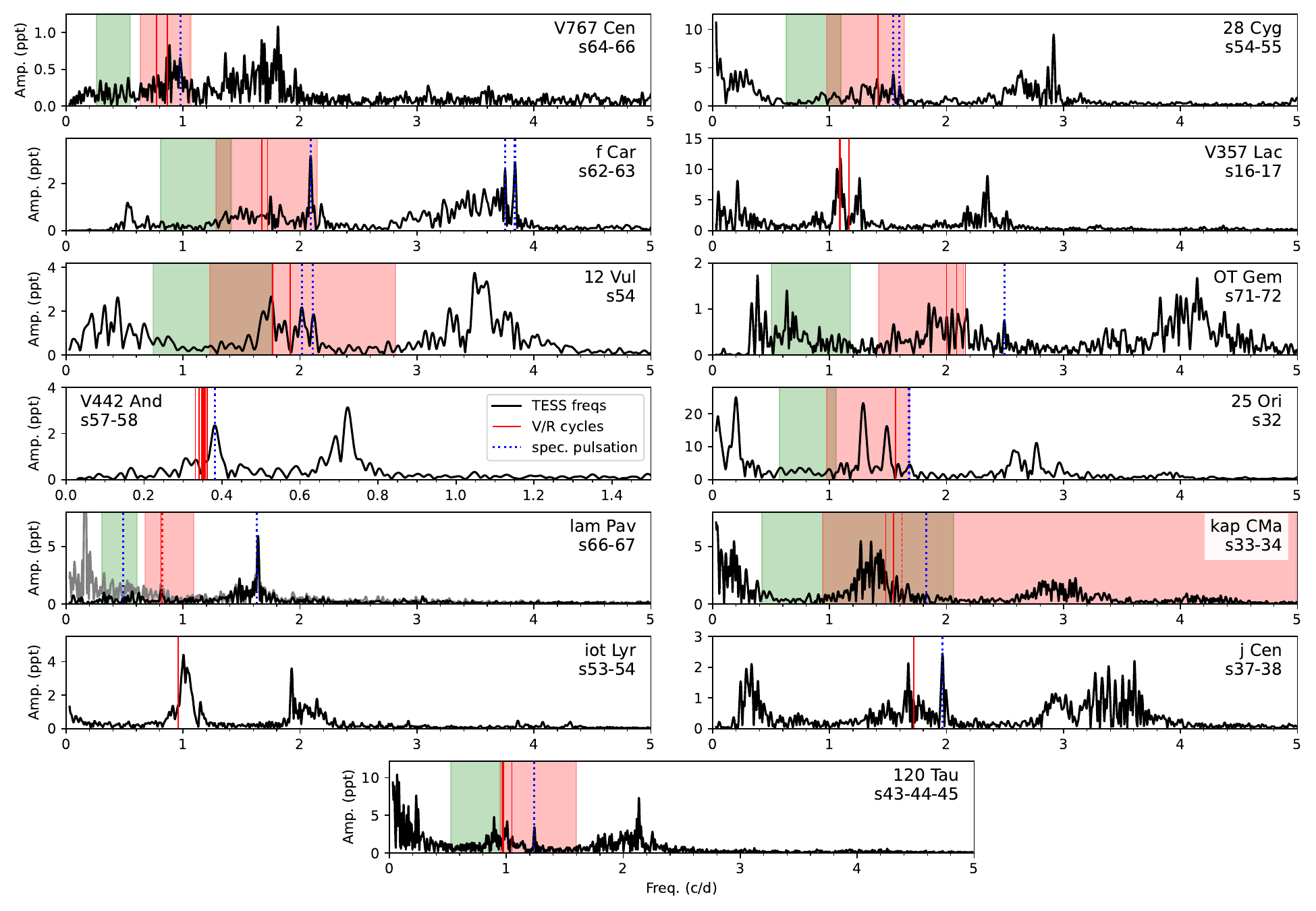}
      \caption{Frequency spectra (black solid curve) from TESS data. The frequencies of the rapid EW$_{\rm V}$/EW$_{\rm R}$\xspace cycles for these stars are indicated by solid red vertical lines, and spectroscopic pulsational frequencies are indicated by dotted blue vertical lines. The green (red) shaded region is the rotation (orbital) frequency range determined from the parameters of \citet{zorec2016}. The red vertical dotted line for $\kappa$\,CMa is the ``secondary frequency'' from \citet{2003A&A...411..229R} that may be circumstellar, and likewise for $\lambda$\,Pav from \citet{2011A&A...533A..75L}.
              }
         \label{fig:FTs_TESS_spec}
   \end{figure*}

\begin{figure}
   \centering
   \includegraphics[width=1\hsize]{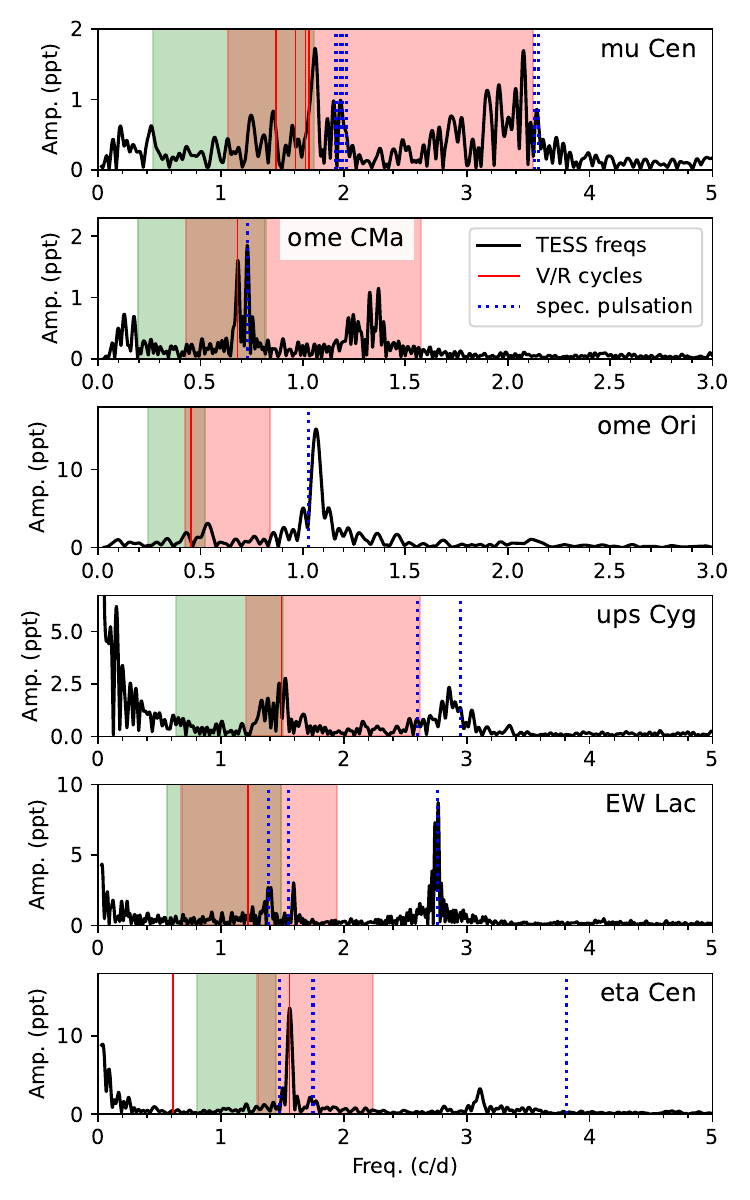}
      \caption{Same as Fig.~\ref{fig:FTs_TESS_spec}, but for stars with V/R cycles found in the literature (see Introduction, Tab.~\ref{tbl:measurements}). 
              }
         \label{fig:FTs_TESS_spec_lit}
   \end{figure}

For each system studied, the measured frequency information (Tab.~\ref{tbl:measurements}) is shown in Fig.~\ref{fig:FTs_TESS_spec}, including the photometric signals from TESS, spectroscopic pulsation frequencies, and the frequencies of the EW$_{\rm V}$/EW$_{\rm R}$\xspace cycles. 
In addition, we plot the rotational frequencies (Eq.~\ref{eq:rot_freq}) as green bands, where the width indicates the 1-$\sigma$ uncertainty.
In all cases, the spectroscopic and photometric data were acquired over the same observing window (except for V442~And, where the TESS observing windows partly cover two of the nine spectroscopically measured events). For most of the sample, our spectra were sufficient to detect one or more pulsation frequencies via line profile variability of photospheric features (as in Fig.~\ref{fig:V767_Cen_puls}). For others, this information was gathered from the literature: V442~And \citep{2021MNRAS.508.2002R}, $\lambda$~Pav \citep{2011A&A...533A..75L}, and 28~Cyg \citep{2000ASPC..214..232T}. The same information is shown in Fig.~\ref{fig:FTs_TESS_spec_lit} for stars with rapid V/R cycles measured in the literature -- in these cases, the spectroscopic quantities (i.e., V/R frequencies and spectroscopically identified pulsation frequencies) were obtained many years prior to the TESS photometry. Note that in $\eta$ Cen and $\omega$ CMa (Fig.~\ref{fig:FTs_TESS_spec_lit}) the indicated V/R frequency may be qualitatively different than the other examples -- it is a ``transient frequency'' that remains phase coherent for long times but still seems to originate in the circumstellar environment. Additionally for $\eta$ Cen, the spectroscopic frequencies reported in \citet{2003A&A...400..599L} and \citet{1995A&A...294..135S, 1998ASPC..135..348S} differ significantly (see Tab.~\ref{tbl:measurements}), with the former frequencies not matching the qualitative patterns seen in the rest of the sample (perhaps the 0.61 d$^{-1}$ circumstellar frequency from \citet{2003A&A...400..599L} is an alias).
 
The TESS frequency information provides a useful backdrop to contextualize the spectroscopic frequencies, and is thus discussed first. 
In most cases the lowest frequency signals in TESS were removed from the data prior to calculating the frequency spectrum for greater visibility of the higher frequency signals. All of these systems have two or more ``frequency groups'' made up of many (often unresolved) peaks (Sect.~\ref{sec:phot}). For example, for V767~Cen, $g1$ is centered at $\sim$0.9 \,d$^{-1}$\xspace, and $g2$ at $\sim$1.7 \,d$^{-1}$\xspace. Such frequency groups, with various amplitudes, are seen in about 85\% of Be stars \citep{JLB2022}. Furthermore, all Be stars in the sample of \citet{JLB2022} that displayed a photometric flicker had one or more frequency groups. 

Regarding the dominant spectroscopic pulsational frequencies, these are typically found on the high-frequency edge of $g1$, and/or $g2$. Usually there is a corresponding photometric signal at the spectroscopically determined pulsational frequency (the most notable exception is $\kappa$ CMa). There are two stars without a spectroscopic pulsational frequency in our sample: V357~Lac and $\iota$~Lyr. For V357~Lac there are clear LPVs consistent with pulsation, but we currently lack sufficient observations of absorption lines to determine a frequency. For $\iota$~Lyr, our data are insufficient for detecting photospheric LPVs. The transient circumstellar V/R and EW$_{\rm V}$/EW$_{\rm R}$\xspace frequencies that occur in the early stages of flickers are generally somewhere within $g1$, and are typically $\sim$10\% to 30\% lower than the nearby spectroscopic pulsational frequency.

The stellar rotation frequencies from Eq.~\ref{eq:rot_freq} are generally below the TESS $g1$ frequency group, but the uncertainties in $f_{\rm rot}$ are large. It is therefore not possible to pinpoint a precise value of $f_{\rm rot}$ for any given system from the \citet{zorec2016} stellar parameters.

\section{Discussion} \label{sec:discussion}

The stage for this study was set already in 1998. \citet{1998ASPC..135..348S} analyzed high-resolution high-cadence echelle spectroscopy for three Be stars -- $\mu$ Cen, $\eta$ Cen, and $\omega$ CMa (see also Fig.~\ref{fig:FTs_TESS_spec_lit}). All of these exhibited transient emission-line oscillations (now called {\v{S}}tefl frequencies) of the same character as the EW$_{\rm V}$/EW$_{\rm R}$\xspace oscillations described in this work. The primary signature of {\v{S}}tefl frequencies are transient rapid EW$_{\rm V}$/EW$_{\rm R}$\xspace and V/R oscillations slightly below the dominant pulsation frequency, originating in the near circumstellar environment during the early outburst stage. However, in the intervening years, only a few other Be stars were the targets of comparable observing campaigns and analysis (see Sect.~\ref{sec:introduction}) but without contemporaneous space photometry. 

\subsection{Summary of key results} \label{sec:discussion_summary}

The analysis of contemporaneous TESS photometry and high-cadence spectroscopy covering 33 events in 13 stars expands on and corroborates results from past studies.  Including the six systems studied in a similar fashion in the literature (Fig.~\ref{fig:FTs_TESS_spec_lit}, Tab.~\ref{tbl:measurements}), 100\% of Be stars observed with sufficiently high cadence during the early stages of a flicker exhibit circumstellar spectroscopic {\v{S}}tefl frequencies in emission lines \citep{1998ASPC..135..348S}, indicating a localized ejection site.
To our knowledge, there are no published exceptions. 

It must be emphasized that these results are specific to relatively short-lived discrete mass ejection events. In some Be stars, the rising phase of outbursts lasts for months or longer \citep[e.g.,][]{2018AJ....155...53L, Bernhard2018, Rimulo2018}. Without high-cadence and longer time-baseline spectroscopy of such events, it is impossible to conclude whether or not the mass is being launched from some localized region. Furthermore, it cannot be determined whether, in these longer events, the disk is fed continuously or through discrete episodes.

The EW$_{\rm V}$/EW$_{\rm R}$\xspace cycle length varied star-by-star, but for those with multiple events the cycle lengths were similar (but not identical) from event-to-event. In all cases the EW$_{\rm V}$/EW$_{\rm R}$\xspace cycles were slower than the ``main'' spectroscopic pulsational frequency usually found on the high frequency side of the first photometric frequency group, $g1$\footnote{Spectroscopic pulsation frequencies are also identified in the second photometric frequency group for some stars.}. However, the EW$_{\rm V}$/EW$_{\rm R}$\xspace cycles do not necessarily correspond to any particular photometric frequency in $g1$. 
In each event, we find only a single EW$_{\rm V}$/EW$_{\rm R}$\xspace frequency for a given emission line.

The EW$_{\rm V}$/EW$_{\rm R}$\xspace cycles usually persist for about five to ten oscillations. In some cases, the amplitude of the EW$_{\rm V}$/EW$_{\rm R}$\xspace cycles is highest at the start of a flicker, and decreases with time. In other cases, the amplitude remains relatively constant until the EW$_{\rm V}$/EW$_{\rm R}$\xspace cycles become incoherent and can no longer be fit with Eq.~\ref{eq:1}. In apparently all cases, allowing the EW$_{\rm V}$/EW$_{\rm R}$\xspace oscillation frequency to linearly vary (i.e., allowing E in Eq.~\ref{eq:1} to be nonzero) did not improve the fit, suggesting that for a given event the EW$_{\rm V}$/EW$_{\rm R}$\xspace frequency is consistent with being constant for the several days over which these cycles occur. Similar EW$_{\rm V}$/EW$_{\rm R}$\xspace patterns with similar cycle lengths are usually seen in multiple emission lines (e.g., H$\alpha$, H$\beta$, \ion{He}{I}\,$\lambda$6678, \ion{Mg}{II}\,$\lambda$4481), with indistinguishable beginning dates but sometimes with different damping timescales (e.g., Fig.~\ref{fig:V767_Cen}).

\subsection{Orbit versus corotation: Ejecta geometry and evolution} \label{sec:orbit_vs_rot}

A qualitative but descriptive picture emerging from our observations is that a Be star ejects a cloud of material from some localized azimuthal region at or near the equator. This outflow may last for only about one to a few days. It is probably not longer than the duration of the rising phase in photometry, but modeling is required to demonstrate this. The ejected cloud then orbits the star with a period that depends on its distance from the star (which may vary slightly from event to event) and the stellar mass and equatorial radius. As it orbits, the cloud gradually expands due to viscosity and any initial velocity dispersion and smears out as a result of orbital phase mixing, until a near-symmetric disk structure is reached after several orbital timescales. This scenario has been proposed for a small number of individual systems studied in this fashion in the literature \citep[e.g.,][]{1998A&A...333..125R, 2002A&A...388..899N, 2000A&A...362.1020F}. A more quantitative description will require models of the initial conditions and evolution of circumstellar material that can be compared to observations. An ongoing effort that combines smoothed-particle hydrodynamics with three-dimensional radiative transfer is underway to model the presented observations more accurately (Rubio et al., in prep.). This approach aims to capture the complex interplay between fluid dynamics and radiative processes, with the end goal of providing a detailed understanding of the underlying physical mechanisms for mass ejection.

   \begin{figure}
   \centering
   \includegraphics[width=01\hsize]{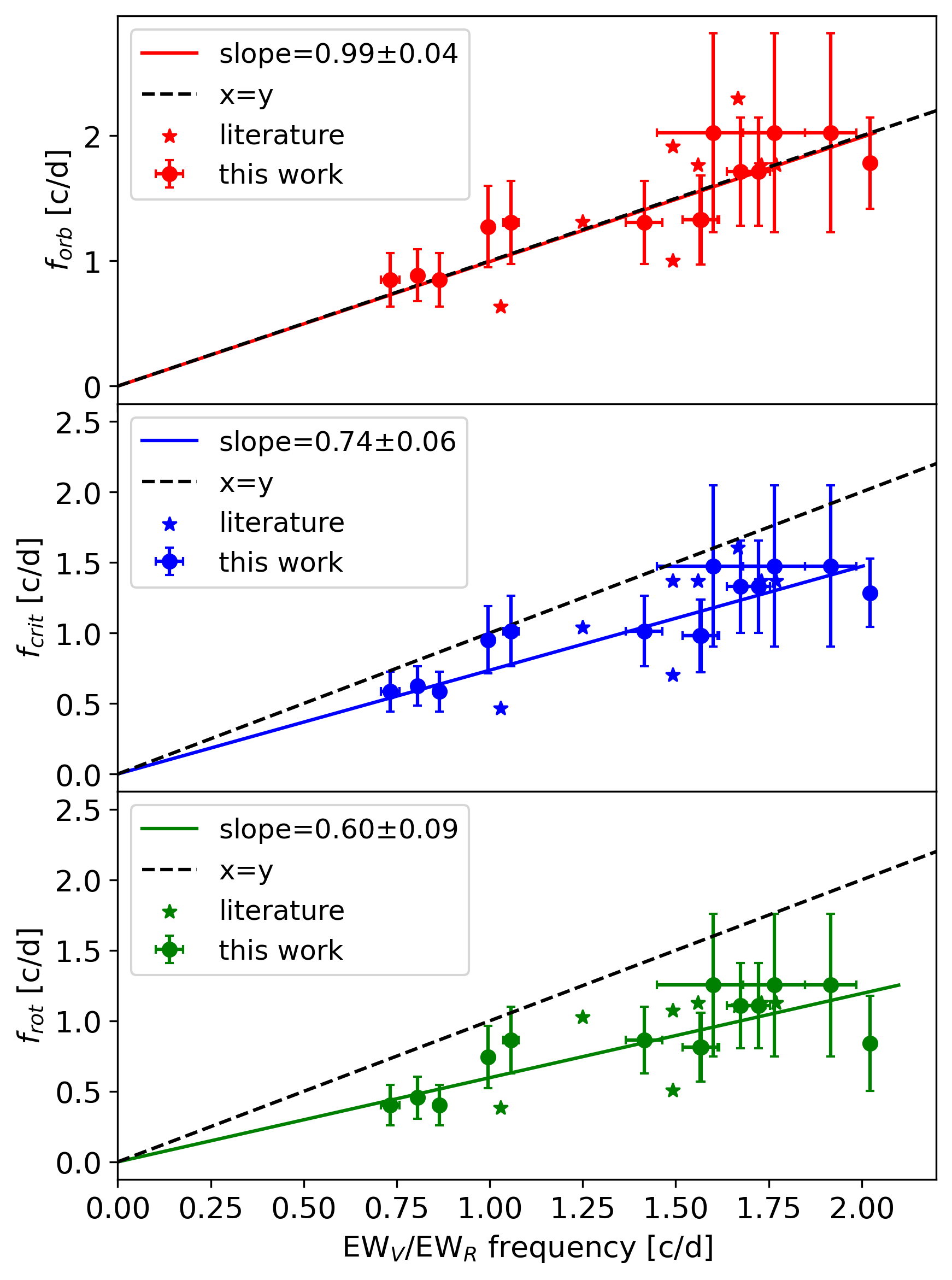}
      \caption{Characteristic stellar frequencies versus EW$_{\rm V}$/EW$_{\rm R}$\xspace frequencies.
      Top: Orbital frequency at the stellar equator (Eq.~\ref{eq:orb_freq}).
      Middle: Critical frequency (Eq.~\ref{eq:crit_freq}).
      Bottom: Rotational frequency (Eq.~\ref{eq:rot_freq}).
      Circles with uncertainties: this work (upper part of Tab.~\ref{tbl:measurements}). 
      Stars: data from the literature (lower part of Tab.~\ref{tbl:measurements}). 
      The solid lines show a linear fit to the points and the dashed line indicates $x=y$. $\kappa$~CMa is omitted, as the derived stellar parameters make it an extreme outlier.    
                            }
         \label{fig:compare_zorec}
   \end{figure}

A rotational explanation of the V/R cycles is disfavored for several reasons. {\v{S}}tefl frequencies can vary slightly from event to event for a given star, but the stellar rotation frequency is a constant \citep[although ${v\sin i}$ may increase by up to 35 km~s$^{-1}$ during mass ejection, ][]{2013A&A...559L...4R}. Corotation would require some anchoring to the star (e.g., via magnetic fields), and in the large majority of Be stars observed with space photometry there is not a harmonic series originating with the stellar rotation frequency as seen in other types of spotted and/or magnetic stars \citep[e.g.,][]{2019MNRAS.487..304D, 2023ApJ...955..123S, 2022MNRAS.516.2812C}. Emission peaks are confined to within, or only slightly outside of, ${v\sin i}$, whereas forced corotation would push emission to higher projected velocities \citep[e.g.,][]{2012MNRAS.419..959O,2015MNRAS.451.2015O}. No large-scale magnetic fields have been observed in any Be star down to a detection limit of about 50 -- 100 Gauss \citep[][in a sample of 85 Be stars]{Wade2016}. Field strengths of only 10 Gauss would significantly disrupt a Be star disk \citep{2018MNRAS.478.3049U}. If there were two confined clouds on opposite sides of the star, as with a dipolar magnetic field as proposed by \citet{1999ApJ...521..407B}, major oscillations in peak separation would dominate over V/R variations. Very rapid solar-like flares (driven by magnetic re-connection) have not been seen in any classical Be star observed with high-cadence space photometry\footnote{The B9/A0Vne star HD~19818 (TIC 207176480) does have solar-like flares, as reported in \citet{2020MNRAS.493.2528B}, but high resolution time-series spectroscopy reveal this to be a close spectroscopic binary with a cool evolved component and not a classical Be star \citep{2020svos.conf..185L}.} (e.g., 2-minute cadence TESS data), nor are solar-like flares observed in X-rays \citep{2018A&A...619A.148N, 2020MNRAS.493.2511N}, and there is no convincing evidence that small-scale magnetic fields play any major role (but neither can these be observationally ruled out). 

In contrast, Fig.~\ref{fig:compare_zorec} provides strong evidence in favor of the orbiting cloud scenario. Each plot compares one of the star’s three characteristic frequencies, as defined in Sect.~\ref{sec:characteristic_freq}, with the observed EW$_{\rm V}$/EW$_{\rm R}$\xspace frequencies. The solid straight lines represent linear fits to each dataset, while the dotted line illustrates a one-to-one relationship between the two quantities plotted. The figure reveals a clear correspondence between the orbital frequency at the stellar equator and the measured line asymmetry frequencies, indicating that these frequencies are closely aligned. In contrast, the critical rotational frequencies shown in the middle panel are systematically lower than the EW$_{\rm V}$/EW$_{\rm R}$\xspace frequencies, while the actual rotational frequencies fall even further below. This pattern suggests that the material responsible for the line asymmetries is orbiting the star at or very near the equatorial surface. These findings also imply that Be stars are subcritical rotators in agreement with other studies \citep[e.g.,][]{2005A&A...440..305F, zorec2016}. If the stars were critical rotators, the critical frequencies would align more closely with the observed line asymmetry frequencies. Thus, the observed differences offer a valuable insight into the rotational properties of Be stars and support that our sample as a whole rotates subcritically (but uncertainties are too large to make strong claims for any star in particular).

\subsection{Flickers in TESS: Amplitudes and timescales} \label{sec:TESS_flicker_amps}

Mass ejection outbursts in photometry have been measured and studied in the literature, usually with timescales ranging from weeks to years. Generally, photometric amplitudes go up to $\sim$0.5 magnitudes in the visible and larger in the infrared, and the dissipation time is on average about twice as long as the build-up time \citep[e.g.,][]{2002A&A...393..887M, 2018AJ....155...53L}. The observed frequency of outbursts and their duration varies significantly, with stars exhibiting between 0 to 20 events per year, and some apparently individual events lasting up to several years \citep[e.g.,][]{2002A&A...393..887M, 2018AJ....155...53L, Bernhard2018, Rimulo2018}. However, studies using ground-based data will likely miss short-lived low-amplitude events, while TESS data is poorly suited for studying events lasting about one month and longer. 

The short-term mass ejection events (of a few days) studied here are qualitatively similar to longer-lasting outbursts. The dissipation times are typically longer than the build-up phase (Tab.~\ref{tbl:measurements}), and there is considerable range in their frequency, duration, and amplitude.

\begin{figure}
   \centering
   \includegraphics[width=1\hsize]{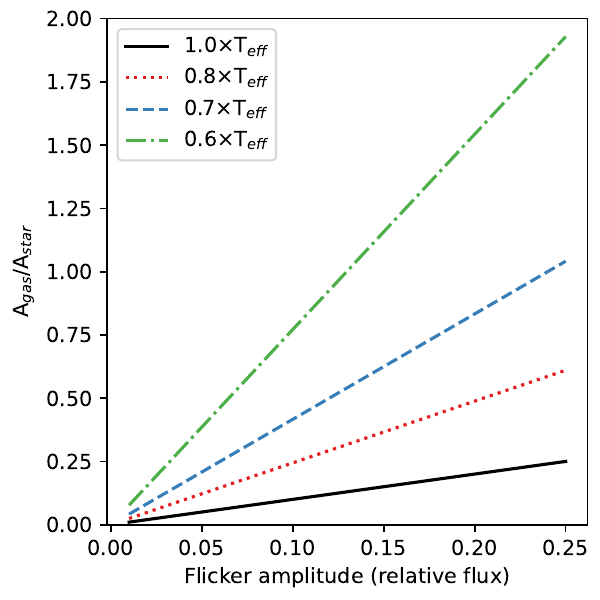}
      \caption{Relative projected area of the circumstellar material versus that of the star ($A_{\rm gas}$/$A_{\rm star}$) as a function of flicker amplitude ($\Delta X$) from Eq.~\ref{eq:relative_area} for four possible values of the gas temperature (in units of $T_{\rm eff}$). 
              }
         \label{fig:compare_area}
   \end{figure}

The photometric amplitude of a flicker is related fundamentally to the density of the material ejected and its geometry (in particular the emitting area projected on the plane of the sky). However, it is also very dependent on the inclination angle of the stars \citep[see Fig.~2 of][]{Rimulo2018}. Therefore, any firm conclusion about the size and density of the ejecta that causes these photometric variations requires a substantial modeling effort that is well beyond the scope of this paper. However, a simple scaling argument reveals that, if the emitting material (cloud) is optically thick and is not seen projected against the stellar disk, the amplitude $\Delta X$ can be expressed in terms of the size of the cloud ($A_{\rm gas}$) and its temperature ($T_{\rm gas}$) as
\begin{equation} \label{eq:relative_area}
\Delta X \approx 
\frac{A_{\rm gas}}{A_{\rm star}}
\frac{T_{\rm gas}^4}{T_{\rm eff}^4}
\,,
\end{equation}
where $A_{\rm star}$ and $T_{\rm eff}$ is the projected stellar area and effective temperature, respectively. 
The temperature of the cloud ($T_{\rm gas}$) is not known, but can be somewhat constrained. $T_{\rm gas}$ should not be above $T_{\rm eff}$ (owing to the surface temperature gradient in rapid rotators, the  equatorial temperature, presumably where the cloud is launched from, is lower than the average $T_{\rm eff}$). From models of a steady-state Be disk \citep[e.g.,][]{carciofi06} the gas temperature close to the star is not below 0.6$\times T_{\rm eff}$. Fig.~\ref{fig:compare_area} shows the relative projected area of the TESS-emitting circumstellar material as a function of flicker amplitude according to Eq.~\ref{eq:relative_area} for values of the gas temperature within the allowed range. For example, a flicker with a TESS amplitude of 10\% ($\sim$0.1 mag) corresponds to a relative projected area of between 10\% and 77\% at the two extremes ($T_{\rm gas} = T_{\rm eff}$ and $0.6\times T_{\rm eff}$). This relation should be most reliable at low inclination angles where the projected area of emitting material is maximum. 

This simple analysis suggests that a flicker with an amplitude of about 0.20\,mag (amplitudes between about 0.15 -- 0.20\,mag were seen in three events in our sample, Tab.~\ref{tbl:measurements}) and a temperature typical of Be stars disks \citep[$0.7\times T_{\rm eff}$,][]{carciofi06}, is associated with an emitting area almost as large as the star itself. This has relevant implications for future models attempting to reproduce this behavior.

\subsection{Photometric signals: Stellar versus circumstellar} \label{sec:stellar_vs_circumstellar}

Through analysis of different spectral lines and velocity ranges, it is possible to distinguish between stellar and circumstellar variations in time-series spectroscopy. Disentangling stellar versus circumstellar variations in photometry is more difficult but can be aided by contemporaneous observations of some other method (e.g., spectroscopy). For instance, when a photometric frequency agrees exactly with a spectroscopic pulsational frequency (Sect.~\ref{sec:distinguishing}), it can be reasonably presumed that the physical origin of that photometric signal is pulsational. In the simplest case, when spectroscopy indicates no evidence of any circumstellar material, all variations in TESS almost certainly originate from the stellar photosphere. On the other hand, when there is disk material present, and especially during mass ejection episodes, photometry may contain circumstellar signals. 

\begin{figure*}
   \centering
   \includegraphics[width=1.0\hsize]{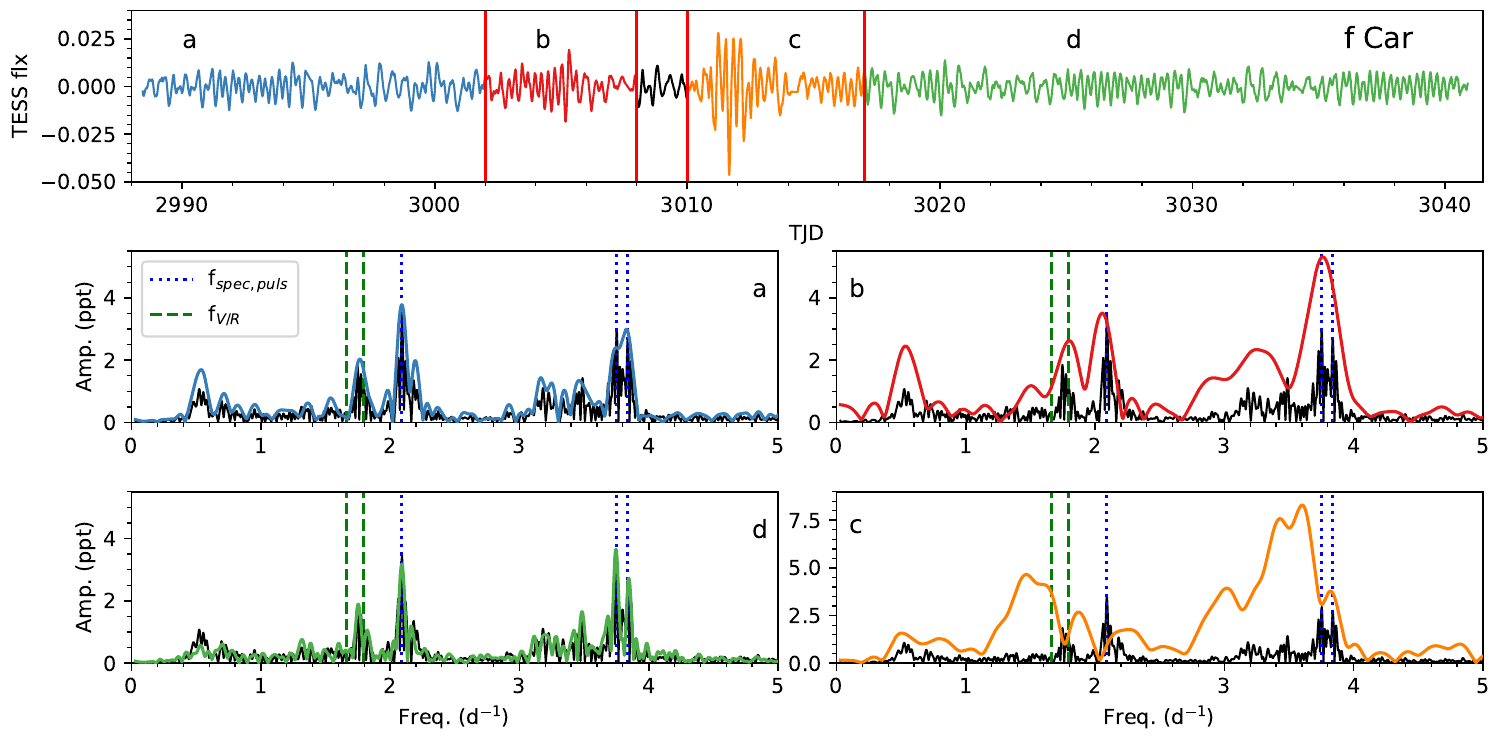}
      \caption{Top panel: TESS light curve from Fig.~\ref{fig:f_Car} after subtracting out the low-frequency variations. It is split into four sections, labeled ``a,'' ``b,'' ``c,'' and ``d,'' corresponding to the pre-outburst disk-less phase (a), the first outburst (b), the second larger outburst (c), and the post-outburst phase (d). The corresponding frequency spectra from these sections are shown in the lower panels (in clockwise order), where the black curve is computed from sections a and d. 
              }
         \label{fig:fCar_FTs}
   \end{figure*}

Fig.~\ref{fig:fCar_FTs} shows the same TESS light curve for f~Car as in Fig.~\ref{fig:f_Car}, but after removing long term trends and split up into four sections in order to compare the frequency spectrum during the disk-less phase, the first and second flickers, and the post-flicker dissipation phase. The pre- and post-flicker phases (``a'' and ``d,'' respectively) are essentially identical in terms of the signals in the frequency spectrum (differences are probably mostly due to different time baselines). This suggests that the mere presence of a disk (during the dissipation phase where material is falling back onto the star) does not seem to impact the frequency spectrum in this case \citep[see also Appendix D in ][]{2024A&A...689A.320N}. However, the epochs where the disk is actively being built up do show significant differences from the inactive state (Sect.~\ref{sec:phot}). The two main frequency groups are slightly enhanced during the first short event in f~Car (panel ``b'' in Fig.~\ref{fig:fCar_FTs}), and are strongly enhanced during the second event (which also had a significantly larger photometric and spectroscopic signature; panel ``c'' in Fig.~\ref{fig:fCar_FTs}) -- are these transient signals stellar or circumstellar? The following subsections focus on this question for Be stars in general.

\subsubsection{Arguments for stellar origins}

In photometry during outbursts, it is evident that many (new) peaks emerge or grow in the frequency spectrum often over a reasonably wide range of frequencies (wider than the span of different values of EW$_{\rm V}$/EW$_{\rm R}$\xspace during different events for a given star). This sort of behavior is typical of outbursting Be stars observed from space \citep[e.g.,][]{2009A&A...506...95H, Semaan2018, 2020MNRAS.498.3171N, 2020MNRAS.494..958N, JLB2022}, including f~Car as discussed above. This may be due to a genuine multiplicity of (unresolved) signals, and/or one or a few signals whose frequencies (and amplitudes) shift, causing the appearance of a broad feature in the frequency spectrum. This enhancement of power in the frequency spectrum is most evident in the lower frequency region of the frequency groups discussed in Sect.~\ref{sec:spec_vs_phot_freqs}, and may be seen in $g1$ and/or $g2$ (and higher order groups).

For all observed outbursts, we find only one single circumstellar EW$_{\rm V}$/EW$_{\rm R}$\xspace frequency for a given spectroscopic feature. This signal may not be perfectly sinusoidal, so there may be one or more harmonics in a frequency analysis of EW$_{\rm V}$/EW$_{\rm R}$\xspace, and, since the amplitude changes over time, a frequency peak will be made wider. In any event, the photometric signals associated with outbursts are wide and multiple, unlike the EW$_{\rm V}$/EW$_{\rm R}$\xspace frequencies. Thus, at least some of the emergent photometric frequencies coinciding with outbursts should be photospheric in origin.

\subsubsection{Contributions from circumstellar material}

It is possible that circumstellar material contributes to the enhanced amplitude of frequency groups especially since the three-dimensional distribution of recently ejected material is hardly constrained. In this scenario, photometric variations near the orbital frequency and its harmonics may appear largely as a line of sight effect, with orbiting material occasionally transiting the stellar surface and passing behind the star, and/or self-obscuration of circumstellar material. If there is any sub-structure within the ejecta this could manifest as multiple frequencies. Indeed, this scenario seems inevitable for high inclination systems, but without a realistic model of the structure and evolution of the newly ejected circumstellar material coupled with radiative transfer calculations it is unclear what the expected observational signals would be in the continuum and line features. 

However, for systems at very low inclination angles, there should be essentially no line of sight effect (assuming that material is mostly in the equatorial plane). Any photometric modulation would then require that the circumstellar material is intrinsically oscillating in brightness, which is not known to occur (but see Sect.~\ref{sec:phot_freq_mechanisms}). In fact, pole-on Be stars typically exhibit qualitatively the same behavior as their intermediate- and high-inclination counterparts, sometimes with extreme increases in amplitude of frequency groups during flickers (see Appendix~\ref{sec:pole_vs_edge}). It then seems likely that circumstellar variations may contribute to photometric frequencies in some cases, but cannot be the sole explanation of increased frequency group amplitudes in general. 

While the above discussion is general and is based on the analysis of the sample presented in this work, there are at least two distinct cases from the literature where circumstellar signals do have a solid photometric counterpart -- $\eta$~Cen and $\omega$~CMa \citep[and perhaps also $\alpha$ Eri, ][]{2011MNRAS.411..162G, 2016A&A...588A..56B}. For $\eta$~Cen, a frequency at 1.56 \,d$^{-1}$\xspace was detected in circumstellar lines but not in purely photospheric lines \citep{2003A&A...411..229R}. This signal was also clearly detected in photometry with variations in amplitude (between about 15 and 35 mmag) and in frequency (shifting by $\pm 3\%$), and can be described as quasi-permanent \citep{2016A&A...588A..56B}. The situation for $\omega$~CMa is similar, with a circumstellar frequency at 0.67 \,d$^{-1}$\xspace  that remains coherent for long times (greater than 1 month) during photometric high states, when the inner disk is most dense \citep{2003A&A...411..167S}, although the photometric counterpart is less obvious compared to $\eta$~Cen. It should be noted that $\eta$~Cen is a shell star, but $\omega$~CMa is observed at a low inclination angle. In the BRITE photometry of $\eta$~Cen, in which the 1.56 \,d$^{-1}$\xspace signal was strong, there is additionally a group of several nearby frequencies (mostly slightly slower) that apparently vary in time \citep{2016A&A...588A..56B}, not dissimilar to the TESS signals typically seen during mass ejection in this work. TESS photometry and archival H$\alpha$ data for $\eta$~Cen and $\omega$~CMa are briefly discussed in Appendix~\ref{sec:muCen}.

\subsubsection{What mechanism(s) can explain the transient photometric signals?} \label{sec:phot_freq_mechanisms}

Especially for short outburst events, these emergent photometric frequencies can appear and disappear suddenly, within one to a few days. The typical growth rates of excited modes are on the order of 10$^{-4}$ \citep[e.g.,][]{2020A&A...644A...9N}, meaning that the required timescale for amplitudes to grow is about 10$^{4}$ times the oscillation period (i.e., thousands of days for typical observed frequencies). It thus seems impossible for these emergent frequencies to be driven by typical internal driving mechanisms (e.g., the $\kappa$ mechanism). 

In (rapidly) rotating stars, Rossby modes (r modes) can couple with the spheroidal motion caused by the Coriolis force, leading to temperature perturbations and a photometric signal \citep{1978MNRAS.182..423P,Saio2018}. These are primarily surface flows and do not necessarily penetrate deep into the stellar interior. \citet{Saio2018} note that ``The hypothesis of mechanical generation of r modes is supported by the presence of r-mode humps in the frequently outbursting Be star KIC 6954724.'' \citet{2023MNRAS.524.4196H} also posit that the hump-and-spike A-type stars have prominent r modes perhaps generated due to (variable) spots on the star disrupting surface flow. Whatever the mechanism that launches material from the surface of a Be star, there must be a disturbance of some sort in and around the region where mass is ejected (especially since the ejection region is not azimuthally symmetric). Considering that emission features appear suddenly at the start of an outburst, any surface disturbances are probably also quick to develop. It seems plausible that this results in the rapid generation of r modes (and/or enhanced visibility) at the surface as suggested by \citet{Saio2018}, and perhaps similar to the spotted hump-and-spike A stars. Some time after mass ejection ceases, the stellar surface should return to its pre-outburst state where without a disturbance to drive the newly generated r modes they dissipate. Note that a physical ``spot'' of some sort is not required for r modes. For instance, r modes are observed in rapidly rotating g mode $\gamma$ Dor pulsators \citep{2016A&A...593A.120V}, which are not in general spotted. Some of the photometric frequencies even in quiescent disk-less epochs of Be stars may be r modes \citep[e.g.,][]{2023MNRAS.521.4765K}.

There is also the notion that stochastically excited g modes can become unstable near the surface during or near to outbursts as discussed in \cite{2020A&A...644A...9N}. These would naturally appear in photometry at near integer multiples of the stellar rotation frequency (i.e., within or near the frequency groups). It is thus difficult to distinguish between temporary g and r modes.

Another possibility is that there are intrinsic oscillations in the brightness of the near circumstellar environment; however, this has not yet been demonstrated to occur. If the inner disk is reasonably dense and physically connected to the star, it is perhaps plausible that stellar pulsation waves can leak into the circumstellar material \citep{2000ASPC..214..288T, 2000MNRAS.319..289T}. An existing inner disk may also be perturbed by the sudden localized injection of new material, where the hydrodynamic response may be (incoherent) oscillations over a range in frequencies. These explanations are speculative, but underscore our current lack of understanding on the star-disk interface. In any event, we do not see evidence for several frequencies being excited in the observed circumstellar emission features for our sample.

Near continuous spectroscopy and/or advanced asteroseismic and circumstellar models may be required to understand the physical process(es) responsible for the links between the relatively sudden enhancement of frequency groups and outbursts. This is complicated by the relative timing between these two features -- pulsations can be enhanced slightly before flickers \citep{2009A&A...506...95H}, primarily during the flicker rising phase (as in the second flicker we observed in f~Car, Fig.~\ref{fig:f_Car}), or even after peak brightness as in V357~Lac (Appendix~\ref{sec:V357_Lac}).

\subsection{Rapidly changing peculiar absorption features} \label{sec:transient_absorptions}

In addition to the emergence of broad emission wings and rapid V/R cycles seen in the first several days of an outburst, \citet{1998A&A...333..125R} identified temporary high-velocity absorption (well outside of ${v\sin i}$) and transient narrow absorptions in photospheric lines during outbursts in $\mu$ Cen \citep[see also ][where in \ion{He}{I}\,$\lambda$6678 high-velocity absorption suddenly appeared in as little as 12 minutes from one spectrum to the next]{1998ApJ...502L..59P}. Qualitatively similar features are evident in some other systems, such as $\omega$ CMa \citep{2003A&A...402..253S}, $\lambda$ Eri \citep{1989ApJS...71..357S}, and EW~Lac \citep{2000A&A...362.1020F}. In $\mu$ Cen the absorptions are all blue-shifted, while in $\omega$ CMa they are redshifted. In $\lambda$ Eri these small absorptions are seen both blue- and red-shifted. In $\omega$ CMa the red-shifted feature is stable over weeks to months, while the blue-shifted feature in $\mu$ Cen lasts for only a few tens of minutes.

Properly noticing and studying these features in our sample falls somewhat outside the scope of this study and would require a detailed analysis of a large number of lines for each star. For example, in $\omega$ CMa the super-${v\sin i}$ absorptions are most prominent in \ion{Mg}{II}\,$\lambda$4481 and \ion{Si}{II}\,$\lambda$6347 \citep{2003A&A...402..253S} -- these lines were not analyzed in our work. On the other hand, in $\mu$ Cen and $\lambda$ Eri, such features are clearly seen in \ion{He}{I}\,$\lambda$6678 \citep{1998A&A...333..125R, 1989ApJS...71..357S}. A preliminary analysis of \ion{He}{I}\,$\lambda$6678 for most of our sample does not reveal strong peculiar features, except for 25~Ori (Appendix~\ref{sec:25_Ori}) and $\kappa$ CMa (Appendix~\ref{sec:kap_CMa}). Thus, at first glance these features do not appear to be universal in outbursting Be stars. However, a more careful analysis of the full optical spectrum for stars in our sample may reveal details that are not seen in the strong emission lines that form the basis of this present study. 
It should also be noted that such features can appear quite suddenly, and may simply have been missed in our observations.

\subsection{Are outbursts heralded by a precursor phase?} \label{sec:precursor}

Outbursts in Be stars are sometimes described in terms of four sequential phases \citep[e.g.,][]{1998A&A...333..125R,2009A&A...506...95H,Semaan2018}: 

\begin{enumerate}
  \item Relative quiescence: The ``default'' configuration of the star where there is no ongoing or recent mass ejection. Disk material may be present, but at very low densities \citep[see, e.g., the residual disk reported in Achernar and characterized by][]{2008ApJ...676L..41C}.
  \item Precursor: Prior to the ejection of material, certain signatures may be seen such as a change in the pulsation properties and/or continuum flux level \citep{2009A&A...506...95H}, or a relatively sudden decrease in the observed emission intensity levels \citep{1998A&A...333..125R}. 
  \item Outburst: Emission levels and broadband flux are increasing (when not observed edge-on) -- the star is unambiguously ejecting material. 
  \item Relaxation: Mass ejection has ceased (or severely diminished), and the disk is dissipating -- continuum and line emission are decreasing.  
\end{enumerate}

The spectroscopic description of the precursor phase requires the presence of a preexisting disk, and is generally more pronounced the stronger the disk. From a purely spectroscopic standpoint, the hallmark of the precursor phase is a rapid (over roughly 10 days) drop in the emission peak intensity relative to the continuum ($E/C$) of all emission lines. At the same time, extended high-velocity wings grow in hydrogen emission lines \citep{1998A&A...333..125R}. This is clearly seen in many cases.

When simultaneous photometry is available, in at least some cases the rapid drop in $E/C$ can be simply explained by a rapid increase in the continuum flux \citep[e.g., Sect.~\ref{sec:V357_Lac}, Fig. 5 in ][]{2018AJ....155...53L}. Thus, when a star begins to eject material, the inner disk quickly increases in density (causing the continuum level to rapidly rise, and high-velocity emission wings to strengthen), and the rate of change in the continuum level is faster than in the emission lines \citep[formed over a larger volume, ][]{Carciofi2011} so that $E/C$ at first decreases. 

In this picture, the spectroscopic precursor phase is not physically distinct from the outburst phase. \citet{1998A&A...333..125R} considered this possibility when defining the precursor phase for $\mu$ Cen, but given the photometric data available at that time such rapid increases in the continuum flux had never been seen. To explain the precursor phase in terms of solely a change in the continuum level, the required brightness increase was $\sim$0.2 magnitudes over about one week. However, in TESS sector 11, the brightness of $\mu$ Cen increased by about 25\% over 15 days (or, during the highest rate of change, by about 17\% in 5 days), making this scenario plausible (see Appendix~\ref{sec:muCen}). 

In terms of photometry, the existence of a distinct precursor phase is not often reported. The most prominent example from the literature is from the Convection, Rotation and planetary Transits (CoRoT) satellite light curve of the B0.5IVe star HD~49330 \citep{2009A&A...506...95H}. Following the spectroscopically motivated scheme of \citet{1998A&A...333..125R}, the CoRoT light curve was split into the four distinct outburst phases listed above. However, in this instance, the precursor phase saw a decrease in the net brightness prior to the brightening event that marked the outburst phase. Nevertheless, the photometric precursor phase for HD~49330 does clearly show changes in the pulsational properties compared to the relative quiescence phase. Unfortunately, without spectroscopy during the CoRoT photometry of HD~49330 it is not possible to know if material had started to be ejected during the photometric precursor phase, or if mass ejection only began during the brightening phase. Precursor phases were labeled and analyzed in CoRoT photometry of three Be stars in \citet{Semaan2018}, but the variability seems the same as that during the quiescence phases. In these three cases, there is no compelling reason to distinguish between precursor and quiescence. 

In our sample, there does not seem to be any systematic signature in photometry or spectroscopy that precedes mass ejection. That is, mass ejection happens fairly suddenly, and without prior announcement. 

Nevertheless, in several cases with strong preexisting disks (V357~Lac, 120~Tau, perhaps 25 Ori), H$\alpha$ $E/C$ does suddenly drop at the beginning of a flicker as the continuum rises rapidly. However, this is happening as mass is already being ejected (i.e., during, but not before, outburst). Again, while this is the case with the relatively short events studied in this work, this does not preclude the existence of precursor phases in other systems and for longer-lived events.

\section{Conclusions} \label{sec:conclusions}

In this work we analyzed high-cadence time-series spectroscopy and contemporaneous TESS space photometry for 13 Be stars that exhibited mass ejection episodes during the observing window. In every mass ejection event, the most prominent observational signatures, which emerge at approximately the same time, are an increase in the photometric brightness and the sudden appearance of high-velocity asymmetric emission (Sect.~\ref{sec:flicker_obs}). For the relatively short events studied here, the build-up phase in photometry (where brightness is increasing) lasted from between about one to several days, although build-up phases can be much longer in general (Sect.~\ref{sec:TESS_flicker_amps}). When there is already an established disk, line emission from the preexisting disk relative to the continuum can decrease as the continuum flux (produced in the inner-most disk) rises rapidly. The emission asymmetry oscillated across the line profile with a frequency consistent with the near-surface Keplerian orbital frequency (Sect.~\ref{sec:orbit_vs_rot}). For a given star these frequencies may vary only slightly even though the duration and amplitude of separate events can be quite different. While the EW$_{\rm V}$/EW$_{\rm R}$\xspace cycles are active, they appear to have a constant frequency until either  EW$_{\rm V}$/EW$_{\rm R}$\xspace gradually becomes flat or EW$_{\rm V}$/EW$_{\rm R}$\xspace continues to vary but no longer appears cyclic.

After about five to ten oscillation cycles, the asymmetry damped out or became incoherent. The characteristic emission asymmetry cycles during the early stages of an outburst for a given star are always slightly lower (by about 10\% to 20\%) than the dominant spectroscopic pulsational frequencies \citep[as initially reported in ][]{1998ASPC..135..348S}, and fall within the $g1$ photometric frequency group but not necessarily corresponding to any particular photometric signal (Fig.~\ref{fig:FTs_TESS_spec}). The above is true whether or not there is a preexisting disk. Six other Be stars from the literature showed behavior consistent with those in our sample (Fig.~\ref{fig:FTs_TESS_spec_lit}). We could find no systems with a high enough cadence where emission features grew during an outburst without emission asymmetry oscillations, neither in our dataset nor in the literature.

The observational evidence points to a scenario where mass is ejected from some localized range in azimuth on the surface of the Be star (presumably at the equator). In all cases, the data are incompatible with an outflow over the entire equator, and are also inconsistent with two corotating clouds on opposite sides of the star as with a dipolar magnetic field as proposed by  \citet{1999ApJ...521..407B, 2020MNRAS.493.2528B}, where major oscillations in peak separation would dominate over V/R variations (Sect.~\ref{sec:orbit_vs_rot}). The ejected material initially orbits close to the star, causing the observed emission asymmetry cycles. Over several orbital timescales, the ejecta evolve toward an axi-symmetric disk configuration (unless interrupted by further mass ejection, which disrupts the asymmetry oscillation cycles). Any viable descriptions of the mass ejection mechanism acting in Be stars must allow for a relatively localized ejection site. Baade et al. (in prep.) found that, at the times of outbursts, up to dozens of pulsation modes are synchronized.  If multimode synchronization is the cause of the outbursts, this explains the brevity of the events.  If the modes have different wavelengths in the photosphere, the synchronization is not just limited in time but also in stellar azimuth. Resolving these networks of frequencies requires far longer time baselines then are available from TESS.

A simple scaling argument suggests that for the higher-amplitude flickers, the projected area of the circumstellar material approaches that of the star itself (Sect.~\ref{sec:TESS_flicker_amps}). In addition to the photometric brightening (or dimming for edge-on systems), there is also an enhancement in the lower-frequency region of the photometric frequency groups (present in all stars in our sample) during nearly all mass ejection events. These enhancements can be present at the start, near the peak, and/or on the declining phase of a flicker. We suggest that these photometric signals that emerge cannot be solely circumstellar, and that during mass ejection oscillations of some sort are excited (or become visible) near or at the stellar surface (Sect.~\ref{sec:stellar_vs_circumstellar}). However, the nature of these quickly excited signals is not yet understood (Sect.~\ref{sec:stellar_vs_circumstellar}).

While the aforementioned transient photometric signals seem to be a consequence of mass ejection, this does not contradict that other relatively stable pulsation modes can contribute to the triggering of mass ejection. For instance, in HD 6226, the outbursts repeat with approximately an 87 day period, equal to the difference between two closely spaced pulsation modes (near 0.71 \,d$^{-1}$\xspace, located in $g2$ in Fig.~\ref{fig:FTs_TESS_spec}). There is additional evidence linking mode coupling of stable frequencies to the triggering of outbursts in several other stars, such as $\mu$ Cen, $\eta$ Cen, 28~Cyg, and 25~Ori \citep{1998ASPC..135..343R,2016A&A...588A..56B, 2018A&A...610A..70B,2018pas8.conf...69B}.

Some systems exhibit high-velocity absorption and/or transient narrow absorptions in photospheric lines during outbursts as in $\mu$ Cen (Sect.~\ref{sec:transient_absorptions}), but these are not universal. The majority of our sample did not exhibit this behavior at the observed epochs. On the short timescales studied here, there were no indications of precursor phases preceding mass ejection, neither in photometry or spectroscopy (Sect.~\ref{sec:precursor}).

Our sample includes one $\gamma$ Cas analog (V767~Cen, Sect.~\ref{sec:V767_Cen}), which are classical Be stars with hard and variable X-rays \citep{2016AdSpR..58..782S}. The characteristics of the photometric and spectroscopic variability in the optical during the outbursts of V767~Cen do not differ from the remainder of the sample. This is in agreement with other works which have found that, except in their X-ray emission, $\gamma$ Cas analogs are indistinguishable from Be stars in general \citep{2020MNRAS.498.3171N, 2021MNRAS.502..242L, 2022MNRAS.510.2286N}.

This first analysis provides the main characteristics of the photometric and spectroscopic behavior of outbursting Be stars and suggests that {\v{S}}tefl frequencies (and therefore also azimuthally localized mass ejection) are universal properties of the Be stars' mass ejection process. Forthcoming papers will examine how to reproduce the observations with SPH models (Rubio et al., in prep), and will focus on detailed studies of specific stars including a more quantitative analysis of the stellar pulsation and examining the circumstellar behavior in all available emission lines.

\section*{Data availability}
The Appendix sections A, C, D, and E are available on Zenodo (\url{https://doi.org/10.5281/zenodo.14990388}). NRES spectroscopy can be downloaded at the LCO science archive (\url{https://archive.lco.global/}), DAO spectra from the DAO science archive (\url{https://www.cadc-ccda.hia-iha.nrc-cnrc.gc.ca/en/dao/}), CHIRON spectra from the NOIRLab Astro data archive (\url{https://astroarchive.noirlab.edu/portal/search/}), and amateur spectra from the BeSS database (\url{http://basebe.obspm.fr}). TESS data can be downloaded from the MAST archive (\url{https://archive.stsci.edu/}).

\phantom{\object{V767 Cen}, \object{f Car}, \object{12 Vul}, \object{V442 And}, \object{lambda Pav}, \object{iot Lyr}, \object{28 Cyg}, \object{V357 Lac}, \object{OT Gem}, \object{25 Ori}, \object{kappa CMa}, \object{j Cen}, \object{120 Tau}, \object{mu Cen}, \object{omega CMa}, \object{omega Ori}, \object{nu Cyg}, \object{EW Lac}, \object{eta Cen}}

\begin{acknowledgements}
The authors thank the anonymous referee for their helpful suggestions. A.C.C. acknowledges support from CNPq (grant 314545/2023-9) and FAPESP (grants 2018/04055-8 and 2019/13354-1). 
A.C.R. acknowledges support from the CAPES grant 88887.464563/2019-00, DAAD grant 57552338, the ESO Studentship program and MPA.
A.L.F. acknowledges support from CAPES (grant 88882.332925/2019-01). 
Y.N. acknowledges support from the Fonds National de la Recherche Scientifique (Belgium), the European Space Agency (ESA) and the Belgian Federal Science Policy Office (BELSPO) in the framework of the PRODEX Programme (contracts linked to XMM-Newton and Gaia).
N.D.R and S.N. are grateful for support from NASA grants 80NSSC23K1049 and 80NSSC24K0229. S.N. acknowledges support from the ERAU Undergraduate Research Institute and the Arizona Space Grant programs. 
C.A. thanks to Fondecyt N1230131. 
The authors are grateful to the amateur spectroscopy community, whose archival data was used in this work, in particular Albert Stiewing, Dong Li, Franck Houpert, Terry Bohlsen, Arnold de Bruin, Arthur Leduc, Joe Daglen, Christian Kreider, and James R. Foster.
This work makes use of observations from the LCOGT network. This paper includes data collected by the TESS mission, which are publicly available from the Mikulski Archive for Space Telescopes (MAST). Funding for the TESS mission is provided by NASA's Science Mission directorate. 
This work makes use of observations obtained at the Dominion Astrophysical Observatory, Herzberg Astronomy and Astrophysics Research Centre, National Research Council of Canada.
This work has made use of data from the European Space Agency (ESA) mission {\it Gaia} (\url{https://www.cosmos.esa.int/gaia}), processed by the {\it Gaia} Data Processing and Analysis Consortium (DPAC \url{https://www.cosmos.esa.int/web/gaia/dpac/consortium}). Funding for the DPAC has been provided by national institutions, in particular the institutions participating in the {\it Gaia} Multilateral Agreement. This research has made use of NASA's Astrophysics Data System. This research has made use of the SIMBAD database, operated at CDS, Strasbourg, France. This work has made use of the BeSS database, operated at LESIA, Observatoire de Meudon, France: http://basebe.obspm.fr. This research made use of Lightkurve, a Python package for Kepler and TESS data analysis \citep{Lightkurve2018}. This research made use of Astropy,\footnote{http://www.astropy.org} a community-developed core Python package for Astronomy \citep{astropy2013, astropy2018}.

\end{acknowledgements}

%
%

\bibliographystyle{aa}
\bibliography{aanda.bib}

\begin{thebibliography}{119}
\expandafter\ifx\csname natexlab\endcsname\relax\def\natexlab#1{#1}\fi

\bibitem[{{Astropy Collaboration} {et~al.}(2018){Astropy Collaboration},
  {Price-Whelan}, {Sip{\H{o}}cz}, {G{\"u}nther}, {Lim}, {Crawford}, {Conseil},
  {Shupe}, {Craig}, {Dencheva}, {Ginsburg}, {Vand erPlas}, {Bradley},
  {P{\'e}rez-Su{\'a}rez}, {de Val-Borro}, {Aldcroft}, {Cruz}, {Robitaille},
  {Tollerud}, {Ardelean}, {Babej}, {Bach}, {Bachetti}, {Bakanov}, {Bamford},
  {Barentsen}, {Barmby}, {Baumbach}, {Berry}, {Biscani}, {Boquien}, {Bostroem},
  {Bouma}, {Brammer}, {Bray}, {Breytenbach}, {Buddelmeijer}, {Burke},
  {Calderone}, {Cano Rodr{\'\i}guez}, {Cara}, {Cardoso}, {Cheedella}, {Copin},
  {Corrales}, {Crichton}, {D'Avella}, {Deil}, {Depagne}, {Dietrich}, {Donath},
  {Droettboom}, {Earl}, {Erben}, {Fabbro}, {Ferreira}, {Finethy}, {Fox},
  {Garrison}, {Gibbons}, {Goldstein}, {Gommers}, {Greco}, {Greenfield},
  {Groener}, {Grollier}, {Hagen}, {Hirst}, {Homeier}, {Horton}, {Hosseinzadeh},
  {Hu}, {Hunkeler}, {Ivezi{\'c}}, {Jain}, {Jenness}, {Kanarek}, {Kendrew},
  {Kern}, {Kerzendorf}, {Khvalko}, {King}, {Kirkby}, {Kulkarni}, {Kumar},
  {Lee}, {Lenz}, {Littlefair}, {Ma}, {Macleod}, {Mastropietro}, {McCully},
  {Montagnac}, {Morris}, {Mueller}, {Mumford}, {Muna}, {Murphy}, {Nelson},
  {Nguyen}, {Ninan}, {N{\"o}the}, {Ogaz}, {Oh}, {Parejko}, {Parley}, {Pascual},
  {Patil}, {Patil}, {Plunkett}, {Prochaska}, {Rastogi}, {Reddy Janga},
  {Sabater}, {Sakurikar}, {Seifert}, {Sherbert}, {Sherwood-Taylor}, {Shih},
  {Sick}, {Silbiger}, {Singanamalla}, {Singer}, {Sladen}, {Sooley},
  {Sornarajah}, {Streicher}, {Teuben}, {Thomas}, {Tremblay}, {Turner},
  {Terr{\'o}n}, {van Kerkwijk}, {de la Vega}, {Watkins}, {Weaver}, {Whitmore},
  {Woillez}, {Zabalza}, \& {Astropy Contributors}}]{astropy2018}
{Astropy Collaboration}, {Price-Whelan}, A.~M., {Sip{\H{o}}cz}, B.~M., {et~al.}
  2018, \aj, 156, 123

\bibitem[{{Astropy Collaboration} {et~al.}(2013){Astropy Collaboration},
  {Robitaille}, {Tollerud}, {Greenfield}, {Droettboom}, {Bray}, {Aldcroft},
  {Davis}, {Ginsburg}, {Price-Whelan}, {Kerzendorf}, {Conley}, {Crighton},
  {Barbary}, {Muna}, {Ferguson}, {Grollier}, {Parikh}, {Nair}, {Unther},
  {Deil}, {Woillez}, {Conseil}, {Kramer}, {Turner}, {Singer}, {Fox}, {Weaver},
  {Zabalza}, {Edwards}, {Azalee Bostroem}, {Burke}, {Casey}, {Crawford},
  {Dencheva}, {Ely}, {Jenness}, {Labrie}, {Lim}, {Pierfederici}, {Pontzen},
  {Ptak}, {Refsdal}, {Servillat}, \& {Streicher}}]{astropy2013}
{Astropy Collaboration}, {Robitaille}, T.~P., {Tollerud}, E.~J., {et~al.} 2013,
  \aap, 558, A33

\bibitem[{{Baade} {et~al.}(2023){Baade}, {Labadie-Bartz}, {Rivinius}, \&
  {Carciofi}}]{2023A&A...678A..47B}
{Baade}, D., {Labadie-Bartz}, J., {Rivinius}, T., \& {Carciofi}, A.~C. 2023,
  \aap, 678, A47

\bibitem[{{Baade} {et~al.}(2018{\natexlab{a}}){Baade}, {Pigulski}, {Rivinius},
  {Carciofi}, {Panoglou}, {Ghoreyshi}, {Handler}, {Kuschnig}, {Moffat},
  {Pablo}, {Popowicz}, {Wade}, {Weiss}, \& {Zwintz}}]{2018A&A...610A..70B}
{Baade}, D., {Pigulski}, A., {Rivinius}, T., {et~al.} 2018{\natexlab{a}}, \aap,
  610, A70

\bibitem[{{Baade} {et~al.}(2017){Baade}, {Rivinius}, {Pigulski}, {Carciofi},
  {Handler}, {Kuschnig}, {Martayan}, {Mehner}, {Moffat}, {Pablo}, {Popowicz},
  {Rucinski}, {Wade}, {Weiss}, \& {Zwintz}}]{Baade2017}
{Baade}, D., {Rivinius}, T., {Pigulski}, A., {et~al.} 2017, in Second
  BRITE-Constellation Science Conference: Small Satellites - Big Science,
  Vol.~5, 196--205

\bibitem[{{Baade} {et~al.}(2016){Baade}, {Rivinius}, {Pigulski}, {Carciofi},
  {Martayan}, {Moffat}, {Wade}, {Weiss}, {Grunhut}, {Handler}, {Kuschnig},
  {Mehner}, {Pablo}, {Popowicz}, {Rucinski}, \&
  {Whittaker}}]{2016A&A...588A..56B}
{Baade}, D., {Rivinius}, T., {Pigulski}, A., {et~al.} 2016, \aap, 588, A56

\bibitem[{{Baade} {et~al.}(2018{\natexlab{b}}){Baade}, {Rivinius}, {Pigulski},
  {Panoglou}, {Carciofi}, {Handler}, {Kuschnig}, {Martayan}, {Mehner},
  {Moffat}, {Pablo}, {Popowicz}, {Rucinski}, {Wade}, {Weiss}, \&
  {Zwintz}}]{2018pas8.conf...69B}
{Baade}, D., {Rivinius}, T., {Pigulski}, A., {et~al.} 2018{\natexlab{b}}, in
  3rd BRITE Science Conference, ed. G.~A. {Wade}, D.~{Baade}, J.~A. {Guzik}, \&
  R.~{Smolec}, Vol.~8, 69--76

\bibitem[{{Balona} \& {Kaye}(1999)}]{1999ApJ...521..407B}
{Balona}, L.~A. \& {Kaye}, A.~B. 1999, \apj, 521, 407

\bibitem[{{Balona} \& {Ozuyar}(2020)}]{2020MNRAS.493.2528B}
{Balona}, L.~A. \& {Ozuyar}, D. 2020, \mnras, 493, 2528

\bibitem[{{Bernhard} {et~al.}(2018){Bernhard}, {Otero}, {H{\"u}mmerich},
  {Kaltcheva}, {Paunzen}, \& {Bohlsen}}]{Bernhard2018}
{Bernhard}, K., {Otero}, S., {H{\"u}mmerich}, S., {et~al.} 2018, \mnras, 479,
  2909

\bibitem[{{Bo{\v{z}}i{\'c} } {et~al.}(1999){Bo{\v{z}}i{\'c} }, {Ru{\v{z}}djak},
  \& {Sudar}}]{1999A&A...350..566B}
{Bo{\v{z}}i{\'c} }, H., {Ru{\v{z}}djak}, D., \& {Sudar}, D. 1999, \aap, 350,
  566

\bibitem[{{Brown} {et~al.}(2013){Brown}, {Baliber}, {Bianco}, {Bowman},
  {Burleson}, {Conway}, {Crellin}, {Depagne}, {De Vera}, {Dilday}, {Dragomir},
  {Dubberley}, {Eastman}, {Elphick}, {Falarski}, {Foale}, {Ford}, {Fulton},
  {Garza}, {Gomez}, {Graham}, {Greene}, {Haldeman}, {Hawkins}, {Haworth},
  {Haynes}, {Hidas}, {Hjelstrom}, {Howell}, {Hygelund}, {Lister}, {Lobdill},
  {Martinez}, {Mullins}, {Norbury}, {Parrent}, {Paulson}, {Petry}, {Pickles},
  {Posner}, {Rosing}, {Ross}, {Sand}, {Saunders}, {Shobbrook}, {Shporer},
  {Street}, {Thomas}, {Tsapras}, {Tufts}, {Valenti}, {Vander Horst}, {Walker},
  {White}, \& {Willis}}]{2013PASP..125.1031B}
{Brown}, T.~M., {Baliber}, N., {Bianco}, F.~B., {et~al.} 2013, \pasp, 125, 1031

\bibitem[{{Carciofi}(2011)}]{Carciofi2011}
{Carciofi}, A.~C. 2011, in IAU Symposium, Vol. 272, Active OB Stars: Structure,
  Evolution, Mass Loss, and Critical Limits, ed. C.~{Neiner}, G.~{Wade},
  G.~{Meynet}, \& G.~{Peters}, 325--336

\bibitem[{{Carciofi} \& {Bjorkman}(2006)}]{carciofi06}
{Carciofi}, A.~C. \& {Bjorkman}, J.~E. 2006, \apj, 639, 1081

\bibitem[{{Carciofi} \& {Bjorkman}(2008)}]{2008ApJ...684.1374C}
{Carciofi}, A.~C. \& {Bjorkman}, J.~E. 2008, \apj, 684, 1374

\bibitem[{{Carciofi} {et~al.}(2008){Carciofi}, {Domiciano de Souza},
  {Magalh{\~a}es}, {Bjorkman}, \& {Vakili}}]{2008ApJ...676L..41C}
{Carciofi}, A.~C., {Domiciano de Souza}, A., {Magalh{\~a}es}, A.~M.,
  {Bjorkman}, J.~E., \& {Vakili}, F. 2008, \apjl, 676, L41

\bibitem[{{Carciofi} {et~al.}(2006){Carciofi}, {Miroshnichenko}, {Kusakin},
  {Bjorkman}, {Bjorkman}, {Marang}, {Kuratov}, {Garc{\'\i}a-Lario},
  {Calder{\'o}n}, {Fabregat}, \& {Magalh{\~a}es}}]{2006ApJ...652.1617C}
{Carciofi}, A.~C., {Miroshnichenko}, A.~S., {Kusakin}, A.~V., {et~al.} 2006,
  \apj, 652, 1617

\bibitem[{{Castor} {et~al.}(1975){Castor}, {Abbott}, \&
  {Klein}}]{1975ApJ...195..157C}
{Castor}, J.~I., {Abbott}, D.~C., \& {Klein}, R.~I. 1975, \apj, 195, 157

\bibitem[{{Chojnowski} {et~al.}(2022){Chojnowski}, {Hubrig}, {Labadie-Bartz},
  {Rivinius}, {Sch{\"o}ller}, {Niemczura}, {Nidever}, {Stutz}, \&
  {Hummel}}]{2022MNRAS.516.2812C}
{Chojnowski}, S.~D., {Hubrig}, S., {Labadie-Bartz}, J., {et~al.} 2022, \mnras,
  516, 2812

\bibitem[{{Chojnowski} {et~al.}(2017){Chojnowski}, {Wisniewski}, {Whelan},
  {Labadie-Bartz}, {Borges Fernandes}, {Lin}, {Majewski}, {Stringfellow},
  {Mennickent}, {Roman-Lopes}, {Tang}, {Hearty}, {Holtzman}, {Pepper}, \&
  {Zasowski}}]{2017AJ....153..174C}
{Chojnowski}, S.~D., {Wisniewski}, J.~P., {Whelan}, D.~G., {et~al.} 2017, \aj,
  153, 174

\bibitem[{{Cochetti} {et~al.}(2019){Cochetti}, {Arcos}, {Kanaan}, {Meilland},
  {Cidale}, \& {Cur{\'e}}}]{2019A&A...621A.123C}
{Cochetti}, Y.~R., {Arcos}, C., {Kanaan}, S., {et~al.} 2019, \aap, 621, A123

\bibitem[{{Cochetti} {et~al.}(2021){Cochetti}, {Arias}, {Kraus}, {Cidale},
  {Torres}, {Granada}, \& {Maryeva}}]{2021A&A...647A.164C}
{Cochetti}, Y.~R., {Arias}, M.~L., {Kraus}, M., {et~al.} 2021, \aap, 647, A164

\bibitem[{{Cur{\'e}} \& {Araya}(2023)}]{2023Galax..11...68C}
{Cur{\'e}}, M. \& {Araya}, I. 2023, Galaxies, 11, 68

\bibitem[{{Cyr} {et~al.}(2020){Cyr}, {Jones}, {Carciofi}, {Steckel}, {Tycner},
  \& {Okazaki}}]{2020MNRAS.497.3525C}
{Cyr}, I.~H., {Jones}, C.~E., {Carciofi}, A.~C., {et~al.} 2020, \mnras, 497,
  3525

\bibitem[{{David-Uraz} {et~al.}(2019){David-Uraz}, {Neiner}, {Sikora},
  {Bowman}, {Petit}, {Chowdhury}, {Handler}, {Pergeorelis}, {Cantiello},
  {Cohen}, {Erba}, {Keszthelyi}, {Khalack}, {Kobzar}, {Kochukhov},
  {Labadie-Bartz}, {Lovekin}, {MacInnis}, {Owocki}, {Pablo}, {Shultz},
  {ud-Doula}, \& {Wade}}]{2019MNRAS.487..304D}
{David-Uraz}, A., {Neiner}, C., {Sikora}, J., {et~al.} 2019, \mnras, 487, 304

\bibitem[{{Floquet} {et~al.}(2000){Floquet}, {Hubert}, {Hirata}, {McDavid},
  {Zorec}, {Gies}, {Hahula}, {Janot-Pacheco}, {Kambe}, {Leister},
  {{\v{S}}tefl}, {Tarasov}, \& {Neiner}}]{2000A&A...362.1020F}
{Floquet}, M., {Hubert}, A.~M., {Hirata}, R., {et~al.} 2000, \aap, 362, 1020

\bibitem[{{Fr{\'e}mat} {et~al.}(2005){Fr{\'e}mat}, {Zorec}, {Hubert}, \&
  {Floquet}}]{2005A&A...440..305F}
{Fr{\'e}mat}, Y., {Zorec}, J., {Hubert}, A.~M., \& {Floquet}, M. 2005, \aap,
  440, 305

\bibitem[{{Ghoreyshi} {et~al.}(2018){Ghoreyshi}, {Carciofi}, {R{\'\i}mulo},
  {Vieira}, {Faes}, {Baade}, {Bjorkman}, {Otero}, \&
  {Rivinius}}]{Ghoreyshi2018}
{Ghoreyshi}, M.~R., {Carciofi}, A.~C., {R{\'\i}mulo}, L.~R., {et~al.} 2018,
  \mnras, 479, 2214

\bibitem[{{G{\l}{\c{e}}bocki} \& {Gnaci{\'n}ski}(2005)}]{2005ESASP.560..571G}
{G{\l}{\c{e}}bocki}, R. \& {Gnaci{\'n}ski}, P. 2005, in ESA Special
  Publication, Vol. 560, 13th Cambridge Workshop on Cool Stars, Stellar Systems
  and the Sun, ed. F.~{Favata}, G.~A.~J. {Hussain}, \& B.~{Battrick}, 571

\bibitem[{{Goss} {et~al.}(2011){Goss}, {Karoff}, {Chaplin}, {Elsworth}, \&
  {Stevens}}]{2011MNRAS.411..162G}
{Goss}, K.~J.~F., {Karoff}, C., {Chaplin}, W.~J., {Elsworth}, Y., \& {Stevens},
  I.~R. 2011, \mnras, 411, 162

\bibitem[{{Hanuschik} {et~al.}(1993){Hanuschik}, {Dachs}, {Baudzus}, \&
  {Thimm}}]{1993A&A...274..356H}
{Hanuschik}, R.~W., {Dachs}, J., {Baudzus}, M., \& {Thimm}, G. 1993, \aap, 274,
  356

\bibitem[{{Harmanec}(1983)}]{1983HvaOB...7...55H}
{Harmanec}, P. 1983, Hvar Observatory Bulletin, 7, 55

\bibitem[{{Harmanec}(2000)}]{2000ASPC..214...13H}
{Harmanec}, P. 2000, in Astronomical Society of the Pacific Conference Series,
  Vol. 214, IAU Colloq. 175: The Be Phenomenon in Early-Type Stars, ed. M.~A.
  {Smith}, H.~F. {Henrichs}, \& J.~{Fabregat}, 13

\bibitem[{{Haubois} {et~al.}(2012){Haubois}, {Carciofi}, {Rivinius}, {Okazaki},
  \& {Bjorkman}}]{Haubois2012}
{Haubois}, X., {Carciofi}, A.~C., {Rivinius}, T., {Okazaki}, A.~T., \&
  {Bjorkman}, J.~E. 2012, \apj, 756, 156

\bibitem[{{Henriksen} {et~al.}(2023){Henriksen}, {Antoci}, {Saio}, {Grundahl},
  {Kjeldsen}, {Van Reeth}, {Bowman}, {P{\'a}pics}, {De Cat}, {Kr{\"u}ger},
  {Andersen}, \& {Pall{\'e}}}]{2023MNRAS.524.4196H}
{Henriksen}, A.~I., {Antoci}, V., {Saio}, H., {et~al.} 2023, \mnras, 524, 4196

\bibitem[{{H{\"o}fner} \& {Olofsson}(2018)}]{2018A&ARv..26....1H}
{H{\"o}fner}, S. \& {Olofsson}, H. 2018, \aapr, 26, 1

\bibitem[{{Horch} {et~al.}(2010){Horch}, {Falta}, {Anderson}, {DeSousa},
  {Miniter}, {Ahmed}, \& {van Altena}}]{2010AJ....139..205H}
{Horch}, E.~P., {Falta}, D., {Anderson}, L.~M., {et~al.} 2010, \aj, 139, 205

\bibitem[{{Horch} {et~al.}(2008){Horch}, {van Altena}, {Cyr}, {Kinsman-Smith},
  {Srivastava}, \& {Zhou}}]{2008AJ....136..312H}
{Horch}, E.~P., {van Altena}, W.~F., {Cyr}, William~M., J., {et~al.} 2008, \aj,
  136, 312

\bibitem[{{Huat} {et~al.}(2009){Huat}, {Hubert}, {Baudin}, {Floquet}, {Neiner},
  {Fr{\'e}mat}, {Guti{\'e}rrez-Soto}, {Andrade}, {de Batz}, {Diago}, {Emilio},
  {Espinosa Lara}, {Fabregat}, {Janot-Pacheco}, {Leroy}, {Martayan}, {Semaan},
  {Suso}, {Auvergne}, {Catala}, {Michel}, \& {Samadi}}]{2009A&A...506...95H}
{Huat}, A.~L., {Hubert}, A.~M., {Baudin}, F., {et~al.} 2009, \aap, 506, 95

\bibitem[{{Hubert} \& {Floquet}(1998)}]{1998A&A...335..565H}
{Hubert}, A.~M. \& {Floquet}, M. 1998, \aap, 335, 565

\bibitem[{{Kee} {et~al.}(2018){Kee}, {Owocki}, \&
  {Kuiper}}]{2018MNRAS.479.4633K}
{Kee}, N.~D., {Owocki}, S., \& {Kuiper}, R. 2018, \mnras, 479, 4633

\bibitem[{{Klement} {et~al.}(2015){Klement}, {Carciofi}, {Rivinius},
  {Panoglou}, {Vieira}, {Bjorkman}, {{\v{S}}tefl}, {Tycner}, {Faes},
  {Kor{\v{c}}{\'a}kov{\'a}}, {M{\"u}ller}, {Zavala}, \&
  {Cur{\'e}}}]{2015A&A...584A..85K}
{Klement}, R., {Carciofi}, A.~C., {Rivinius}, T., {et~al.} 2015, \aap, 584, A85

\bibitem[{{Klement} {et~al.}(2024){Klement}, {Rivinius}, {Gies}, {Baade},
  {M{\'e}rand}, {Monnier}, {Schaefer}, {Lanthermann}, {Anugu}, {Kraus}, \&
  {Gardner}}]{2024ApJ...962...70K}
{Klement}, R., {Rivinius}, T., {Gies}, D.~R., {et~al.} 2024, \apj, 962, 70

\bibitem[{{Koen} \& {Eyer}(2002)}]{2002MNRAS.331...45K}
{Koen}, C. \& {Eyer}, L. 2002, \mnras, 331, 45

\bibitem[{{Kroll} \& {Hanuschik}(1997)}]{1997ASPC..121..494K}
{Kroll}, P. \& {Hanuschik}, R.~W. 1997, in Astron. Soc. Pacific Conf. Ser.,
  Vol. 121, IAU Colloq. 163: Accretion Phenomena and Related Outflows, ed.
  D.~T. {Wickramasinghe}, G.~V. {Bicknell}, \& L.~{Ferrario}, 494

\bibitem[{{Kurtz} {et~al.}(2023){Kurtz}, {Jayaraman}, {Sowicka}, {Handler},
  {Saio}, {Labadie-Bartz}, \& {Lee}}]{2023MNRAS.521.4765K}
{Kurtz}, D.~W., {Jayaraman}, R., {Sowicka}, P., {et~al.} 2023, \mnras, 521,
  4765

\bibitem[{{Labadie-Bartz} {et~al.}(2021){Labadie-Bartz}, {Baade}, {Carciofi},
  {Rubio}, {Rivinius}, {Borre}, {Martayan}, \& {Siverd}}]{2021MNRAS.502..242L}
{Labadie-Bartz}, J., {Baade}, D., {Carciofi}, A.~C., {et~al.} 2021, \mnras,
  502, 242

\bibitem[{{Labadie-Bartz} \& {Carciofi}(2020)}]{2020svos.conf..185L}
{Labadie-Bartz}, J. \& {Carciofi}, A.~C. 2020, in Stars and their Variability
  Observed from Space, ed. C.~{Neiner}, W.~W. {Weiss}, D.~{Baade}, R.~E.
  {Griffin}, C.~C. {Lovekin}, \& A.~F.~J. {Moffat}, 185--187

\bibitem[{{Labadie-Bartz} {et~al.}(2022){Labadie-Bartz}, {Carciofi}, {Henrique
  de Amorim}, {Rubio}, {Luiz Figueiredo}, {Ticiani dos Santos}, \&
  {Thomson-Paressant}}]{JLB2022}
{Labadie-Bartz}, J., {Carciofi}, A.~C., {Henrique de Amorim}, T., {et~al.}
  2022, \aj, 163, 226

\bibitem[{{Labadie-Bartz} {et~al.}(2018){Labadie-Bartz}, {Chojnowski},
  {Whelan}, {Pepper}, {McSwain}, {Borges Fernandes}, {Wisniewski},
  {Stringfellow}, {Carciofi}, {Siverd}, {Glazier}, {Anderson}, {Caravello},
  {Stassun}, {Lund}, {Stevens}, {Rodriguez}, {James}, \&
  {Kuhn}}]{2018AJ....155...53L}
{Labadie-Bartz}, J., {Chojnowski}, S.~D., {Whelan}, D.~G., {et~al.} 2018, \aj,
  155, 53

\bibitem[{{Labadie-Bartz} {et~al.}(2023){Labadie-Bartz}, {H{\"u}mmerich},
  {Bernhard}, {Paunzen}, \& {Shultz}}]{2023A&A...676A..55L}
{Labadie-Bartz}, J., {H{\"u}mmerich}, S., {Bernhard}, K., {Paunzen}, E., \&
  {Shultz}, M.~E. 2023, \aap, 676, A55

\bibitem[{{Labadie-Bartz} {et~al.}(2017){Labadie-Bartz}, {Pepper}, {McSwain},
  {Bjorkman}, {Bjorkman}, {Lund}, {Rodriguez}, {Stassun}, {Stevens}, {James},
  {Kuhn}, {Siverd}, \& {Beatty}}]{2017AJ....153..252L}
{Labadie-Bartz}, J., {Pepper}, J., {McSwain}, M.~V., {et~al.} 2017, \aj, 153,
  252

\bibitem[{{Lee} {et~al.}(1991){Lee}, {Osaki}, \& {Saio}}]{1991MNRAS.250..432L}
{Lee}, U., {Osaki}, Y., \& {Saio}, H. 1991, \mnras, 250, 432

\bibitem[{{Lenz} \& {Breger}(2005)}]{Lenz2005}
{Lenz}, P. \& {Breger}, M. 2005, Communications in Asteroseismology, 146, 53

\bibitem[{{Levenhagen} {et~al.}(2011){Levenhagen}, {Leister}, \&
  {K{\"u}nzel}}]{2011A&A...533A..75L}
{Levenhagen}, R.~S., {Leister}, N.~V., \& {K{\"u}nzel}, R. 2011, \aap, 533, A75

\bibitem[{{Levenhagen} {et~al.}(2003){Levenhagen}, {Leister}, {Zorec},
  {Janot-Pacheco}, {Hubert}, \& {Floquet}}]{2003A&A...400..599L}
{Levenhagen}, R.~S., {Leister}, N.~V., {Zorec}, J., {et~al.} 2003, \aap, 400,
  599

\bibitem[{{Lightkurve Collaboration} {et~al.}(2018){Lightkurve Collaboration},
  {Cardoso}, {Hedges}, {Gully-Santiago}, {Saunders}, {Cody}, {Barclay}, {Hall},
  {Sagear}, {Turtelboom}, {Zhang}, {Tzanidakis}, {Mighell}, {Coughlin}, {Bell},
  {Berta-Thompson}, {Williams}, {Dotson}, \& {Barentsen}}]{Lightkurve2018}
{Lightkurve Collaboration}, {Cardoso}, J.~V.~d.~M., {Hedges}, C., {et~al.}
  2018, {Lightkurve: Kepler and TESS time series analysis in Python},
  Astrophysics Source Code Library

\bibitem[{{Marr} {et~al.}(2021){Marr}, {Jones}, {Carciofi}, {Rubio}, {Mota},
  {Ghoreyshi}, {Hatfield}, \& {R{\'\i}mulo}}]{2021ApJ...912...76M}
{Marr}, K.~C., {Jones}, C.~E., {Carciofi}, A.~C., {et~al.} 2021, \apj, 912, 76

\bibitem[{{Mennickent} {et~al.}(2002){Mennickent}, {Pietrzy{\'n}ski}, {Gieren},
  \& {Szewczyk}}]{2002A&A...393..887M}
{Mennickent}, R.~E., {Pietrzy{\'n}ski}, G., {Gieren}, W., \& {Szewczyk}, O.
  2002, \aap, 393, 887

\bibitem[{{Monin} {et~al.}(2014){Monin}, {Saddlemyer}, \&
  {Bohlender}}]{2014RMxAC..45...69M}
{Monin}, D., {Saddlemyer}, L., \& {Bohlender}, D. 2014, in Revista Mex. Astron.
  Astrof. Conf. Ser., Vol.~45, Revista Mex. Astron. Astrof. Conf. Ser., 69

\bibitem[{{Naz{\'e}} {et~al.}(2024){Naz{\'e}}, {Britavskiy}, \&
  {Labadie-Bartz}}]{2024A&A...689A.320N}
{Naz{\'e}}, Y., {Britavskiy}, N., \& {Labadie-Bartz}, J. 2024, \aap, 689, A320

\bibitem[{{Naz{\'e}} \& {Motch}(2018)}]{2018A&A...619A.148N}
{Naz{\'e}}, Y. \& {Motch}, C. 2018, \aap, 619, A148

\bibitem[{{Naz{\'e}} {et~al.}(2020{\natexlab{a}}){Naz{\'e}}, {Motch}, {Rauw},
  {Kumar}, {Robrade}, {Lopes de Oliveira}, {Smith}, \&
  {Torrej{\'o}n}}]{2020MNRAS.493.2511N}
{Naz{\'e}}, Y., {Motch}, C., {Rauw}, G., {et~al.} 2020{\natexlab{a}}, \mnras,
  493, 2511

\bibitem[{{Naz{\'e}} {et~al.}(2020{\natexlab{b}}){Naz{\'e}}, {Pigulski},
  {Rauw}, \& {Smith}}]{2020MNRAS.494..958N}
{Naz{\'e}}, Y., {Pigulski}, A., {Rauw}, G., \& {Smith}, M.~A.
  2020{\natexlab{b}}, \mnras, 494, 958

\bibitem[{{Naz{\'e}} {et~al.}(2022{\natexlab{a}}){Naz{\'e}}, {Rauw}, {Bohlsen},
  {Heathcote}, {Mc Gee}, {Cacella}, \& {Motch}}]{2022MNRAS.512.1648N}
{Naz{\'e}}, Y., {Rauw}, G., {Bohlsen}, T., {et~al.} 2022{\natexlab{a}}, \mnras,
  512, 1648

\bibitem[{{Naz{\'e}} {et~al.}(2022{\natexlab{b}}){Naz{\'e}}, {Rauw}, {Czesla},
  {Smith}, \& {Robrade}}]{2022MNRAS.510.2286N}
{Naz{\'e}}, Y., {Rauw}, G., {Czesla}, S., {Smith}, M.~A., \& {Robrade}, J.
  2022{\natexlab{b}}, \mnras, 510, 2286

\bibitem[{{Naz{\'e}} {et~al.}(2020{\natexlab{c}}){Naz{\'e}}, {Rauw}, \&
  {Pigulski}}]{2020MNRAS.498.3171N}
{Naz{\'e}}, Y., {Rauw}, G., \& {Pigulski}, A. 2020{\natexlab{c}}, \mnras, 498,
  3171

\bibitem[{{Neiner} {et~al.}(2011){Neiner}, {de Batz}, {Cochard}, {Floquet},
  {Mekkas}, \& {Desnoux}}]{BeSS}
{Neiner}, C., {de Batz}, B., {Cochard}, F., {et~al.} 2011, \aj, 142, 149

\bibitem[{{Neiner} {et~al.}(2005){Neiner}, {Floquet}, {Hubert}, {Fr{\'e}mat},
  {Hirata}, {Masuda}, {Gies}, {Buil}, \& {Martayan}}]{2005A&A...437..257N}
{Neiner}, C., {Floquet}, M., {Hubert}, A.~M., {et~al.} 2005, \aap, 437, 257

\bibitem[{{Neiner} {et~al.}(2002){Neiner}, {Hubert}, {Floquet}, {Jankov},
  {Henrichs}, {Foing}, {Oliveira}, {Orlando}, {Abbott}, {Baldry}, {Bedding},
  {Cami}, {Cao}, {Catala}, {Cheng}, {Domiciano de Souza}, {Janot-Pacheco},
  {Hao}, {Kaper}, {Kaufer}, {Leister}, {Neff}, {O'Toole}, {Sch{\"a}fer},
  {Smartt}, {Stahl}, {Telting}, {Tubbesing}, \& {Zorec}}]{2002A&A...388..899N}
{Neiner}, C., {Hubert}, A.~M., {Floquet}, M., {et~al.} 2002, \aap, 388, 899

\bibitem[{{Neiner} {et~al.}(2020){Neiner}, {Lee}, {Mathis}, {Saio}, {Lovekin},
  \& {Augustson}}]{2020A&A...644A...9N}
{Neiner}, C., {Lee}, U., {Mathis}, S., {et~al.} 2020, \aap, 644, A9

\bibitem[{{Okazaki}(1997)}]{1997A&A...318..548O}
{Okazaki}, A.~T. 1997, \aap, 318, 548

\bibitem[{{Oksala} {et~al.}(2015){Oksala}, {Kochukhov}, {Krti{\v{c}}ka},
  {Townsend}, {Wade}, {Prv{\'a}k}, {Mikul{\'a}{\v{s}}ek}, {Silvester}, \&
  {Owocki}}]{2015MNRAS.451.2015O}
{Oksala}, M.~E., {Kochukhov}, O., {Krti{\v{c}}ka}, J., {et~al.} 2015, \mnras,
  451, 2015

\bibitem[{{Oksala} {et~al.}(2012){Oksala}, {Wade}, {Townsend}, {Owocki},
  {Kochukhov}, {Neiner}, {Alecian}, \& {Grunhut}}]{2012MNRAS.419..959O}
{Oksala}, M.~E., {Wade}, G.~A., {Townsend}, R.~H.~D., {et~al.} 2012, \mnras,
  419, 959

\bibitem[{{Owocki}(2006)}]{2006ASPC..355..219O}
{Owocki}, S. 2006, in Astronomical Society of the Pacific Conference Series,
  Vol. 355, Stars with the B[e] Phenomenon, ed. M.~{Kraus} \& A.~S.
  {Miroshnichenko}, 219

\bibitem[{{Owocki} {et~al.}(1988){Owocki}, {Castor}, \&
  {Rybicki}}]{1988ApJ...335..914O}
{Owocki}, S.~P., {Castor}, J.~I., \& {Rybicki}, G.~B. 1988, \apj, 335, 914

\bibitem[{{Pablo} {et~al.}(2016){Pablo}, {Whittaker}, {Popowicz}, {Mochnacki},
  {Kuschnig}, {Grant}, {Moffat}, {Rucinski}, {Matthews},
  {Schwarzenberg-Czerny}, {Handler}, {Weiss}, {Baade}, {Wade},
  {Zoc{\l}o{\'n}ska}, {Ramiaramanantsoa}, {Unterberger}, {Zwintz}, {Pigulski},
  {Rowe}, {Koudelka}, {Orlea{\'n}ski}, {Pamyatnykh}, {Neiner}, {Wawrzaszek},
  {Marciniszyn}, {Romano}, {Wo{\'z}niak}, {Zawistowski}, \&
  {Zee}}]{2016PASP..128l5001P}
{Pablo}, H., {Whittaker}, G.~N., {Popowicz}, A., {et~al.} 2016, \pasp, 128,
  125001

\bibitem[{{Panoglou} {et~al.}(2018){Panoglou}, {Faes}, {Carciofi}, {Okazaki},
  {Baade}, {Rivinius}, \& {Borges Fernandes}}]{2018MNRAS.473.3039P}
{Panoglou}, D., {Faes}, D.~M., {Carciofi}, A.~C., {et~al.} 2018, \mnras, 473,
  3039

\bibitem[{{Papaloizou} \& {Pringle}(1978)}]{1978MNRAS.182..423P}
{Papaloizou}, J. \& {Pringle}, J.~E. 1978, \mnras, 182, 423

\bibitem[{{Pavlovski} {et~al.}(1997){Pavlovski}, {Harmanec}, {Bozic},
  {Koubsky}, {Hadrava}, {Kriiz}, {Ruzic}, \& {Stefl}}]{1997A&AS..125...75P}
{Pavlovski}, K., {Harmanec}, P., {Bozic}, H., {et~al.} 1997, \aaps, 125, 75

\bibitem[{{Pereira} {et~al.}(2024){Pereira}, {Janot-Pacheco}, {Andrade}, \&
  {Emilio}}]{2024A&A...691L...4P}
{Pereira}, A.~W., {Janot-Pacheco}, E., {Andrade}, L., \& {Emilio}, M. 2024,
  \aap, 691, L4

\bibitem[{{Peters}(1998)}]{1998ApJ...502L..59P}
{Peters}, G.~J. 1998, \apjl, 502, L59

\bibitem[{{Puls} {et~al.}(2008){Puls}, {Vink}, \&
  {Najarro}}]{2008A&ARv..16..209P}
{Puls}, J., {Vink}, J.~S., \& {Najarro}, F. 2008, \aapr, 16, 209

\bibitem[{{Richardson} {et~al.}(2021){Richardson}, {Thizy}, {Bjorkman},
  {Carciofi}, {Rubio}, {Thomas}, {Bjorkman}, {Labadie-Bartz}, {Genaro},
  {Wisniewski}, {Wang}, {Gies}, {Chojnowski}, {Daly}, {Edwards}, {Fowler},
  {Gullingsrud}, {Habel}, {James}, {Kehoe}, {Kuchta}, {Lane}, {Miroshnichenko},
  {Mishra}, {Pablo}, {Peploski}, {Pepper}, {Rodriguez}, {Siverd}, {Stassun},
  {Stevens}, {Trucks}, {Windsor}, {Wood}, {Bertrand}, {Broussat}, {Bryssinck},
  {Buil}, {Charbonnel}, {de Bruin}, {Daglen}, {Desnoux}, {Dull}, {Garde},
  {Graham}, {Gurney}, {Halsey}, {Fosanelli}, {Guarro Fl{\'o}}, {Houpert},
  {James}, {Kreider}, {Leadbeater}, {Lester}, {Li}, {Maetz}, {Stiewing},
  {Somogyi}, {Terry}, {Ubaud}, \& {Waldschlaeger}}]{2021MNRAS.508.2002R}
{Richardson}, N.~D., {Thizy}, O., {Bjorkman}, J.~E., {et~al.} 2021, \mnras,
  508, 2002

\bibitem[{{Ricker} {et~al.}(2015){Ricker}, {Winn}, {Vanderspek}, {Latham},
  {Bakos}, {Bean}, {Berta-Thompson}, {Brown}, {Buchhave}, {Butler}, {Butler},
  {Chaplin}, {Charbonneau}, {Christensen-Dalsgaard}, {Clampin}, {Deming},
  {Doty}, {De Lee}, {Dressing}, {Dunham}, {Endl}, {Fressin}, {Ge}, {Henning},
  {Holman}, {Howard}, {Ida}, {Jenkins}, {Jernigan}, {Johnson}, {Kaltenegger},
  {Kawai}, {Kjeldsen}, {Laughlin}, {Levine}, {Lin}, {Lissauer}, {MacQueen},
  {Marcy}, {McCullough}, {Morton}, {Narita}, {Paegert}, {Palle}, {Pepe},
  {Pepper}, {Quirrenbach}, {Rinehart}, {Sasselov}, {Sato}, {Seager},
  {Sozzetti}, {Stassun}, {Sullivan}, {Szentgyorgyi}, {Torres}, {Udry}, \&
  {Villasenor}}]{Ricker2015}
{Ricker}, G.~R., {Winn}, J.~N., {Vanderspek}, R., {et~al.} 2015, Journal of
  Astronomical Telescopes, Instruments, and Systems, 1, 014003

\bibitem[{{R{\'\i}mulo} {et~al.}(2018){R{\'\i}mulo}, {Carciofi}, {Vieira},
  {Rivinius}, {Faes}, {Figueiredo}, {Bjorkman}, {Georgy}, {Ghoreyshi}, \&
  {Soszy{\'n}ski}}]{Rimulo2018}
{R{\'\i}mulo}, L.~R., {Carciofi}, A.~C., {Vieira}, R.~G., {et~al.} 2018,
  \mnras, 476, 3555

\bibitem[{{Rivinius} {et~al.}(2016){Rivinius}, {Baade}, \&
  {Carciofi}}]{2016A&A...593A.106R}
{Rivinius}, T., {Baade}, D., \& {Carciofi}, A.~C. 2016, \aap, 593, A106

\bibitem[{{Rivinius} {et~al.}(2013{\natexlab{a}}){Rivinius}, {Baade},
  {Townsend}, {Carciofi}, \& {{\v{S}}tefl}}]{2013A&A...559L...4R}
{Rivinius}, T., {Baade}, D., {Townsend}, R.~H.~D., {Carciofi}, A.~C., \&
  {{\v{S}}tefl}, S. 2013{\natexlab{a}}, \aap, 559, L4

\bibitem[{{Rivinius} {et~al.}(2003){Rivinius}, {Baade}, \&
  {{\v{S}}tefl}}]{2003A&A...411..229R}
{Rivinius}, T., {Baade}, D., \& {{\v{S}}tefl}, S. 2003, \aap, 411, 229

\bibitem[{{Rivinius} {et~al.}(2001{\natexlab{a}}){Rivinius}, {Baade},
  {{\v{S}}tefl}, \& {Maintz}}]{2001A&A...379..257R}
{Rivinius}, T., {Baade}, D., {{\v{S}}tefl}, S., \& {Maintz}, M.
  2001{\natexlab{a}}, \aap, 379, 257

\bibitem[{{Rivinius} {et~al.}(1998{\natexlab{a}}){Rivinius}, {Baade},
  {{\v{S}}tefl}, {Stahl}, {Wolf}, \& {Kaufer}}]{1998ASPC..135..343R}
{Rivinius}, T., {Baade}, D., {{\v{S}}tefl}, S., {et~al.} 1998{\natexlab{a}}, in
  Astronomical Society of the Pacific Conference Series, Vol. 135, A Half
  Century of Stellar Pulsation Interpretation, ed. P.~A. {Bradley} \& J.~A.
  {Guzik}, 343

\bibitem[{{Rivinius} {et~al.}(1998{\natexlab{b}}){Rivinius}, {Baade},
  {{\v{S}}tefl}, {Stahl}, {Wolf}, \& {Kaufer}}]{1998A&A...333..125R}
{Rivinius}, T., {Baade}, D., {{\v{S}}tefl}, S., {et~al.} 1998{\natexlab{b}},
  \aap, 333, 125

\bibitem[{{Rivinius} {et~al.}(1998{\natexlab{c}}){Rivinius}, {Baade},
  {{\v{S}}tefl}, {Stahl}, {Wolf}, \& {Kaufer}}]{1998A&A...336..177R}
{Rivinius}, T., {Baade}, D., {{\v{S}}tefl}, S., {et~al.} 1998{\natexlab{c}},
  \aap, 336, 177

\bibitem[{{Rivinius} {et~al.}(2001{\natexlab{b}}){Rivinius}, {Baade},
  {{\v{S}}tefl}, {Townsend}, {Stahl}, {Wolf}, \&
  {Kaufer}}]{2001A&A...369.1058R}
{Rivinius}, T., {Baade}, D., {{\v{S}}tefl}, S., {et~al.} 2001{\natexlab{b}},
  \aap, 369, 1058

\bibitem[{{Rivinius} {et~al.}(2013{\natexlab{b}}){Rivinius}, {Carciofi}, \&
  {Martayan}}]{Rivinius2013}
{Rivinius}, T., {Carciofi}, A.~C., \& {Martayan}, C. 2013{\natexlab{b}}, \aapr,
  21, 69

\bibitem[{{Saio} {et~al.}(2018){Saio}, {Kurtz}, {Murphy}, {Antoci}, \&
  {Lee}}]{Saio2018}
{Saio}, H., {Kurtz}, D.~W., {Murphy}, S.~J., {Antoci}, V.~L., \& {Lee}, U.
  2018, \mnras, 474, 2774

\bibitem[{{Semaan} {et~al.}(2018){Semaan}, {Hubert}, {Zorec},
  {Guti{\'e}rrez-Soto}, {Fr{\'e}mat}, {Martayan}, {Fabregat}, \&
  {Eggenberger}}]{Semaan2018}
{Semaan}, T., {Hubert}, A.~M., {Zorec}, J., {et~al.} 2018, \aap, 613, A70

\bibitem[{{Shen} {et~al.}(2023){Shen}, {Li}, {Abdusamatjan}, {Fu}, {Zhu}, {Yu},
  {Zhang}, {L{\"u}}, {Zhai}, \& {Liu}}]{2023ApJ...955..123S}
{Shen}, D.-X., {Li}, G., {Abdusamatjan}, I., {et~al.} 2023, \apj, 955, 123

\bibitem[{{Sigut} \& {Ghafourian}(2023)}]{2023ApJ...948...34S}
{Sigut}, T.~A.~A. \& {Ghafourian}, N.~R. 2023, \apj, 948, 34

\bibitem[{{Smith}(1989)}]{1989ApJS...71..357S}
{Smith}, M.~A. 1989, \apjs, 71, 357

\bibitem[{{Smith} {et~al.}(2016){Smith}, {Lopes de Oliveira}, \&
  {Motch}}]{2016AdSpR..58..782S}
{Smith}, M.~A., {Lopes de Oliveira}, R., \& {Motch}, C. 2016, Advances in Space
  Research, 58, 782

\bibitem[{{Solar} {et~al.}(2022){Solar}, {Arcos}, {Cur{\'e}}, {Levenhagen}, \&
  {Araya}}]{2022MNRAS.511.4404S}
{Solar}, M., {Arcos}, C., {Cur{\'e}}, M., {Levenhagen}, R.~S., \& {Araya}, I.
  2022, \mnras, 511, 4404

\bibitem[{{Stoeckley}(1968)}]{1968MNRAS.140..141S}
{Stoeckley}, T.~R. 1968, \mnras, 140, 141

\bibitem[{{Tokovinin} {et~al.}(2013){Tokovinin}, {Fischer}, {Bonati},
  {Giguere}, {Moore}, {Schwab}, {Spronck}, \&
  {Szymkowiak}}]{2013PASP..125.1336T}
{Tokovinin}, A., {Fischer}, D.~A., {Bonati}, M., {et~al.} 2013, \pasp, 125,
  1336

\bibitem[{{Townsend}(2000{\natexlab{a}})}]{2000ASPC..214..288T}
{Townsend}, R. 2000{\natexlab{a}}, in Astronomical Society of the Pacific
  Conference Series, Vol. 214, IAU Colloq. 175: The Be Phenomenon in Early-Type
  Stars, ed. M.~A. {Smith}, H.~F. {Henrichs}, \& J.~{Fabregat}, 288

\bibitem[{{Townsend}(2000{\natexlab{b}})}]{2000MNRAS.319..289T}
{Townsend}, R.~H.~D. 2000{\natexlab{b}}, \mnras, 319, 289

\bibitem[{{Townsend} {et~al.}(2004){Townsend}, {Owocki}, \&
  {Howarth}}]{2004MNRAS.350..189T}
{Townsend}, R.~H.~D., {Owocki}, S.~P., \& {Howarth}, I.~D. 2004, \mnras, 350,
  189

\bibitem[{{Tubbesing} {et~al.}(2000){Tubbesing}, {Rivinius}, {Wolf}, \&
  {Kaufer}}]{2000ASPC..214..232T}
{Tubbesing}, S., {Rivinius}, T., {Wolf}, B., \& {Kaufer}, A. 2000, in
  Astronomical Society of the Pacific Conference Series, Vol. 214, IAU Colloq.
  175: The Be Phenomenon in Early-Type Stars, ed. M.~A. {Smith}, H.~F.
  {Henrichs}, \& J.~{Fabregat}, 232

\bibitem[{{ud-Doula} {et~al.}(2018){ud-Doula}, {Owocki}, \&
  {Kee}}]{2018MNRAS.478.3049U}
{ud-Doula}, A., {Owocki}, S.~P., \& {Kee}, N.~D. 2018, \mnras, 478, 3049

\bibitem[{{Van Reeth} {et~al.}(2016){Van Reeth}, {Tkachenko}, \&
  {Aerts}}]{2016A&A...593A.120V}
{Van Reeth}, T., {Tkachenko}, A., \& {Aerts}, C. 2016, \aap, 593, A120

\bibitem[{{Vieira} {et~al.}(2017){Vieira}, {Carciofi}, {Bjorkman}, {Rivinius},
  {Baade}, \& {R{\'\i}mulo}}]{2017MNRAS.464.3071V}
{Vieira}, R.~G., {Carciofi}, A.~C., {Bjorkman}, J.~E., {et~al.} 2017, \mnras,
  464, 3071

\bibitem[{{{\v{S}}tefl} {et~al.}(1995){{\v{S}}tefl}, {Baade}, {Harmanec}, \&
  {Balona}}]{1995A&A...294..135S}
{{\v{S}}tefl}, S., {Baade}, D., {Harmanec}, P., \& {Balona}, L.~A. 1995, \aap,
  294, 135

\bibitem[{{{\v{S}}tefl} {et~al.}(2003{\natexlab{a}}){{\v{S}}tefl}, {Baade},
  {Rivinius}, {Otero}, {Stahl}, {Budovi{\v{c}}ov{\'a}}, {Kaufer}, \&
  {Maintz}}]{2003A&A...402..253S}
{{\v{S}}tefl}, S., {Baade}, D., {Rivinius}, T., {et~al.} 2003{\natexlab{a}},
  \aap, 402, 253

\bibitem[{{{\v{S}}tefl} {et~al.}(2003{\natexlab{b}}){{\v{S}}tefl}, {Baade},
  {Rivinius}, {Stahl}, {Budovi{\v{c}}ov{\'a}}, {Kaufer}, \&
  {Maintz}}]{2003A&A...411..167S}
{{\v{S}}tefl}, S., {Baade}, D., {Rivinius}, T., {et~al.} 2003{\natexlab{b}},
  \aap, 411, 167

\bibitem[{{{\v{S}}tefl} {et~al.}(1998){{\v{S}}tefl}, {Baade}, {Rivinius},
  {Stahl}, {Wolf}, \& {Kaufer}}]{1998ASPC..135..348S}
{{\v{S}}tefl}, S., {Baade}, D., {Rivinius}, T., {et~al.} 1998, in Astronomical
  Society of the Pacific Conference Series, Vol. 135, A Half Century of Stellar
  Pulsation Interpretation, ed. P.~A. {Bradley} \& J.~A. {Guzik}, 348

\bibitem[{{Wade} {et~al.}(2016){Wade}, {Petit}, {Grunhut}, {Neiner}, \& {MiMeS
  Collaboration}}]{Wade2016}
{Wade}, G.~A., {Petit}, V., {Grunhut}, J.~H., {Neiner}, C., \& {MiMeS
  Collaboration}. 2016, in Astronomical Society of the Pacific Conference
  Series, Vol. 506, Bright Emissaries: Be Stars as Messengers of Star-Disk
  Physics, ed. T.~A.~A. {Sigut} \& C.~E. {Jones}, 207

\bibitem[{{Walker} {et~al.}(2005){Walker}, {Kuschnig}, {Matthews}, {Cameron},
  {Saio}, {Lee}, {Kambe}, {Masuda}, {Guenther}, {Moffat}, {Rucinski},
  {Sasselov}, \& {Weiss}}]{Walker2005}
{Walker}, G.~A.~H., {Kuschnig}, R., {Matthews}, J.~M., {et~al.} 2005, \apjl,
  635, L77

\bibitem[{{Wang} {et~al.}(2018){Wang}, {Gies}, \&
  {Peters}}]{2018ApJ...853..156W}
{Wang}, L., {Gies}, D.~R., \& {Peters}, G.~J. 2018, \apj, 853, 156

\bibitem[{{Weiss} {et~al.}(2014){Weiss}, {Rucinski}, {Moffat},
  {Schwarzenberg-Czerny}, {Koudelka}, {Grant}, {Zee}, {Kuschnig}, {Mochnacki},
  {Matthews}, {Orleanski}, {Pamyatnykh}, {Pigulski}, {Alves}, {Guedel},
  {Handler}, {Wade}, \& {Zwintz}}]{2014PASP..126..573W}
{Weiss}, W.~W., {Rucinski}, S.~M., {Moffat}, A.~F.~J., {et~al.} 2014, \pasp,
  126, 573

\bibitem[{{Yudin}(2001)}]{2001A&A...368..912Y}
{Yudin}, R.~V. 2001, \aap, 368, 912

\bibitem[{{Zorec} {et~al.}(2016){Zorec}, {Fr{\'e}mat}, {Domiciano de Souza},
  {Royer}, {Cidale}, {Hubert}, {Semaan}, {Martayan}, {Cochetti}, {Arias},
  {Aidelman}, \& {Stee}}]{zorec2016}
{Zorec}, J., {Fr{\'e}mat}, Y., {Domiciano de Souza}, A., {et~al.} 2016, \aap,
  595, A132

\end{thebibliography}

\onecolumn
\begin{appendix} 

\section{Tables} \label{sec:appendix_tbls}

This appendix section includes tables describing the sample and their properties. Table~\ref{tbl:sample} gives the target list with information from the literature and catalogs. Table~\ref{tbl:measurements} provides information for the mass ejection events that were sampled in this work, including the emission asymmetry frequencies. Table~\ref{tbl:zorec} includes stellar properties and the orbital, critical, and rotational frequencies as derived from \citep{zorec2016}, and Tbl.~\ref{tbl:obs_log} lists the observing log.

\begin{table*}[!h]
\caption{Target list. }
\label{tbl:sample}      
\centering                                      
\begin{tabular}{c c c c c c c}          
\hline\hline
 ID        & HD        & TESS ID   & Spectral & Vmag   & ${v \sin i}$     & $i$           \\    
           &           &           & Type     &        & (km s$^{-1}$) &  ($^{\circ}$) \\
\hline  
\multicolumn{7}{c}{This work}\\
\hline
\object{V767 Cen}   &  120991 & 307225534    & B2Ve     &  6.1   & 106 $\pm$ 11 (1)  &  $25^{\circ} \pm 7^{\circ}$ (1)\\
\object{f Car}      &   75311  & 46028220    & B3Vne    &  4.49  & 268 $\pm$ 18 (1)  &  $66^{\circ} \pm 16^{\circ}$ (1)   \\
\object{12 Vul}     &  187811 & 299630806    & B2.5Ve   &  4.96  & 264 $\pm$ 25 (1)  &  $52^{\circ} \pm 13^{\circ}$ (1) \\
           &           &            &          &        &                     &   $60^{\circ} \pm 5^{\circ,}$ (2) \\
\object{V442 And}   &  6226   & 196501216    & B2.5IIIe &  6.82  & $57 \substack{+42 \\ -18}$ (3)  & $13.4^{\circ} \substack{+2.5 \\ -3.0}$ (3) \\
\object{$\lambda$ Pav}    &  173948 & 375232307   & B2Ve     &  4.21  & 145 $\pm$ 12 (1)  & $36^{\circ} \pm 10^{\circ}$ (1) \\
           &           &            &          &        &  130 $\pm$ 15 (5) &   \\
\object{$\iota$ Lyr}    & 178475 & 120967488    & B6IVe    &  5.25  &  ...        &  intermediate/high       \\
\object{28 Cyg}     &  191610 & 42360166     & B2.5Ve   &  4.93  & 314 $\pm$ 33 (1) & $69^{\circ} \pm 17^{\circ}$ (1) \\
           &           &              &          &        &                    &  $40^{\circ} \pm 5^{\circ}$ (2)\\
\object{V357 Lac}   & 212044 & 431116093   & B2e      &  6.98  &  ...           & low   \\
\object{OT Gem}     & 58050  & 14498757     & B2Ve     &  6.41  & 138 $\pm$ 12 (1)  & $31^{\circ} \pm 12^{\circ}$ (1) \\
           &           &            &          &        &                     &  $21^{\circ} \pm 6^{\circ}$ (2) \\
\object{25 Ori}     & 35439  & 264459943   & B1Vne    &  4.96  & 284 $\pm$ 28 (1)  & $58^{\circ} \pm 14^{\circ}$ (1)  \\
           &           &              &          &        &                     &  $63^{\circ} \pm 5^{\circ}$ (2) \\
           &           &              &          &        &                     &   $55^{\circ} \pm 5^{\circ}$ (4) \\
\object{$\kappa$ CMa}    & 50013  & 52982382     & B1.5Ve   &  3.89  & 231 $\pm$ 25 (1)  & $50^{\circ} \pm 17^{\circ}$ (1)\\
           &           &             &          &        &                     & $57^{\circ} \pm 13^{\circ}$ (2) \\
           &           &              &          &        &                     & $53^{\circ} \pm 10^{\circ}$ (4) \\
\object{j Cen}      &  102776 & 325170579   & B3Ve     &  4.31  &  ...        &  intermediate      \\
\object{120 Tau}    &  36576  & 437790952    & B2IV-Ve  &  5.69  & 282 $\pm$ 29 (1) & $57^{\circ} \pm 14^{\circ}$ (1) \\
           &           &              &          &        &                    &  $61^{\circ} \pm 9^{\circ}$ (2) \\
           &           &              &          &        &                    &  $60^{\circ} \pm 5^{\circ}$ (4) \\
\hline
\multicolumn{7}{c}{Literature sample}\\
\hline
\object{$\mu$ Cen}  &  120324 & 243647020   & B2Vnpe   &  3.43  &  149 $\pm$ 13 (1)  & $30^{\circ} \pm 11^{\circ}$ (1);  $38^{\circ} \pm 11^{\circ}$ (2)\\
\object{$\omega$ CMa}  &   56139 & 65903024    & B2.5Ve  &  3.82  &  94 $\pm$ 13 (1)  & $24^{\circ} \pm 8^{\circ}$ (1); $37^{\circ} \pm 5^{\circ}$ (2)\\
\object{$\omega$ Ori}  &  37490 & 281047621    & B3Ve   &  4.59  &  190 $\pm$ 21 (1)  & $50^{\circ} \pm 12^{\circ}$ (1); $59^{\circ} \pm 10^{\circ}$ (2) \\
\object{$\upsilon$ Cyg}  &  202904 & 117146835   & B2Vne   &  4.42  &  193 $\pm$ 23 (1)  & $38^{\circ} \pm 9^{\circ}$ (1); $50^{\circ} \pm 13^{\circ}$ (2) \\
\object{EW Lac}    &   217050 & 252500679   & B4IIIpe   &  5.43  &  358 $\pm$ 37 (1)  & $72^{\circ} \pm 18^{\circ}$ (1) \\
\object{$\eta$ Cen}    &  127972 & 128116539    & B2Ve     &  2.31  &  328 $\pm$ 33 (1)  & $67^{\circ} \pm 16^{\circ}$ (1); $77^{\circ} \pm 5^{\circ}$ (2)  \\
\hline      
\end{tabular}
\tablefoot{References for ${v \sin i}$ and $i$ are given in parenthesis. For stars without numerical inclination angles, the inclination is qualitatively described based on the emission line morphology. }
\tablebib{
(1)~\citet{zorec2016}; (2) \citet{2023ApJ...948...34S}; (3) \citet{2021MNRAS.508.2002R}; (4) \citet{2019A&A...621A.123C};
(5) \citet{2011A&A...533A..75L}.
}
\end{table*}

\begin{table*}
\caption{Information about the well-sampled flickers. }
\label{tbl:measurements}      
\centering                                      
\begin{tabular}{c c c c c c c c}          
\hline\hline
 ID        & Flicker   & Em. osc.            & Method             & Num.     & t$_{b}$/t$_{d}$ &   Amp.      & Spec puls              \\    
           & types     & freq (d$^{-1}$)     &                    & Cycles   &   (d)           &   (\% flx)  & freq (d$^{-1}$)     \\
\hline  
\multicolumn{8}{c}{This work}\\
\hline
V767 Cen   &  Clear    & 0.773 $\pm$ 0.025   &  H$\alpha$ wings EW$_{\rm V}$/EW$_{\rm R}$\xspace  &  3$^*$  &  4.3/...    &  7.5   & 0.97867     \\
           &  Clear    & 0.865 $\pm$ 0.012   &  H$\alpha$ wings EW$_{\rm V}$/EW$_{\rm R}$\xspace  &  7  &  9.3/...    &  14.9  &         \\
f Car      &  Pristine & 1.674 $\pm$ 0.036   &  H$\alpha$ EW$_{\rm V}$/EW$_{\rm R}$\xspace        &  5  &  0.9/3.1   &  1.8   & 2.0898, 3.7520, 3.8370        \\
           &  Pristine & 1.722 $\pm$ 0.032   &  H$\alpha$ EW$_{\rm V}$/EW$_{\rm R}$\xspace        &  6  &  2.5/4.7   &  8.3   &         \\
12 Vul     &  Complex  & 1.765 $\pm$ 0.083   &  H$\alpha$ EW$_{\rm V}$/EW$_{\rm R}$\xspace        &  2$^*$  &  1.3/...    &  3.4   & 2.0158, 2.1077     \\
           &  Complex  & 1.916 $\pm$ 0.069   &  H$\alpha$ EW$_{\rm V}$/EW$_{\rm R}$\xspace        &  4  &  2.1/...    &  4.8   &         \\
           &  Complex  & 1.60 $\pm$ 0.15     &  H$\alpha$ EW$_{\rm V}$/EW$_{\rm R}$\xspace        &  1$^{**}$  &  ...     &  ...    &         \\
V442 And   &  Pristine & 0.348 $\pm$ 0.007   &  H$\alpha$ EW$_{\rm V}$/EW$_{\rm R}$\xspace       &  6  & ...        &  ...   & 0.382332 (11)    \\
           &  Pristine & 0.353 $\pm$ 0.004   &  H$\alpha$ EW$_{\rm V}$/EW$_{\rm R}$\xspace        &  9  & ...        &  ...   &         \\
           &  Clear    & 0.358 $\pm$ 0.005   &  H$\alpha$ EW$_{\rm V}$/EW$_{\rm R}$\xspace        &  9  & ...        &  ...   &         \\
           &  Pristine & 0.356 $\pm$ 0.004   &  H$\alpha$ EW$_{\rm V}$/EW$_{\rm R}$\xspace        &  10 & ...        &  ...   &         \\
           &  Clear    & 0.351 $\pm$ 0.004   &  H$\alpha$ EW$_{\rm V}$/EW$_{\rm R}$\xspace        &  10 & ...        &  ...   &         \\
           &  Pristine & 0.352 $\pm$ 0.007   &  H$\alpha$ EW$_{\rm V}$/EW$_{\rm R}$\xspace        &  6  & ...        &  ...   &         \\
           &  Pristine & 0.362 $\pm$ 0.005   &  H$\alpha$ EW$_{\rm V}$/EW$_{\rm R}$\xspace        &  10 & 10.7/...    &  4.4   &         \\
           &  Pristine & 0.332 $\pm$ 0.009   &  H$\alpha$ EW$_{\rm V}$/EW$_{\rm R}$\xspace        &  4  & ...        &  ...   &         \\
           &  Pristine & 0.341 $\pm$ 0.005   &  H$\alpha$ EW$_{\rm V}$/EW$_{\rm R}$\xspace        &  9  & ...        &  ...   &         \\
$\lambda$ Pav    &  Pristine & 0.810 $\pm$ 0.014   &  H$\alpha$ EW$_{\rm V}$/EW$_{\rm R}$\xspace        &   7$^{**}$  &   ...   &  ...   & 0.49, 0.82, 1.63 (12)       \\
$\iota$ Lyr    &  Pristine &  0.983 $\pm$ 0.016  & H$\alpha$ EW$_{\rm V}$/EW$_{\rm R}$\xspace    &  6  &  ...   &  ...    &           \\
28 Cyg     &  Clear    & 1.415 $\pm$ 0.049   &  H$\alpha$ EW$_{\rm V}$/EW$_{\rm R}$\xspace        &  3  & 2.0/7.5   &  5.7   & 1.54562, 1.59726 (13)       \\
V357 Lac   &  Clear    & 1.085 $\pm$ 0.033   &  H$\alpha$ wings EW$_{\rm V}$/EW$_{\rm R}$\xspace  & 6$^*$  &    &     &         \\
           & Complex   & 1.092 $\pm$ 0.062   &  H$\alpha$ wings EW$_{\rm V}$/EW$_{\rm R}$\xspace  & 4$^{**}$  & 3.0/...   &  10.5   &         \\
           & Clear     & 1.165 $\pm$ 0.028   &  H$\alpha$ wings EW$_{\rm V}$/EW$_{\rm R}$\xspace  & 5$^{**}$  & 4.3/...   &  17.5   &         \\
OT Gem     &  Clear    & 2.161 $\pm$ 0.025   &  H$\alpha$ wings EW$_{\rm V}$/EW$_{\rm R}$\xspace   &   9  &  7.6/...    &  20.0   &  2.494          \\
           &  Clear    & 2.000 $\pm$ 0.029   &  H$\alpha$ wings EW$_{\rm V}$/EW$_{\rm R}$\xspace   &  6$^{**}$   &  1.6/...   &   1.2   &            \\
           &  Clear    & 2.086 $\pm$ 0.034   &  H$\alpha$ wings EW$_{\rm V}$/EW$_{\rm R}$\xspace   &  9  &  4.9/3.9  &    13.0 &            \\
25 Ori     &  Complex  & 1.568 $\pm$ 0.049   &  H$\alpha$ wings EW$_{\rm V}$/EW$_{\rm R}$\xspace   & 3.5 &  ...   &  ...    & 1.6793        \\
           &  Complex  & 1.565 $\pm$ 0.047   &  H$\alpha$ wings EW$_{\rm V}$/EW$_{\rm R}$\xspace   & 3.5$^{**}$ &  ...   &  ...   &         \\
$\kappa$ CMa    &  Complex  &  1.546 $\pm$ 0.020  &  H$\alpha$ wings EW$_{\rm V}$/EW$_{\rm R}$\xspace   &  8   & ...   &  ...   &  1.8256       \\
                &  Complex  &  1.480 $\pm$ 0.022  &  H$\alpha$ wings EW$_{\rm V}$/EW$_{\rm R}$\xspace   &  6   & ...   &  ...   &               \\
j Cen      &  Complex  &  1.719 $\pm$ 0.016  &   H$\alpha$ wings EW$_{\rm V}$/EW$_{\rm R}$\xspace &    12? &  ...  &  ...    & 1.9650       \\
120 Tau    &  Complex  & 0.966 $\pm$ 0.016   & H$\alpha$ wings EW$_{\rm V}$/EW$_{\rm R}$\xspace   &   7  &   ...   &  ...   &  1.2400      \\
           &  Complex  & 1.048 $\pm$ 0.022   &  H$\alpha$ wings EW$_{\rm V}$/EW$_{\rm R}$\xspace  &   5$^*$  &   ...   &  ...   &              \\
           &  Complex  & 0.975 $\pm$ 0.024   &  H$\alpha$ wings EW$_{\rm V}$/EW$_{\rm R}$\xspace  &   5$^{**}$  &   ...   &  ...   &              \\
\hline
\multicolumn{8}{c}{From the literature}\\
\hline
$\mu$ Cen       &    ...       &  1.608 $\pm$ 0.013 (1)  & \ion{He}{I} V/R   & ... & ...  & ... & 1.98837, 1.97037, 2.02215 (2)    \\
                &    ...    &  1.686 $\pm$ 0.026 (1)  & \ion{He}{I} V/R   & ... & ... & ... & 3.55360, 3.58247 (2)    \\
                &    ...    &  1.72 (1)  &   H$\beta$ V/R   &  ...  &  ...    &  ...    &        \\
                &    ...    &  1.4 (1)  &   \ion{He}{I}\,$\lambda$ 6678 V/R   &  ...  &   ...   &  ...    &        \\
$\omega$ CMa    &    ...        &  0.680 (3)  &  \ion{Fe}{II}\, lines V/R    & ...   &  ...    &  ...    & 0.7289 (4)       \\
$\omega$ Ori    &   ...        & 0.46 (5)   &  \ion{He}{I}\,$\lambda \lambda$ 5876, 6678 V/R    & ...   &  ...    &  ...    &  1.03 (5)       \\
$\upsilon$ Cyg  &    ...       &  1.5 (6) &  \ion{He}{I}\,$\lambda$ 6678 V/R     & ...   &  ...    &  ...    & 2.95, 2.6 (6)       \\
EW Lac          &    ...       &  1.22 (7)  &  \ion{He}{I}\,$\lambda$ 6678 V/R    & ...   &  ...    &   ...   & 1.39, 1.55, 2.76 (7)      \\
$\eta$ Cen      &    ...       &  1.56 (8,9)  &  \ion{Si}{III}\,$\lambda$ 4553    &  ...  &  ...    &   ...   & 1.75 (8,9)       \\
                &    ...       &  0.61 (10)  &  \ion{He}{I}, \ion{Fe}{II}, \ion{Mg}{II}, \ion{Si}{II}    &  ...  &   ...   &   ...   & 1.48, 3.81, 5.31 (10)       \\
\hline      
\end{tabular}
\tablefoot{The columns are as follows. \textit{Flicker types}: `Pristine' refers to events where there was no pre-existing disk. `Clear' indicates flickers where there was a pre-existing disk but the flicker is well defined. `Complex' refers to mass ejections where it is difficult to determine exactly when events start and/or end. \textit{Em. osc. freq}: The frequency of the emission asymmetry oscillations. \textit{Method}: The method used to measure the emission asymmetry oscillations, where `H$\alpha$ EW$_{\rm V}$/EW$_{\rm R}$\xspace' made use of the full line profile and `H$\alpha$ wings EW$_{\rm V}$/EW$_{\rm R}$\xspace' used just the high-velocity emission wings. \textit{Num. Cycles}: The approximate number of emission asymmetry cycles sampled by the data for a given event. The superscript $^*$ indicates that the event was interrupted by another event, and $^{**}$ indicates a gap in observations (i.e., the reported cycle numbers are a lower limit). \textit{t$_{b}$/t$_{d}$}: The build-up and dissipation times from TESS for isolated and well-defined flickers are given in days whenever it was possible to measure these quantities. \textit{Amp.}: The maximum photometric amplitude of the flicker whenever measurements were possible. \textit{Spec puls freq}: Spectroscopic pulsation frequencies determined from absorption line profile variations. Each row corresponds to a different flicker event. References for quantities from the literature are given in parenthesis. }
\tablebib{
(1)~\citet{1998A&A...333..125R}; (2) \citet{1998A&A...336..177R}; (3) \citet{2003A&A...402..253S}; (4) \citet{2003A&A...411..167S};
(5) \citet{2002A&A...388..899N}; (6) \citet{2005A&A...437..257N}; (7) \citet{2000A&A...362.1020F}; (8) \citet{1998ASPC..135..348S}; (9) \citet{1995A&A...294..135S}; (10) \citet{2003A&A...400..599L}; (11) \citet{2021MNRAS.508.2002R}; (12) \citet{2011A&A...533A..75L}; (13) \citet{2000ASPC..214..232T}.
}
\end{table*}

\begin{table*}
\caption{Fundamental parameters and frequencies computed for the stars in Tab.~\ref{tbl:sample} with corresponding data in \citet{zorec2016}.}
\label{tbl:zorec}      
\centering     
\begin{tabular}{c c c c c c c}          
\hline\hline
 ID & Mass & $R_{\rm e}$ & $R_{\rm p}$ & 
 $f_{\rm orb}$ & $f_{\rm crit}$ & $f_{\rm rot}$
 \\    
 & ($\rm M_\sun$) & ($\rm R_\sun$) & ($\rm R_\sun$) & (d$^{-1}$\xspace) & (d$^{-1}$\xspace) & (d$^{-1}$\xspace)\\
\hline  
V767 Cen & 18.0 $\pm$ 2.5 & 12.3 $\pm$ 2.7 & 10.5 $\pm$ 2.2 & 0.85 $\pm$ 0.22 & 0.58 $\pm$ 0.14 & 0.40 $\pm$ 0.14 \\
f Car & 7.60 $\pm$ 0.30 & 5.8 $\pm$ 1.3 & 4.6 $\pm$ 1.0 & 1.71 $\pm$ 0.43 & 1.33 $\pm$ 0.33 & 1.11 $\pm$ 0.30 \\
12 Vul & 8.10 $\pm$ 0.30 & 5.3 $\pm$ 1.9 & 4.3 $\pm$ 1.5 & 2.02 $\pm$ 0.79 & 1.47 $\pm$ 0.57 & 1.25 $\pm$ 0.51 \\
$\lambda$ Pav & 12.90 $\pm$ 0.80 & 10.7 $\pm$ 2.2 & 9.0 $\pm$ 1.8 & 0.88 $\pm$ 0.21 & 0.62 $\pm$ 0.14 & 0.46 $\pm$ 0.15 \\
28 Cyg & 10.50 $\pm$ 0.50 & 7.7 $\pm$ 1.7 & 6.1 $\pm$ 1.3 & 1.31 $\pm$ 0.33 & 1.01 $\pm$ 0.25 & 0.86 $\pm$ 0.24 \\
OT Gem & 10.70 $\pm$ 0.80 & 6.3 $\pm$ 1.1 & 5.2 $\pm$ 0.9 & 1.78 $\pm$ 0.36 & 1.28 $\pm$ 0.24 & 0.84 $\pm$ 0.34 \\
25 Ori & 12.80 $\pm$ 0.90 & 8.1 $\pm$ 1.9 & 6.6 $\pm$ 1.5 & 1.33 $\pm$ 0.36 & 0.98 $\pm$ 0.26 & 0.81 $\pm$ 0.24 \\
$\kappa$ CMa & 13.1 $\pm$ 2.9 & 4.8 $\pm$ 2.9 & 4.3 $\pm$ 2.6 & 2.96 $\pm$ 2.02 & 1.91 $\pm$ 1.29 & 1.24 $\pm$ 0.82 \\
120 Tau & 15.5 $\pm$ 1.0 & 8.9 $\pm$ 2.0 & 7.2 $\pm$ 1.6 & 1.27 $\pm$ 0.32 & 0.95 $\pm$ 0.24 & 0.74 $\pm$ 0.22 \\
\hline  
$\mu$ Cen & 10.8 $\pm$ 1.3 & 5.3 $\pm$ 2.6 & 4.5 $\pm$ 2.2 & 2.30 $\pm$ 1.24 & 1.60 $\pm$ 0.86 & 1.10 $\pm$ 0.65 \\
$\omega$ CMa & 9.90 $\pm$ 0.20 & 9.0 $\pm$ 4.6 & 7.6 $\pm$ 3.9 & 1.00 $\pm$ 0.57 & 0.70 $\pm$ 0.40 & 0.51 $\pm$ 0.31 \\
$\omega$ Ori & 11.10 $\pm$ 0.70 & 12.7 $\pm$ 3.7 & 10.4 $\pm$ 3.0 & 0.63 $\pm$ 0.21 & 0.47 $\pm$ 0.15 & 0.39 $\pm$ 0.14 \\
$\nu$ Cyg & 9.50 $\pm$ 0.50 & 5.8 $\pm$ 1.9 & 4.8 $\pm$ 1.6 & 1.91 $\pm$ 0.71 & 1.37 $\pm$ 0.50 & 1.07 $\pm$ 0.43 \\
EW Lac & 8.80 $\pm$ 0.60 & 7.2 $\pm$ 3.1 & 5.6 $\pm$ 2.4 & 1.31 $\pm$ 0.63 & 1.04 $\pm$ 0.50 & 1.03 $\pm$ 0.47 \\
$\eta$ Cen & 10.20 $\pm$ 0.50 & 6.2 $\pm$ 1.5 & 4.9 $\pm$ 1.1 & 1.77 $\pm$ 0.47 & 1.37 $\pm$ 0.36 & 1.13 $\pm$ 0.32 \\
\hline      
\end{tabular}
\end{table*}

\begin{table*}
\caption{Observing log. }
\label{tbl:obs_log}      
\centering     
\begin{tabular}{c c c c c}          
\hline\hline
ID              & Num. spectra             &   TESS     & Num. spectra   & Date range total    \\    
                & during TESS              &   Sec.     &  total         &   (TJD)        \\    
\hline  
V767 Cen        & 249 (N), 23 (B)          &   64, 65   & 293 (N), 26 (B)        & 3025 -- 3115   \\
f Car           & 225 (N)                  &   62, 63   & 236 (N)        & 2979 -- 3043   \\
12 Vul          & 323 (N), 27 (D), 53 (B)  &     54     & 364 (N), 68 (D), 94 (B)        & 2745 -- 2807   \\
V442 And  &  6 (N), 321 (B)          & 17, 57, 58 & 6 (N), 1217 (B)        &  956 -- 3374   \\
$\lambda$ Pav   & 136 (N), 262 (C)         &     67     & 154 (N), 518 (C)        & 3101 -- 3196   \\
$\iota$ Lyr     & 48 (N), 63 (D), 55 (B)   &   53, 54   & 51 (N), 72 (D), 59 (B)        & 2713 -- 2802   \\
28 Cyg          & 83 (N), 63 (B)           &   54, 55   & 83 (N), 14 (D), 64 (B)        & 2708 -- 2825   \\
V357 Lac        & 24 (N), 83 (B)           &   16, 17   & 24 (N), 98 (B)        & 1700 -- 1850   \\
OT Gem          & 133 (N)                  &   71, 72   & 137 (N)        & 3235 -- 3294   \\
25 Ori          & 170 (N)                  &     32     & 198 (N)        & 2096 -- 2237   \\
$\kappa$ CMa    & 279 (N)                  &   33, 34   & 289 (N)        & 2197 -- 2246   \\
j Cen           & 157 (N)                  &   37, 38   & 179 (N)        & 2285 -- 2362   \\
120 Tau         & 130 (N)                  &   44, 45   & 142 (N)        & 2487 -- 2552   \\
\hline      
\end{tabular}
\tablefoot{The second column lists the number of spectra taken during the relevant TESS sectors, printed in the third column. The letters in parenthesis after the number of spectra indicate the source of the spectra: N = NRES, B = BeSS, D = DAO, and C = CHIRON. The number of spectra total, in the fourth column, give the quantity of observations surrounding and including the dates of the TESS observations, corresponding to the date range total given in the last column. For V442 And, the number of spectra in the `TESS' column corresponds to the spectra used to measure the EW$_{\rm V}$/EW$_{\rm R}$\xspace frequencies in the nine events shown in Fig.~\ref{fig:V442_And}, and the number in the next column is the total number of observations as shown in the top two panels of Fig.~\ref{fig:V442_And}. The TESS sectors (\url{https://tess.mit.edu/observations/}) are those used in this study with contemporaneous spectroscopy, but additional TESS data are available for all targets.}
\end{table*}

\clearpage

\section{Asymmetry cycles in the rest of the sample} \label{sec:appendix_b}

An analysis of the photometric and spectroscopic quantities are presented here for the remaining 10 stars in the sample. They are roughly ordered according to the observed behavior. V442~And, $\lambda$ Pav, and $\iota$ Lyr present ``pristine'' flickers where there was little to no preexisting disk. 28~Cyg, V357~Lac, OT~Gem, exhibit ``clear'' flickers that are well defined but with a preexisting disk. The remainder, 25~Ori, $\kappa$~CMa, j~Cen, and 120~Tau show complex variability that likely includes mass ejection but with discrete events being difficult to pinpoint.

\subsection{V442 And (HD 6226; B2.5IIIe)} \label{sec:V442_And}

This system has a low inclination angle, determined to be $i = 13.4^{\circ} \substack{+2.5 \\ -3.0}$, and with $v \sin i$ = $57 \substack{+42 \\ -18}$ km s$^{-1}$ \citep{2021MNRAS.508.2002R}. The behavior of V442 And has been described in detail by \citet{2021MNRAS.508.2002R}, and has many convenient properties for studying various aspects of Be stars and their disks. Its outbursts are cyclic, almost periodic, repeating every $\sim$87 days (but sometimes skipping a cycle). Often, the H$\alpha$ emission from a previous event has fully dissipated by the time the next outburst begins, providing a pristine environment for the new ejecta. 

\begin{figure*}
   \centering
   \includegraphics[width=1.0\hsize]{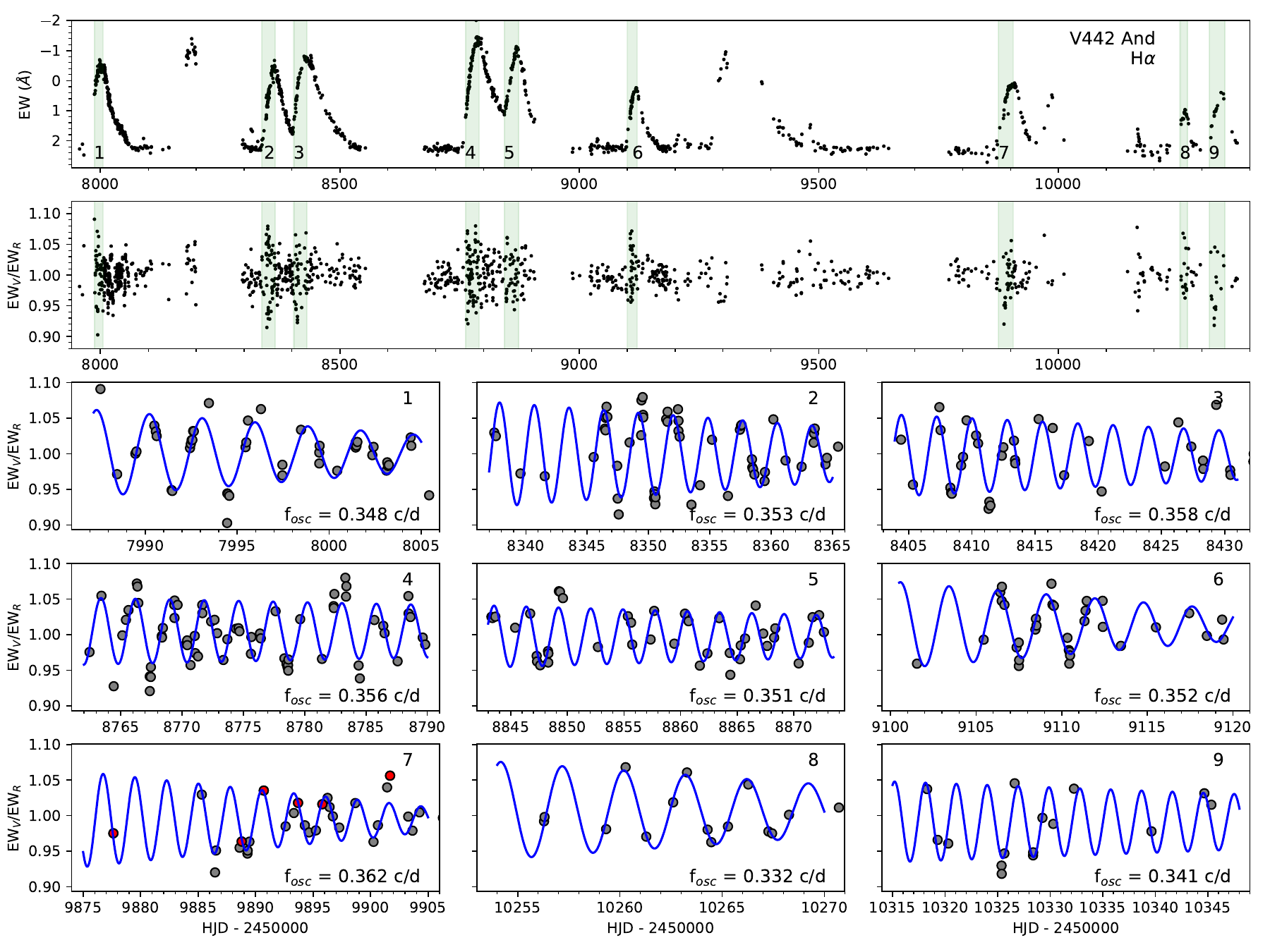}
      \caption{Measurements of H$\alpha$ observations primarily from BeSS of the Be star V442~And. Top panel: EW. Second panel: EW$_{\rm V}$/EW$_{\rm R}$\xspace. Bottom panels: First many days of the outburst events 1 -- 9 marked in the top panel with time coverage corresponding to the vertical shaded regions (NRES spectra plotted as red circles). The frequency of the EW$_{\rm V}$/EW$_{\rm R}$\xspace oscillations are given in the bottom-right corner of these panels. 
              }
         \label{fig:V442_And}
         
   \end{figure*}

Figure~\ref{fig:V442_And} shows measurements of the H$\alpha$ profiles mostly from the BeSS database. Nine outburst events have sufficient spectroscopic coverage of the first $\sim$15 -- 30 days such that the evolution of the EW$_{\rm V}$/EW$_{\rm R}$\xspace can be inspected. The EW and EW$_{\rm V}$/EW$_{\rm R}$\xspace values were calculated between -250 km s$^{-1}$ to +250 km s$^{-1}$. All of these events displayed the same EW$_{\rm V}$/EW$_{\rm R}$\xspace patterns, which map to the photometric frequencies seen in TESS and the stable pulsation frequencies in the same manner as the other systems in our sample. The main difference is that all observed frequencies are relatively low compared to the rest of the sample -- the EW$_{\rm V}$/EW$_{\rm R}$\xspace frequencies are $\sim$0.35 \,d$^{-1}$\xspace, and the dominant pulsation frequency is about 0.38 \,d$^{-1}$\xspace. 

\begin{figure*}
   \centering
   \includegraphics[width=1.0\hsize]{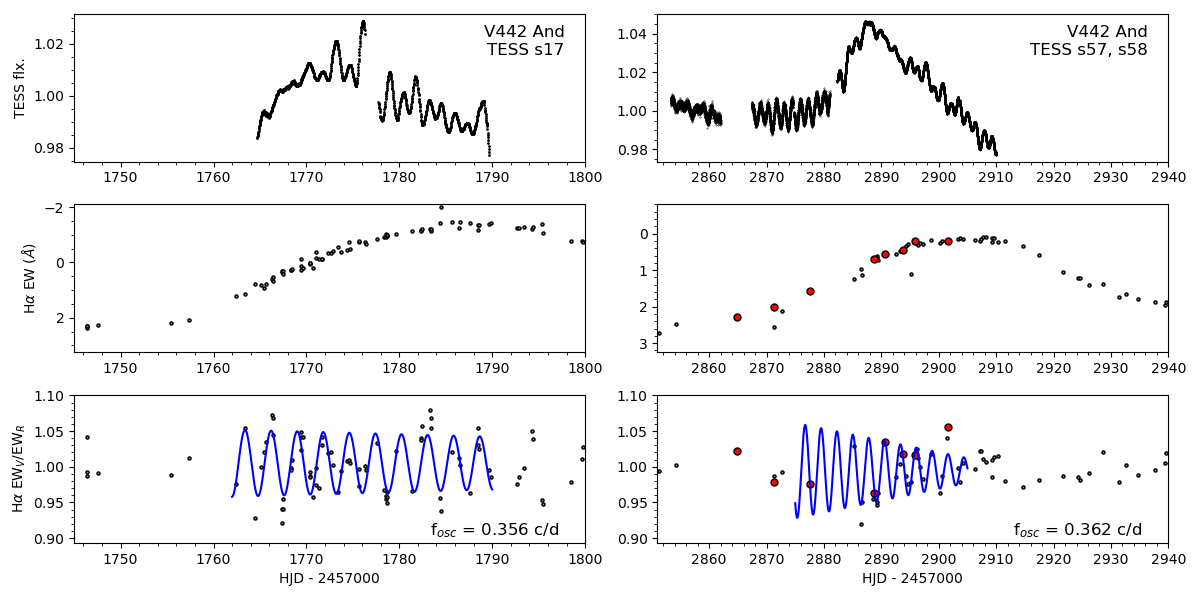}
      \caption{Reproduction of the H$\alpha$ EW and EW$_{\rm V}$/EW$_{\rm R}$\xspace measurements for events 4 and 7 in Fig.~\ref{fig:V442_And}, with the contemporaneous TESS light curves shown in the top panels. 
              }
         \label{fig:V442_And2}
   \end{figure*}

Fig.~\ref{fig:V442_And2} reproduces data from Fig.~\ref{fig:V442_And} for the two events that coincided with TESS observations. In TESS sector 17, only part of the flicker was observed by TESS, but the spectroscopic data are particularly dense. In TESS sectors 57+58, the flicker was more completely sampled and shows a more typical shape.  Interestingly, in sector 58 the TESS flux during the dissipation phase seems to dip below the pre-flicker brightness. Although the TESS time baselines are short compared to the duration of these events, it is interesting to note that in sector 17 the enhancement in high frequency variation occurred mostly after peak brightness, while in sectors 57+58 the opposite is true.


\subsection{$\lambda$ Pav (HD 173948; B2Ve)} \label{sec:lam_Pav}

The disk of $\lambda$ Pav has varied from being at times absent, to other times being of moderate strength \citep[e.g.,][]{2011A&A...533A..75L}. \citet{zorec2016} found ${v\sin i}$ = 145 $\pm$ 12 km s$^{-1}$ and ${i = 36^{\circ} \pm 10^{\circ}}$. \citet{2011A&A...533A..75L} derived ${v\sin i = 130 \pm 15\,\rm km\,s^{-1}}$ and ${i = 30^{\circ} \pm 2^{\circ}}$, in agreement, within errors, with the values of \citet{zorec2016}.

\citet{2003A&A...411..229R} found LPVs in the spectra of $\lambda$\,Pav, but with only 11 spectra over 70 days periodicity could not be established. Multiple frequencies were found in a time-series analysis of line profile variations, including at 0.17 $\pm$ 0.02 \,d$^{-1}$\xspace, 0.49 $\pm$ 0.05 \,d$^{-1}$\xspace, 0.82 $\pm$ 0.03 \,d$^{-1}$\xspace, and 1.63 $\pm$ 0.04 \,d$^{-1}$\xspace \citep{2011A&A...533A..75L}. A 0.16 \,d$^{-1}$\xspace signal (P = 6.18 d) was found in analysis of Hipparcos photometry \citep{2002MNRAS.331...45K}. In all three available sectors of TESS data, the 0.16 \,d$^{-1}$\xspace signal is dominant, which corresponds to the difference between two higher frequency (pulsational) signals evident in TESS (at 1.641 \,d$^{-1}$\xspace and 1.481 \,d$^{-1}$\xspace). This 0.16 \,d$^{-1}$\xspace signal appears phase coherent between the two available TESS timeseries (sectors 13 and 66+67) separated by about four years.

\begin{figure*}
   \centering
   \includegraphics[width=1.0\hsize]{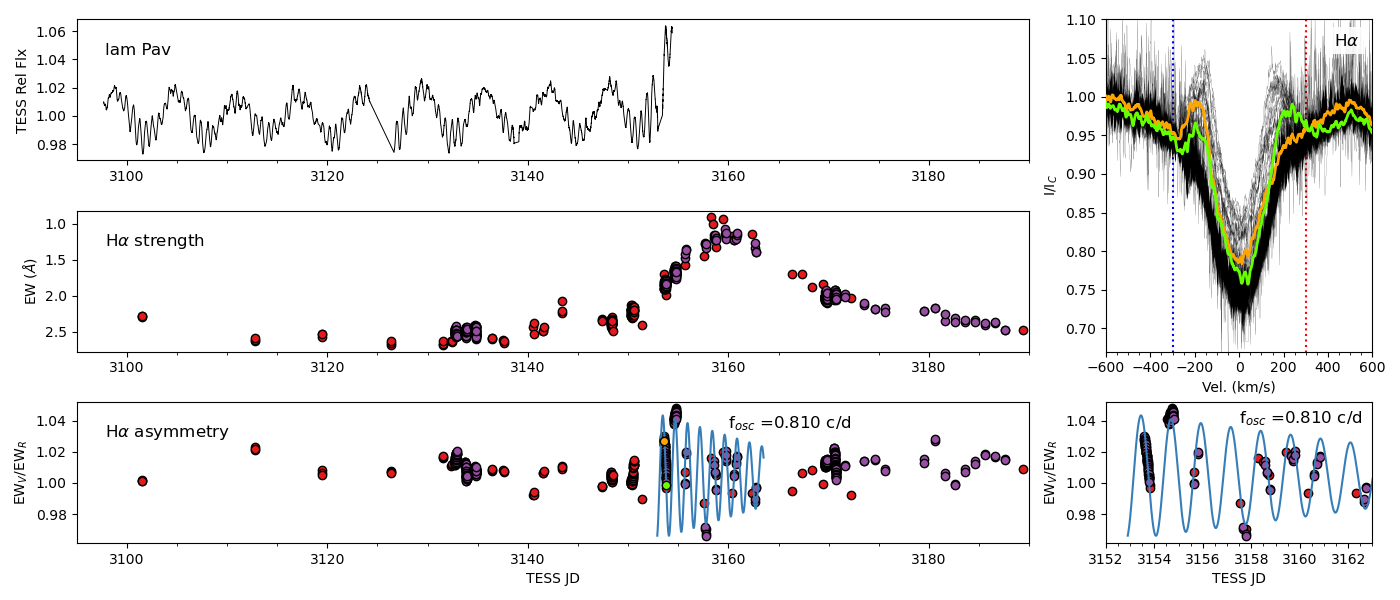}
      \caption{TESS light curve and measurements of H$\alpha$ observations of the Be star $\lambda$\,Pav, similar to Fig.~\ref{fig:f_Car}. The red points are from NRES, and purple from the CHIRON spectrograph. Bottom-right panel: Zoom in on the H$\alpha$ EW$_{\rm V}$/EW$_{\rm R}$\xspace values fit by Eq.~\ref{eq:1}. The orange and green dots correspond to the orange and green line profiles in the upper-right panel.
              }
         \label{fig:lamPav}
   \end{figure*}

This star initially had no disk during the observing window. A mass outburst started in the last few days of the TESS observing window, which was quickly noticed in spectroscopy and so spectroscopic monitoring continued in the weeks after the end of the TESS baseline. A fit to the H$\alpha$ EW$_{\rm V}$/EW$_{\rm R}$\xspace measurements over the first 7 days of the outburst gives a frequency of 0.810 $\pm$ 0.014 \,d$^{-1}$\xspace. The photometry and spectroscopic quantities are shown in Fig.~\ref{fig:lamPav}. \citet{2011A&A...533A..75L} interpreted the 0.82 d$^{-1}$\xspace frequency as being circumstellar, which is very similar to the frequency found in our H$\alpha$ EW$_{\rm V}$/EW$_{\rm R}$\xspace data. 

The TESS frequency spectrum (Fig.~\ref{fig:FTs_TESS_spec}) is somewhat unusual compared to the rest of our sample, in that only one frequency group is present. Likewise, the circumstellar frequency is considerably lower than the dominant pulsation frequency at 1.641\,d$^{-1}$\xspace. In fact, the circumstellar frequency is equal to half of the dominant pulsation frequency -- whether this is a coincidence or has some physical meaning is not yet clear. There is a photometric signal in TESS at 0.815 \,d$^{-1}$\xspace (i.e., consistent with the circumstellar frequency) in the sector 66+67 data even when excluding data after TJD 3140 (i.e., well before the emergence of emission). The purported pulsation frequency of 0.49\,d$^{-1}$\xspace from \citet{2011A&A...533A..75L} does not appear in the TESS data and is perhaps a 1\,d alias of the TESS-detected signal at 1.48\,d$^{-1}$\xspace.

\subsection{$\iota$ Lyr (HD 178475; B6IVe)} \label{sec:iot_Lyr}

For the B6e star $\iota$ Lyr, \citet{2001A&A...368..912Y} determined ${v\sin i}$ = 234 $\pm$ 20\,km\,s$^{-1}$, and \citet{2005ESASP.560..571G} found ${v\sin i}$ = 230\,km\,s$^{-1}$. These are probably underestimated, since gravity darkening was not considered. The inclination angle does not appear in the literature, but based on the spectra presented here and in historical BeSS spectra there can be mild shell absorption suggesting a relatively high inclination angle.

$\iota$ Lyr has a companion about 1.5 magnitudes fainter at a $\sim$0.07 arcsecond separation \citep[e.g.,][]{2008AJ....136..312H,2010AJ....139..205H}. Relatively weak and narrow absorption features are also seen in some of the lines in our spectra (e.g., He I, Mg II). At such a wide separation, the companion does not have any appreciable affect on the Be star or its disk. 

\begin{figure*}
   \centering
   \includegraphics[width=1.0\hsize]{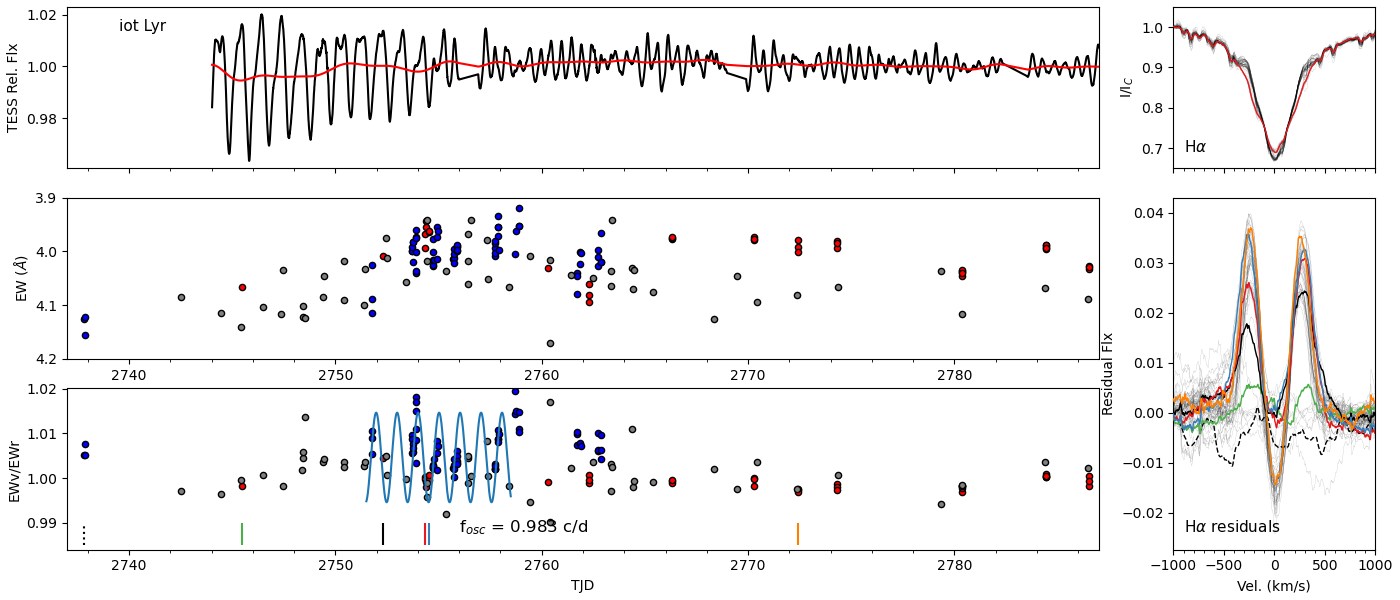}
      \caption{
      Similar to Fig.~\ref{fig:f_Car}, but for $\iota$ Lyr. Top-left panel: TESS light curve (black) with the low-frequency trends ($<$0.5 d$^{-1}$) in red. Top-right panel: All NRES spectra (black curves), with the pre-outburst average profile in red. Bottom-right panel: NRES and DAO emission H$\alpha$ profiles after subtracting off the pre-outburst average. Middle- and bottom-left panels:  H$\alpha$ EW and EW$_{\rm V}$/EW$_{\rm R}$\xspace, respectively with gray markers indicating amateur data, blue from DAO, and red from NRES. In the bottom-left panel, six vertical lines indicate epochs that are highlighted in the corresponding colors in the bottom-right panel. The dashed line is pre-outburst, and the green is the first professional spectrum after the start of the event. 
              }
         \label{fig:iotLyr}
   \end{figure*}

An outburst started sometime shortly before, or very near the beginning of, the TESS observations (see Fig.~\ref{fig:iotLyr}). Due to the (at the time) lower-cadence spectroscopic monitoring, the outburst can only be constrained to have started between TJD 2738 (where there was no excess emission; black dotted line in bottom-right panel of Fig.~\ref{fig:iotLyr}) and TJD 2745 (where there was very slight excess emission; green solid line in bottom-right panel of Fig.~\ref{fig:iotLyr}). TESS monitoring began on TJD 2744. Unlike the other systems studied here, there is little to no net change in the brightness during the outburst, but there is an obvious enhancement in the higher frequency signals. There was no sign of any excess emission prior to TJD 2745 (i.e., the star was apparently disk-less). The central depression in H$\alpha$ deepened over time with emission increasing at higher velocities, consistent with a system seen at a high inclination angle.

Figure~\ref{fig:iotLyr} shows the TESS light curve, H$\alpha$ line profiles, and EW$_{\rm V}$/EW$_{\rm R}$\xspace measurements. Due to the low amount of excess emission during this event, it is relatively difficult to be confident as to whether the velocity distribution is asymmetric or not. Nevertheless, fits to the EW$_{\rm V}$/EW$_{\rm R}$\xspace values of the professional spectra (from both DAO and NRES) find a best-fit frequency of 0.959 \,d$^{-1}$\xspace, which fits into the same pattern as the other systems. Measurements of the H$\alpha$ emission peak intensities find virtually the same frequency for the V/R ratio. The emission H$\alpha$ profiles (after subtraction of the photospheric profile), highlighted in the bottom-right panel of Fig.~\ref{fig:iotLyr}, show rapid V/R variations. The amateur spectra are less homogeneous, and especially due to tellurics (and their different widths for different sites and instruments) such low amplitude features were difficult to measure, but nevertheless the EW$_{\rm V}$/EW$_{\rm R}$\xspace measurements mostly fall onto the same fit determined from the professional spectra. Indeed, the amateur spectra provided the first hint of growing emission, allowing us to increase the cadence for better sampling of the event. 

$\iota$ Lyr's spectral type is the latest (B6IVe) of this sample. It is well known that the observational disk signatures become weaker toward later spectral types, which are generally regarded as being less active and with lower-density disks \citep{2017MNRAS.464.3071V,2017AJ....153..252L}. The low activity level of later Be stars may be partly an observational bias -- if relatively short events like in the B6e star $\iota$ Lyr have weak observational signatures then they are easy to overlook.

\subsection{28 Cyg (HD 191610; B2.5Ve)} \label{sec:28_Cyg}

28 Cyg has been the subject of numerous studies. Space photometry from BRITE and the Solar Mass Ejection Imager (SMEI) revealed a rich pulsational frequency spectrum, including combination frequencies that seem related to the timing of mass ejection events \citep{2018A&A...610A..70B}. Two pulsation frequencies near 1.54\,d$^{-1}$\xspace were recovered from spectroscopy in \citet{2000ASPC..214..232T}. 

\citet{zorec2016} found ${v\sin i}$ = 314 $\pm$ 33 km s$^{-1}$ and $i = 69^{\circ} \pm 17^{\circ}$ by fitting gravitationally darkened absorption lines, but an inclination angle of $i = 40^{\circ} \pm 5^{\circ}$ was derived from fitting the H$\alpha$ emission line \citep{2023ApJ...948...34S}. 28 Cyg is also a Be+sdO binary, where the sdO companion was first detected in the International Ultraviolet Explorer (IUE) UV spectra \citep{2018ApJ...853..156W} and was later confirmed with IR interferometry, with a reported orbital period of about one year \citep{2024ApJ...962...70K}. 

One outburst event occurred during the monitoring window, with the first spectrum to sample the event being taken $\sim$2 days after the  photometric onset of the event. There was a relatively stable and symmetric H$\alpha$ emission peaking at a normalized amplitude of 1.8 prior to the outburst, with the emission level having slowly decreased in the prior $\sim$60 days. Probably, the first few EW$_{\rm V}$/EW$_{\rm R}$\xspace cycles were missed. The variations in H$\alpha$ are subtle, but the emission variations in H$\gamma$ are more pronounced and unambiguously show the same type of asymmetric emission variation seen in other stars. The frequencies measured in the EW$_{\rm V}$/EW$_{\rm R}$\xspace measurements from H$\alpha$ and H$\gamma$ differ significantly. For H$\alpha$, only about three EW$_{\rm V}$/EW$_{\rm R}$\xspace cycles could be fit, with a frequency of 1.415\,d$^{-1}$\xspace. For H$\gamma$, about eight EW$_{\rm V}$/EW$_{\rm R}$\xspace cycles were sampled with a best-fit frequency of 1.057\,d$^{-1}$\xspace. The H$\gamma$ observations are sparser, and the measured frequency is very close to 1.0\, d$^{-1}$\xspace and is possibly affected by the observing cadence and is thus not reported in Tab.~\ref{tbl:measurements}.

   \begin{figure*}
   \centering
   \includegraphics[width=1.0\hsize]{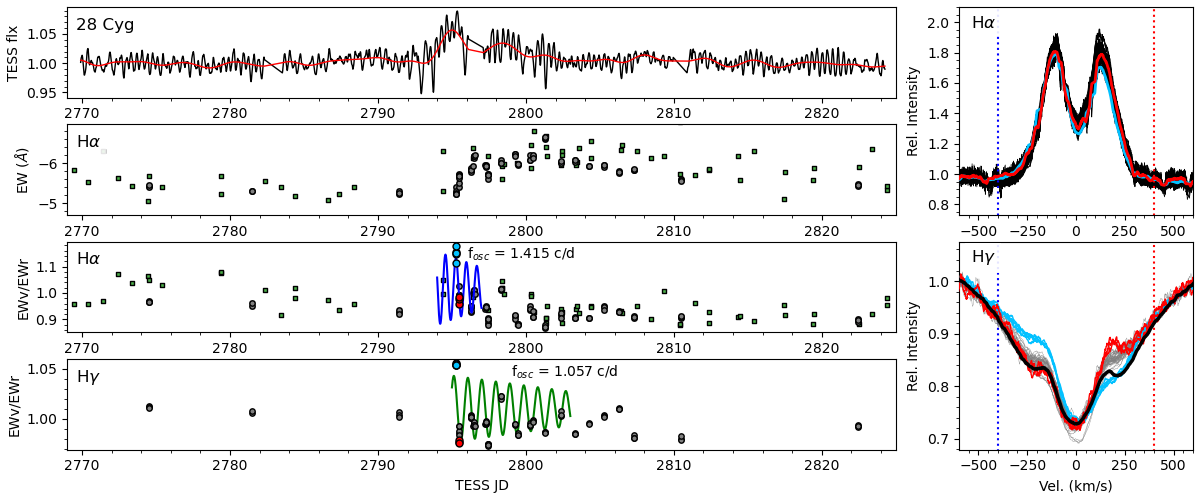}
      \caption{Measurements of H$\alpha$ and H$\gamma$ observations of the Be star 28 Cyg, similar to Fig.~\ref{fig:f_Car}. Top-left panel: TESS data (black), with the red curve tracing the low frequency variability ($<$0.5 d$^{-1}$). Green squares in the middle two left panels (for H$\alpha$) represent measurements from amateur data. The EW$_{\rm V}$/EW$_{\rm R}$\xspace was calculated out to $\pm$400 km s$^{-1}$ for both lines. The blue and red curves plotted for H$\alpha$ and H$\gamma$ are taken about 5.5 hours apart (with corresponding colored markers in the EW$_{\rm V}$/EW$_{\rm R}$\xspace panels).               }
         \label{fig:28Cyg}
   \end{figure*}

\subsection{V357 Lac (HD 212044; B2e)} \label{sec:V357_Lac}

In \citet{2017AJ....153..174C} the authors note that V357 Lac (internal ID ABE-164) ``represents the extreme end of equivalent width variability in the direction of weakening emission,'' with the EW of Br11 changing from -14.0 $\AA$ to -7.2 $\AA$ over 25 days. This may be  due to a rapid increase in the continuum level during this time period, rather than a genuinely weakening line emission.  The star is possibly an RV variable, but this conclusion was based on only three 
Apache Point Observatory Galactic Evolution Experiment (APOGEE) spectra \citep{2017AJ....153..174C}, thereby requiring further confirmation. This system is exceptionally photometrically active, spending very little time in a quiescent state but instead punctuated by numerous, usually fairly rapid, outburst events with peak-to-peak variations of about 0.4 magnitudes in the $\sim$R broad-band Kilodegree Extremely Little Telescope ($KELT$) filter \citep{2018AJ....155...53L}. Over the course of $\sim$15 years, the H$\alpha$ EW varied between -1.2 and -23.5 $\AA$ \citep{2018AJ....155...53L}. Further, outbursts roughly repeated with a cycle length of 59.5 days, but this behavior is clearly not strictly periodic \citep{2018AJ....155...53L}. The TESS baseline, however, is too short to confirm this but instead suggests outbursts can occur more frequently. In summary, this is an early-type Be star that is very active in terms of frequent punctuated outbursts with high amplitude, and usually there is a strong disk present. There are no measurements of the inclination angle or ${v\sin i}$ available in the literature, but based on the emission profiles and photometric variability, the inclination angle is relatively low.

\begin{figure*}
   \centering
   \includegraphics[width=1.0\hsize]{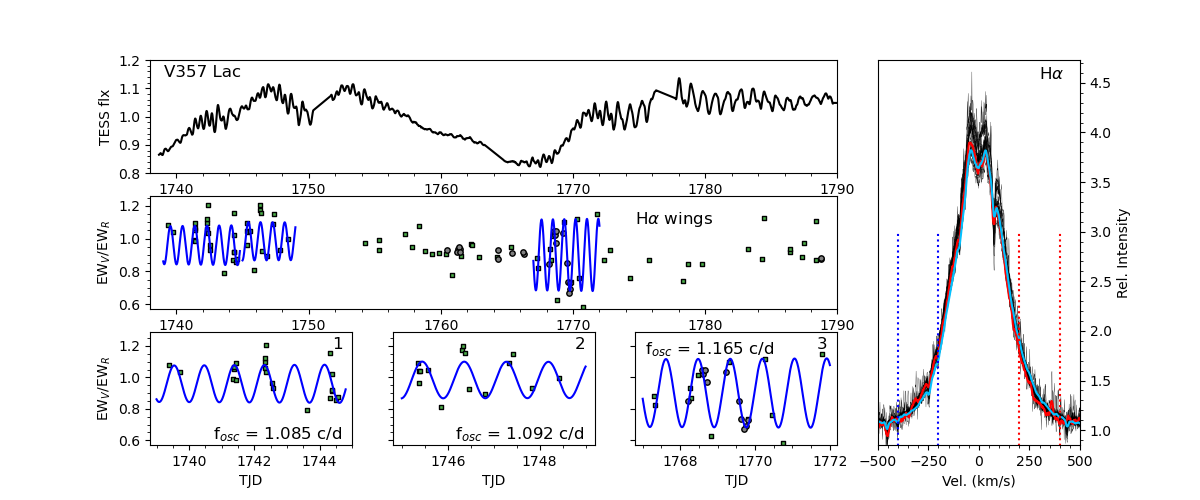}
      \caption{Measurements of H$\alpha$ from the BeSS database (green squares) and NRES (gray circles) of the Be star V357\,Lac, with three sections of EW$_{\rm V}$/EW$_{\rm R}$\xspace values fit with Eq.~\ref{eq:1}. The EW$_{\rm V}$/EW$_{\rm R}$\xspace measurements are from the line wings (between 200 and 400 km s$^{-1}$) in H$\alpha$, but similar results are obtained for the full line profile.
              }
         \label{fig:V357Lac}
   \end{figure*}

Figure~\ref{fig:V357Lac} shows the TESS light curve and corresponding H$\alpha$ measurements (from NRES and BeSS). The disk is fairly strong during this time period, with H$\alpha$ reaching about four times the continuum level.  Because of the strong preexisting disk, the EW$_{\rm V}$/EW$_{\rm R}$\xspace was measured using the H$\alpha$ wings (between $\pm$200 and $\pm$400 km s$^{-1}$). Although ${v\sin i}$ is not reported in the literature, the photospheric lines extend out to approximately $\pm$200 km s$^{-1}$. \ion{He}{I}\,$\lambda$6678 is also highly variable in its asymmetry, but the line is weak and is only observed well in the somewhat sparse NRES data. EW$_{\rm V}$/EW$_{\rm R}$\xspace cycles at frequencies near 1\,d$^{-1}$\xspace are seen in three sections of the data. Given the high level of activity of this star, it is not always clear exactly when outbursts start and end. The EW$_{\rm V}$/EW$_{\rm R}$\xspace values between TJD 1740 and 1750 are fit better when split into two sections and fit separately (but with approximately the same frequency), compared to a single longer fit. In other words, there seems to be a phase shift introduced at TJD $\sim$1747, which may indicate that this is a ``compound outburst'' with two discrete ejection events spaced by only a few days. 

The event near TJD 1770 is more favorably sampled in spectroscopy. Unlike the outbursts discussed so far, the TESS photometry quickly reaches a plateau and remains at a steady relatively high flux level instead of a peak followed by a decrease. This may indicate more sustained mass ejection, but the spectroscopy during the plateau stage is too sparse to determine whether or not EW$_{\rm V}$/EW$_{\rm R}$\xspace cycles continued. 

This is one of only two stars in our sample without a spectroscopically measured pulsational frequency (the other being $\iota$ Lyr). Our dataset is insufficient for this (containing mostly H$\alpha$, and only a few echelle spectra), and there do not seem to be any reports of such in the literature. However, there are clear line profile variations in photospheric lines (e.g., \ion{He}{I}\,$\lambda$4388, \ion{Si}{III}\,$\lambda$4553) typical of pulsating Be stars. The EW$_{\rm V}$/EW$_{\rm R}$\xspace frequencies fit the same pattern as the other stars compared to the frequencies in TESS (they are within the $g1$ frequency group). 

\citet{2024A&A...691L...4P} noted the strong anticorrelation between TESS flux and H$\alpha$ emission strength during three sets of TESS observations, including during sectors 16+17 as shown in Figure~\ref{fig:V357Lac}. \citet{2024A&A...691L...4P} also emphasize the strong variability in the amplitude of photometric frequency groups associated with brightening events. As mentioned in the first paragraph of this subsection, and as seen in other systems with relatively strong preexisting disks (Appendix~\ref{sec:25_Ori},~\ref{sec:kap_CMa},~\ref{sec:120_Tau}), an anticorrelation between continuum flux and emission strength on short timescales can be readily explained by the fact that the EW is measured relative to the continuum. When the continuum level rises rapidly, due to increased density in the inner disk, the absolute line emission may increase slightly, but the line emission relative to the continuum (the EW) will drop (with EW numerically increasing following the convention of EW being negative when in emission).

\subsection{OT Gem (HD 58050; B2Ve)} \label{sec:OT_Gem}

An inclination angle of $i = 21^{\circ} \pm 6^{\circ}$ was derived from fitting the H$\alpha$ emission line \citep{2023ApJ...948...34S}, and $i = 31^{\circ} \pm 12^{\circ}$ (with ${v\sin i}$ = 138 $\pm$ 12\,km\,s$^{-1}$) by fitting gravitationally darkened photospheric lines \citep{zorec2016}. OT Gem displayed rather sharp and punctuated outbursts in the first year of the TESS mission \citep{2021MNRAS.502..242L}. It has a history of both photometrically quiet and active phases, with peak-to-peak variations up to $\sim$0.4 magnitudes in the $V$ band. It has also exhibited strong spectroscopic variations since at least 1943 \citep{1999A&A...350..566B}.

\begin{figure*}
   \centering
   \includegraphics[width=1.0\hsize]{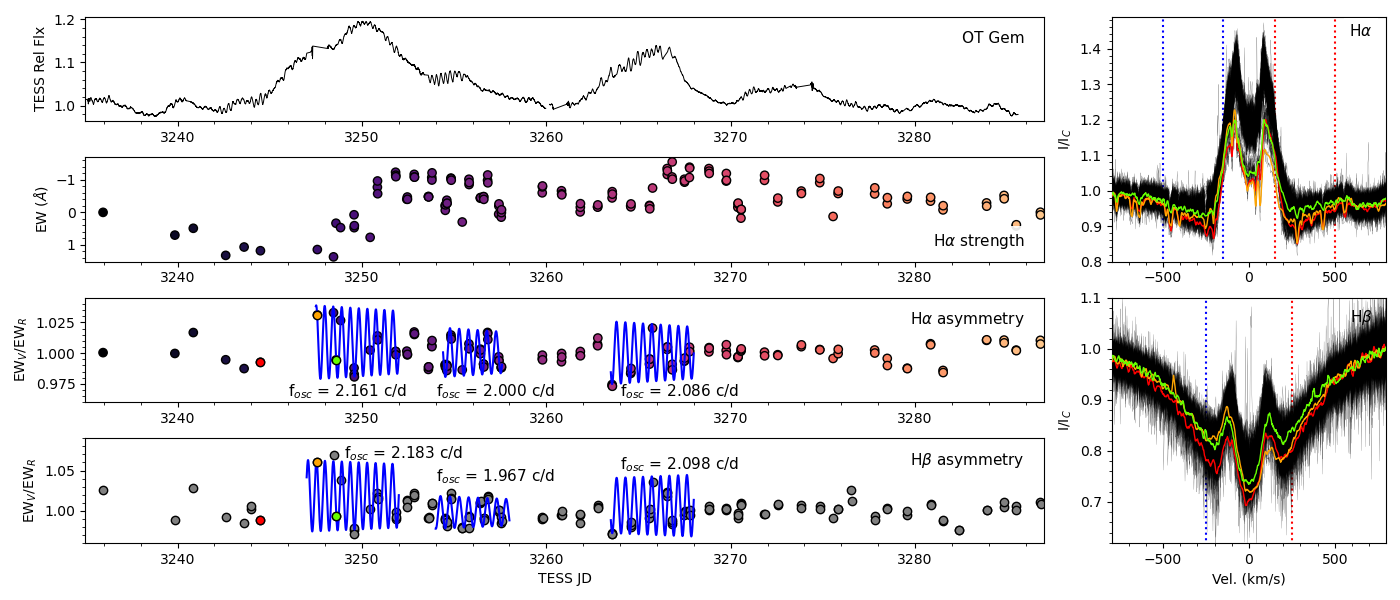}
      \caption{TESS photometry and measurements of H$\alpha$ and H$\beta$ from NRES of the Be star OT Gem. The EW$_{\rm V}$/EW$_{\rm R}$\xspace measurements are made from the full H$\alpha$ profile (between $\pm$500 km s$^{-1}$), and for H$\beta$ between $\pm$250 km s$^{-1}$. 
              }
         \label{fig:OTGem}
   \end{figure*}

Spectroscopic and photometric data for OT Gem are shown in Fig.~\ref{fig:OTGem}. There was a weak preexisting disk at the time of our observations. EW$_{\rm V}$/EW$_{\rm R}$\xspace measurements were made between $\pm$250 km s$^{-1}$ for H$\beta$, and between $\pm$500 km s$^{-1}$ for H$\alpha$. The wings of H$\alpha$ (between $\pm$150 and $\pm$500 km s$^{-1}$) were also measured, with indistinguishable differences in the best-fit frequency compared to the full line profile. The first major event in photometry began near TJD 3245. The spectrum at TJD 3244.5 does not show any obvious signatures of an outburst (the red curves in the right panels of Fig.~\ref{fig:OTGem}), but the next spectrum (TJD 3247.5, the orange curves in Fig.~\ref{fig:OTGem}) shows a clear enhancement on the V side of the emission lines. Over the next several days, the emission strength increased and EW$_{\rm V}$/EW$_{\rm R}$\xspace oscillated (with a frequency of 2.16 \,d$^{-1}$\xspace), with the continuum flux peaking at TJD 3250 and then decreasing. At TJD 3254 a slight increase in flux is seen for $\sim$ 2 days, along with higher amplitude oscillations, probably indicating another (small) mass ejection. The EW$_{\rm V}$/EW$_{\rm R}$\xspace values for the next five days are fit well with a frequency of 2.00 \,d$^{-1}$\xspace. The third event has a longer build-up time in photometry compared to the dissipation phase, and the EW$_{\rm V}$/EW$_{\rm R}$\xspace measurements have a frequency of 2.09 \,d$^{-1}$\xspace. 

The emission oscillation cycles have the same frequencies to within the errors for both H$\alpha$ and H$\beta$. 
The EW$_{\rm V}$/EW$_{\rm R}$\xspace cycles are the fastest among our sample (at about 2 \,d$^{-1}$\xspace), but still fit the same pattern as the other stars (i.e., the photometric and spectroscopic frequencies are also proportionally higher than average).

\subsection{25 Ori (HD 35439; B1Vne)} \label{sec:25_Ori}

25 Ori has reported inclination angles of $i = 63^{\circ} \pm 5^{\circ}$ derived from fitting the H$\alpha$ emission line \citep{2023ApJ...948...34S}, and $i = 58^{\circ} \pm 14^{\circ}$ (with ${v\sin i}$ = 284 $\pm$ 28 km s$^{-1}$) from fitting gravitationally darkened photospheric lines \citep{zorec2016}. From spectro-interferometry (with the VLTI/AMBER interferometer), an inclination angle of $i = 55^{\circ} \pm 5^{\circ}$ was found for the disk \citep{2019A&A...621A.123C}.

\begin{figure*}
   \centering
   \includegraphics[width=1.0\hsize]{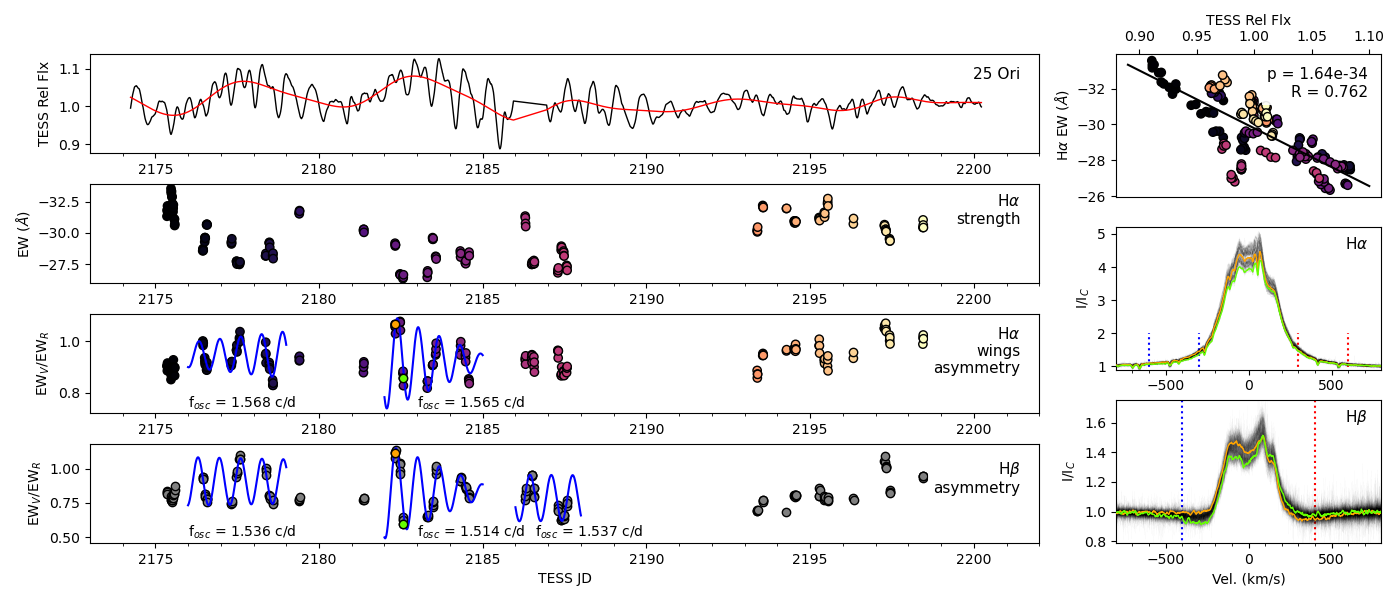}
      \caption{Measurements of H$\alpha$ and H$\beta$ observations of the Be star 25 Ori, similar to Fig.~\ref{fig:f_Car}. For H$\alpha$, the EW$_{\rm V}$/EW$_{\rm R}$\xspace ratio was calculated using the wings of H$\alpha$, as indicated in the middle-right panel (between -600 and -300 km s$^{-1}$ and +300 and +600 km s$^{-1}$, i.e., outside of ${v\sin i}$). However, the same behavior is seen in H$\alpha$ even when using the entire line profile. Similar behavior is seen in H$\beta$ for EW$_{\rm V}$/EW$_{\rm R}$\xspace (calculated using the full line profile out to $\pm$400 km s$^{-1}$), including a third epoch of EW$_{\rm V}$/EW$_{\rm R}$\xspace variations that are less evident in H$\alpha$. Top-right panel: Anticorrelation between TESS brightness and H$\alpha$ emission strength.
              }
         \label{fig:25Ori}
   \end{figure*}

25 Ori already had a strong preexisting disk at the time of our observations, with H$\alpha$ reaching $\sim$5 times the continuum level. The H$\alpha$ EW and the TESS brightness are inversely correlated -- that is, H$\alpha$ emission is weaker when the system is brighter. The TESS light curve, H$\alpha$ EW, and their correlation are shown in Fig.~\ref{fig:25Ori}, along with the H$\alpha$ and H$\beta$ line profiles and EW$_{\rm V}$/EW$_{\rm R}$\xspace measurements. The anticorrelation between brightness and emission strength is probably explained by the strong disk, where the large H$\alpha$-emitting area does not change significantly on short timescales. The continuum flux (as sampled by TESS) does however change quickly, and since EW is measured relative to the local (and variable) continuum, the EW is thus mostly modulated by changes in the continuum, and not the line emission. A similar anticorrelation is seen in 120~Tau, and to a slightly lesser extent in $\kappa$~CMa. 

The photometric amplitudes of the higher frequency variations are relatively high (up to $\sim$20 ppt) compared to the rest of the sample. The TESS frequency spectrum is similar to what was observed in earlier sequences of space photometry with BRITE and SMEI \citep{2018pas8.conf...69B}, including a strong difference frequency at about 0.2 \,d$^{-1}$\xspace (slightly different from the value of 0.1777 \,d$^{-1}$\xspace found in the longer BRITE and SMEI data strings).  

In order to better sample the high-velocity and relatively dense material close to the star, the wings of H$\alpha$ were isolated and the EW of the blue- versus red-shifted wings were compared (between $\pm$300 km s$^{-1}$ and $\pm$600 km s$^{-1}$). However, using the EW$_{\rm V}$/EW$_{\rm R}$\xspace of the full line profile traced similar behavior. Likewise, the EW$_{\rm V}$/EW$_{\rm R}$\xspace behavior of H$\beta$ (using the full line) was similar to that of  H$\alpha$. Fits of Eq.~\ref{eq:1} give approximately the same frequencies for H$\alpha$ and H$\beta$ for two sections of the data (between TJD 2176 and 2179, and between TJD 2182 and 2185). Considering also the increase in TESS brightness during these sections, the behavior is consistent with mass ejection. Similar cycles are seen in H$\beta$ near TJD 2187, but this timespan is not well sampled and H$\alpha$ does not seem to show the same type of variations.

It is worth noting that the average line profiles of H$\alpha$ and H$\beta$ (and other features) are asymmetric during the observing window. Such structure is often caused by $m=1$ density waves in the disk \citep{1997A&A...318..548O}, which generally have cycle lengths of years and longer. Thus, this average asymmetry should have no impact on the asymmetry variations on sub-day timescales.

\begin{figure}
   \centering
   \includegraphics[width=0.5\hsize]{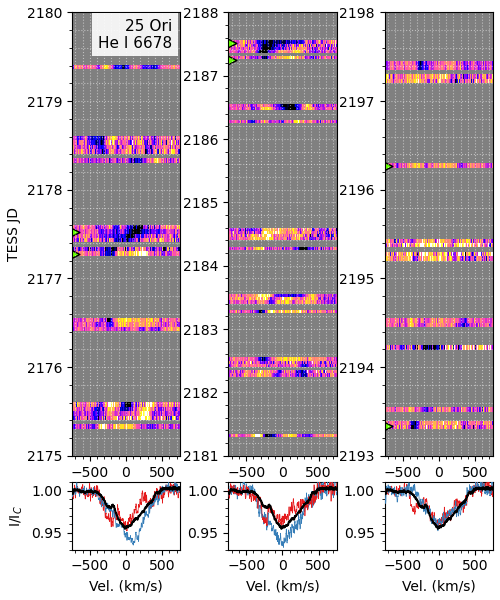}
      \caption{Dynamical spectra of the mean-subtracted \ion{He}{I}\,$\lambda$6678 line during the epochs indicated in Fig.~\ref{fig:25Ori} (top panels). Bottom panels: Mean line profile in black, and two selected epochs (indicated by green triangles in the top panels) to emphasize the strong and fast line profile variations especially in the first half of the dataset (the red line profile precedes the blue).
              }
         \label{fig:25Ori_dyn}
   \end{figure}

Additional supporting evidence for active mass ejections during the fit sections of EW$_{\rm V}$/EW$_{\rm R}$\xspace measurements in Fig.~\ref{fig:25Ori} are found in the \ion{He}{I}\,$\lambda$6678 line. Figure~\ref{fig:25Ori_dyn} shows the mean-subtracted dynamical spectra of \ion{He}{I}\,$\lambda$6678. Significantly stronger variations are seen approximately in the first half of the dataset (including during the possible flickers), but these changes are not obviously periodic or even cyclic. Whether these are stellar, circumstellar, or both is unclear, but they do not resemble the typical line profile variations caused by pulsation. The bottom row in Fig.~\ref{fig:25Ori_dyn} shows the average \ion{He}{I}\,$\lambda$6678 line profile, and two selected epochs (indicated by green triangles in the dynamical spectra) to emphasize the unusual but quickly evolving features. An advantage of \ion{He}{I}\,$\lambda$6678 is that while there is almost certainly some circumstellar contribution, it is far less than the hydrogen lines and thus its variability should be only marginally influenced by variations in the nearby continuum that, in any case, cannot explain changing features within the line profile. Qualitatively similar LPVs are sometimes seen in other Be stars actively ejecting mass \citep[e.g., in $\mu$ Cen and $\lambda$ Eri;][]{1998A&A...333..125R,1989ApJS...71..357S}.

\subsection{$\kappa$ CMa (HD 50013; B1.5Ve)} \label{sec:kap_CMa}

An inclination angle of $i = 57^{\circ} \pm 13^{\circ}$ was derived from fitting the H$\alpha$ emission line \citep{2023ApJ...948...34S}, and $i = 50^{\circ} \pm 17^{\circ}$ (with ${v\sin i}$ = 231 $\pm$ 25\,km\,s$^{-1}$) by fitting gravitationally darkened photospheric lines \citep{zorec2016}. From spectro-interferometry (with the VLTI/AMBER interferometer), an inclination angle of $i = 53^{\circ} \pm 10^{\circ}$ was derived \citep{2019A&A...621A.123C}. 

\citet{2003A&A...411..229R} reported a clear NRP frequency at 1.825 \,d$^{-1}$\xspace, but with unusual characteristics compared to NRP patterns in most Be stars, which does not resemble the $l=m=2$ patterns in most of their sample (the pulsation in $\kappa$ CMa could be tesseral (i.e., $l\neq|m|$). Our spectra also contain the same pulsational signal at 1.825 \,d$^{-1}$\xspace, but which is apparently absent in TESS. They also note a secondary period at 0.617 d (1.621 \,d$^{-1}$\xspace), which is transient and similar to {\v{S}}tefl frequencies, but it seems like this signal is phase coherent over at least weeks to months. They find that $\kappa$ CMa is ``extraordinarily active on short timescales.'' In \citet{2016A&A...588A..56B} the authors note that for $\kappa$ CMa, in regards to the 1.621 \,d$^{-1}$ {\v{S}}tefl frequency, the velocity range of features substantially exceeds ${\pm v\sin i}$, but also that quasi-periodic line profile-crossing features are associated with this variability. They propose the qualitative idea of an ejected cloud with an orbit that has not yet been circularized before merging with the disk. 

\begin{figure*}
   \centering
   \includegraphics[width=1.0\hsize]{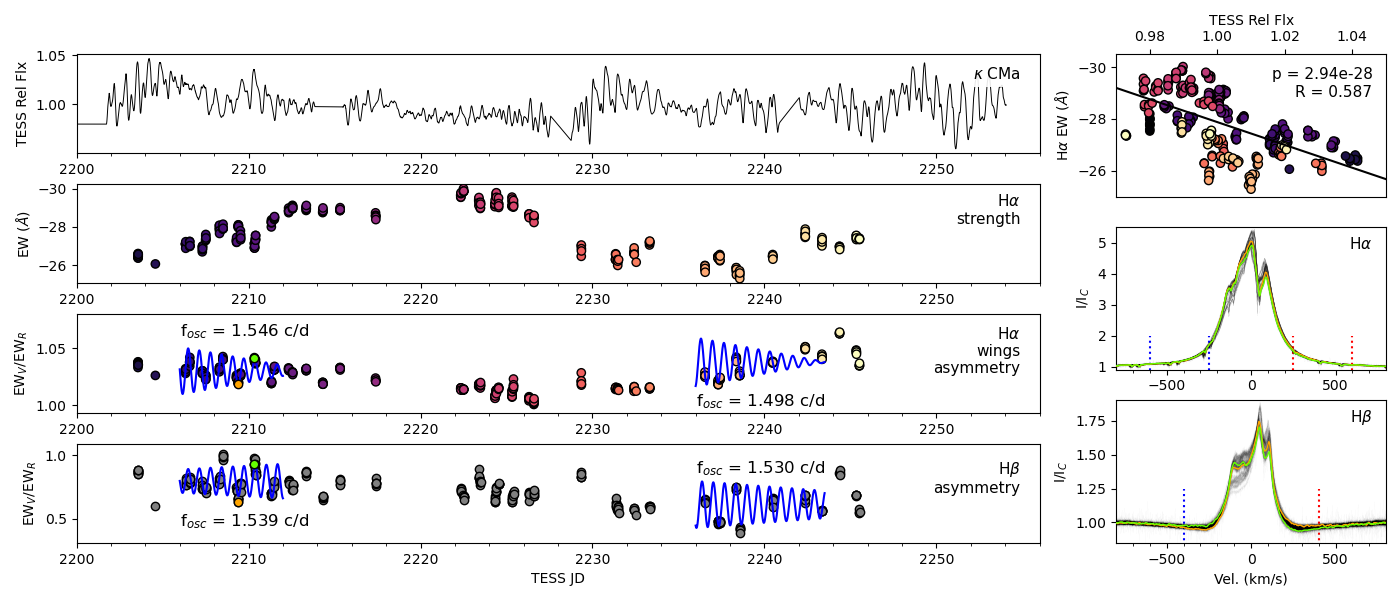}
      \caption{Similar to Fig.~\ref{fig:25Ori} for $\kappa$~CMa.  Outburst events are even less clearly delineated than in 25 Ori, but two sections of EW$_{\rm V}$/EW$_{\rm R}$\xspace measurements could be fit with Eq.~\ref{eq:1}. The full timeseries is not fit well with a single frequency (e.g., treating the full timeseries as one long ``event'').
              }
         \label{fig:kapCMa}
   \end{figure*}

Our data for $\kappa$ CMa is shown in Fig.~\ref{fig:kapCMa}. Like 25 Ori, a strong disk (H$\alpha$ is about 5 times the continuum level) with complex structure was present. While there is variability on slower timescales (days to weeks) in the photometry, it is almost entirely unclear when (or if) outbursts occur. The changes in H$\alpha$ EW hardly clarify the situation. The continuum brightness and emission strength are globally anticorrelated, but less tightly so than in 120 Tau and 25 Ori. Two sections of the H$\alpha$ (using the wings between $\pm$231 km s$^{-1}$ and $\pm$600 km s$^{-1}$) and H$\beta$ (using the full line profile out to $\pm$400 km s$^{-1}$) EW$_{\rm V}$/EW$_{\rm R}$\xspace measurements were fit with Eq.~\ref{eq:1}, resulting in frequencies of about 1.5 \,d$^{-1}$\xspace, slightly slower than that found by \citet{2003A&A...411..229R}. In our dataset, the EW$_{\rm V}$/EW$_{\rm R}$\xspace frequency does not seem to be stable over weeks to months, although it is difficult to determine with some sections of sparse sampling, and also the complex photometric variability that makes it unclear when outbursts start and end.

Overall, it seems like there was mass ejection activity during our observing baseline, but the details are unclear. Nevertheless, there are windows of time where asymmetry cycles are seen which seems consistent with the behavior reported in \citet{2016A&A...588A..56B}.

\subsection{j Cen (HD 102776; B3Ve)} \label{sec:jCen}

There is not much information in the literature on j Cen despite its brightness ($V_{\rm mag} = 4.3$). The morphology of its emission features suggest an intermediate inclination angle, but there are no reports of $i$ or ${v\sin i}$ in the literature. There is neither a correlation nor anticorrelation between H$\alpha$ EW and TESS brightness during sectors 37 and 38.

\begin{figure*}
   \centering
   \includegraphics[width=1.0\hsize]{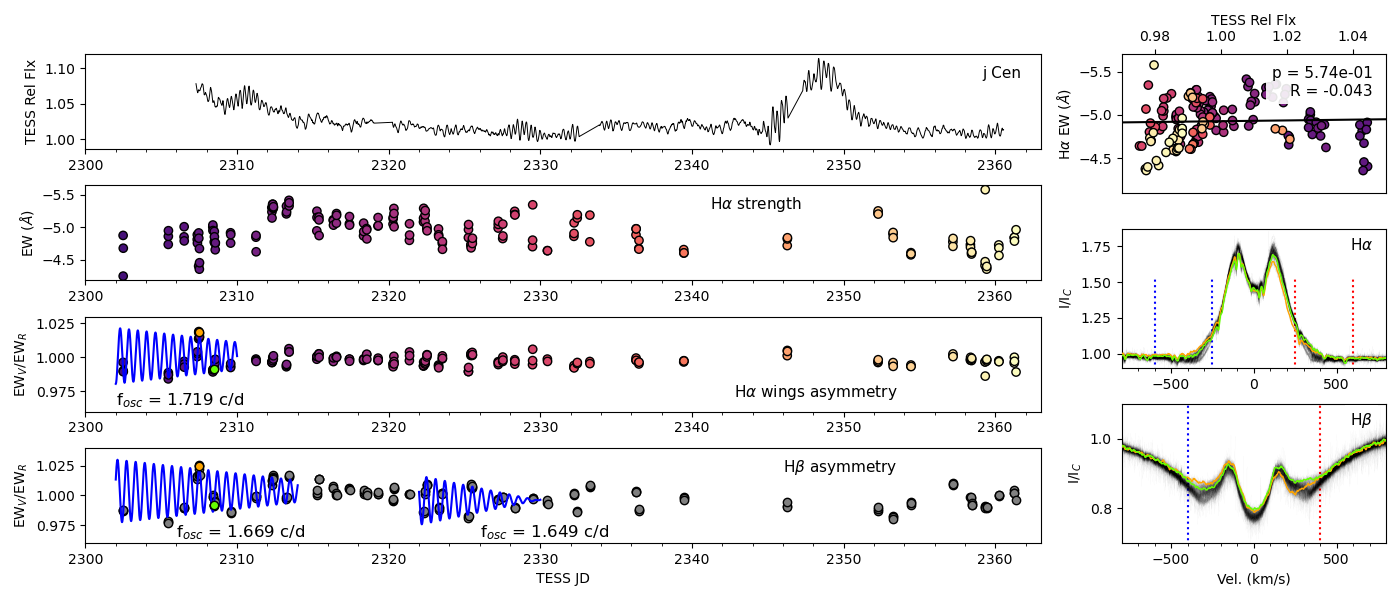}
      \caption{Similar to Fig.~\ref{fig:25Ori} for j\,Cen. For H$\alpha$, only one section of data could be fit with Eq.~\ref{eq:1}. For H$\beta$, two sections were fit.
              }
         \label{fig:jCen}
   \end{figure*}

A mass ejection event probably happened sometime shortly before the observing window, as evidenced by the decreasing TESS flux (and the presence of H$\alpha$ emission; Fig.~\ref{fig:jCen}). During the first few days of the TESS baseline, there were asymmetry oscillations in the emission features. This section of data was fit well with a frequency of 1.72 \,d$^{-1}$\xspace, although the spectroscopic cadence was low and it is not known when the flicker started and peaked. Nevertheless, the two emphasized epochs for H$\alpha$ and H$\beta$ (bottom right panels of Fig.~\ref{fig:jCen}) show clear rapid asymmetry variations. A second section of EW$_{\rm V}$/EW$_{\rm R}$\xspace measurements for H$\beta$ (but not H$\alpha$) were fit with a frequency of 1.65 \,d$^{-1}$\xspace, but despite the asymmetry variations it is unclear whether or not this samples a flicker as there was no photometric indication of an event. The EW$_{\rm V}$/EW$_{\rm R}$\xspace variations are more evident in H$\beta$ compared to H$\alpha$. A clear flicker occurred near the end of the TESS observing baseline, but unfortunately this event was hardly sampled by spectroscopy and no EW$_{\rm V}$/EW$_{\rm R}$\xspace frequency can be determined.

\subsection{120 Tau (HD 36576; B2IV-Ve)} \label{sec:120_Tau}

An inclination angle of $i = 61^{\circ} \pm 9^{\circ}$ was derived from fitting the H$\alpha$ emission line \citep{2023ApJ...948...34S}, and $i = 57^{\circ} \pm 14^{\circ}$ (with ${v\sin i}$ = 282 $\pm$ 29\,km\,s$^{-1}$) from fitting gravitationally darkened photospheric lines \citep{zorec2016}. From spectro-interferometry (with the VLTI/AMBER interferometer), an inclination angle of $i = 60^{\circ} \pm 5^{\circ}$ was derived for the disk \citep{2019A&A...621A.123C}. Ground-based photometry revealed 120 Tau as a likely multimode pulsator \citep{1997A&AS..125...75P}, which is confirmed in the new TESS light curves. 

\begin{figure*}[h]
   \centering
   \includegraphics[width=1.0\hsize]{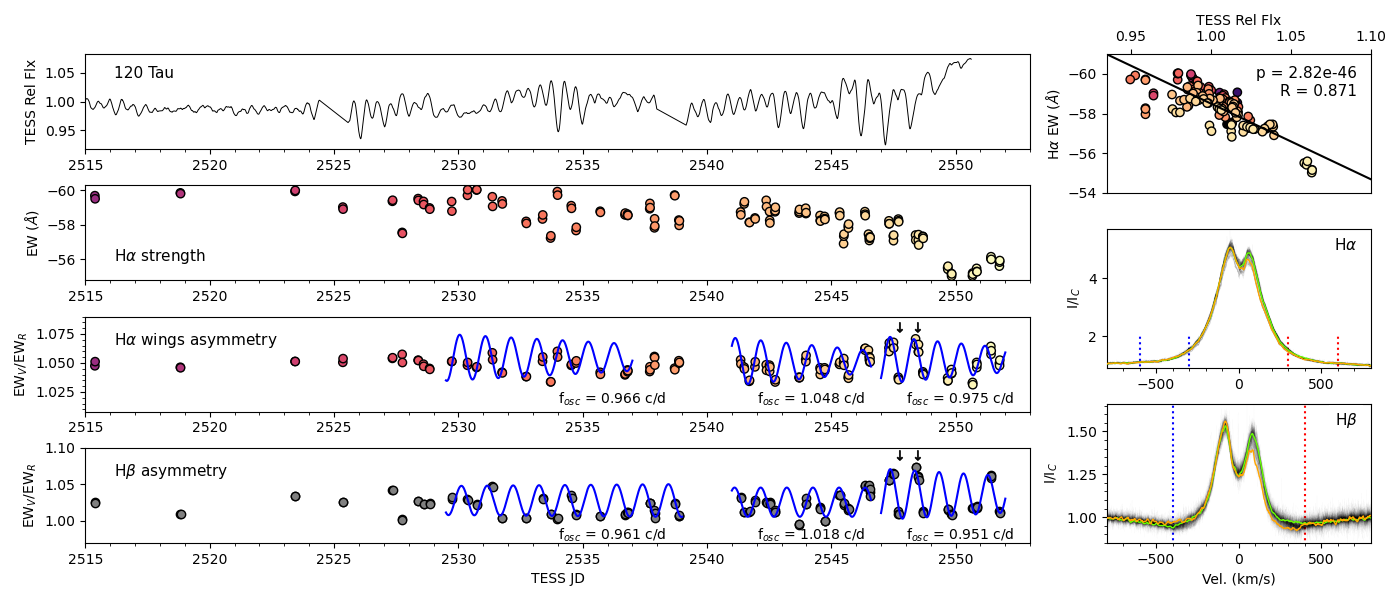}
      \caption{Measurements of TESS brightness and H$\alpha$ and H$\beta$ observations of the Be star 120 Tau, similar to Fig.~\ref{fig:f_Car}. 
              }
         \label{fig:120Tau}
   \end{figure*}

120 Tau had a strong preexisting disk as evidenced by the double-peaked H$\alpha$ profile reaching a normalized amplitude of $\sim$5. The only obvious outburst event occurred near the end of the observing window, and only the early stage (prior to maximum brightness) is sampled. A EW$_{\rm V}$/EW$_{\rm R}$\xspace frequency of 0.975 \,d$^{-1}$\xspace fits well to the H$\alpha$ data, while a frequency of 0.951 \,d$^{-1}$\xspace better describes the H$\beta$ asymmetries. Two slightly earlier sections of data were fit in the same way, with consistent EW$_{\rm V}$/EW$_{\rm R}$\xspace frequencies in both H$\alpha$ and H$\beta$. The photometry and H$\alpha$ and H$\beta$ measurements are shown in Fig.~\ref{fig:120Tau}. The EW$_{\rm V}$/EW$_{\rm R}$\xspace measurements for H$\alpha$ shown in Fig.~\ref{fig:120Tau} use only the high-velocity emission wings (between $\pm$300 and 600 km s$^{-1}$), but the same behavior is seen using the entire line profile. For H$\beta$, the EW$_{\rm V}$/EW$_{\rm R}$\xspace measurements use the full line profile. Due probably to the strong preexisting disk, it is unclear if these earlier sections correspond to discrete mass outbursts because neither the EW or the net brightness changes significantly. The EW$_{\rm V}$/EW$_{\rm R}$\xspace cycles are $\sim$20\% slower than the stable pulsational frequency of 1.2400 \,d$^{-1}$\xspace. Unfortunately, both TESS and spectroscopic observations ended as a stronger outburst may have been starting near TJD 2550.

\clearpage

\section{Flickers at low inclination angles} \label{sec:pole_vs_edge}

Two Be stars at low inclination angles, HD~35345 and PQ~Pup, are briefly discussed to emphasize the increase in brightness oscillations that often coincide with mass ejection. The photometric behavior is qualitatively the same as in the higher inclination angle cases, but the increase in power of the frequency groups can be much more stark, and there should be no obscuration of the star due to ejected material. Although there do not seem to be measurements of the inclination angle in the literature for either system, they are certainly low based on the emission line morphology and narrowness of photospheric absorption lines. These examples bolster the argument that the enhancement in photometric frequency groups during outburst are at least partially photospheric (Sect.~\ref{sec:stellar_vs_circumstellar}). 

Due to geometric cancellation, the photometric amplitude of sectoral modes ($l = |m|$) decreases with decreasing inclination. The much higher amplitude signals that emerge during outburst may be of a different nature than the `standard' sectoral modes commonly observed in Be stars. 

\subsection{HD~35345 (B1Vpe)}
\begin{figure*}[h!]
   \centering
   \includegraphics[width=1.0\hsize]{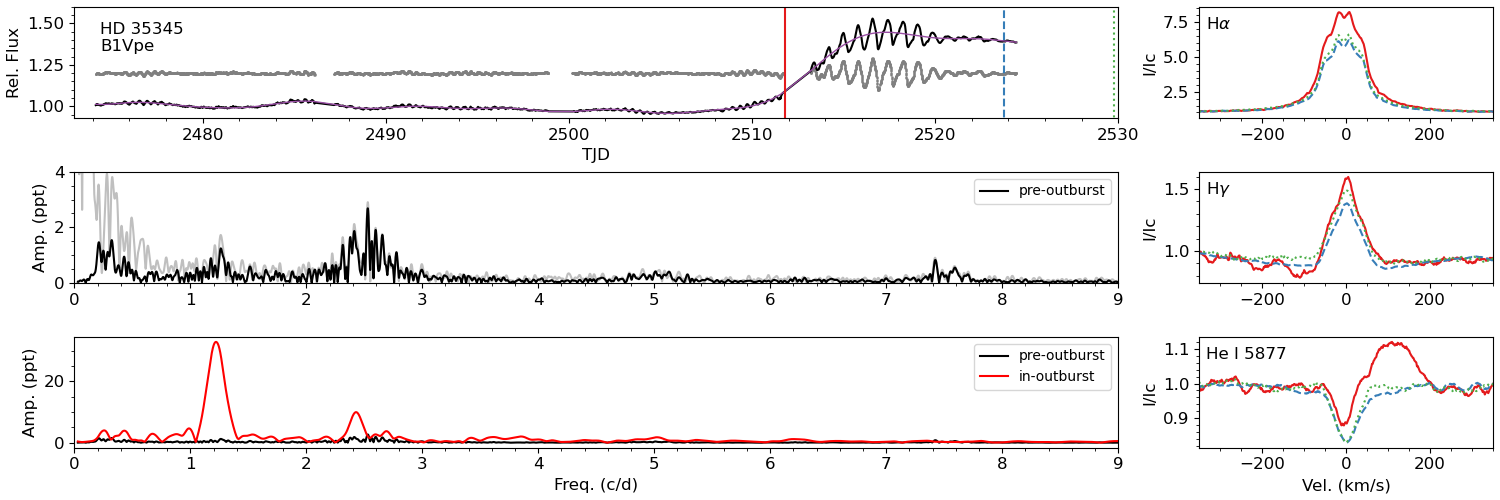}
      \caption{Photometric and spectroscopic information for HD~35345. The top left panel shows the TESS light curve from sectors 43 and 44 (black), with a fit to the low frequency variation (purple). Subtracting this fit results in the flattened light curve (plus a vertical offset, grey points). The middle left panel shows the frequency spectrum computed from the pre-outburst data, prior to TJD 2510 (the grey curve is computed from the original data, and black from the flattened light curve). The bottom panel shows the frequency spectrum of the in-outburst data (red, after TJD 2510) compared to the pre-outburst data (black). The right panels show three NRES epochs for H$\alpha$, H$\gamma$, and \ion{He}{I}\,$\lambda$5877. The line color and style corresponds to the observation date (matching the vertical lines in the top-left panel). 
              }
         \label{fig:HD_35345}
   \end{figure*}

This system experienced frequent, high-amplitude, short-lived flickers in nine years of ground-based KELT photometry and in all spectroscopic observations displayed strong single-peaked emission lines with H$\alpha$ reaching up to $\sim$10 times the continuum level \citep[][their Fig.~25]{2018AJ....155...53L}. TESS observed HD~35345 in sectors 43 and 44. For the first $\sim$35 days of these observations, the system exhibited typical low-amplitude variability. The last $\sim$15 days saw the system brighten by about 45\%, with a significant increase in amplitude of the photometric oscillations. Fig.~\ref{fig:HD_35345} shows the TESS light curve, and the frequency spectrum computed from the pre-outburst data (prior to TJD 2510) and the in-outburst data (after TJD 2510). The amplitude of signals in the $g1$ frequency group increased by a factor of about 30, and in $g2$ by a factor of about 4.

One NRES spectrum was acquired at the beginning of the event at TJD 2512, and a second and third 12 and 18 days later, respectively (right panels of Fig.~\ref{fig:HD_35345}). In the first spectrum, high-velocity asymmetric emission is strongly detected in \ion{He}{I} lines, consistent with asymmetric mass ejection. 

\subsection{PQ~Pup (B3Ve)}
TESS observations of this star in sectors 34 and 35 showed a high-amplitude ($\sim$20\%) outburst also with a strong increase in amplitude of the more rapid oscillations. TESS observations in sectors 61 and 62 did not show any sign of mass ejection, although low-level relatively slow variations are present in addition to the typical frequency groups. The frequency spectrum computed from the sector 61 and 62 data is compared to the frequency spectrum computed from the in-outburst data (after TJD 2241 in sectors 34 and 35). The strongest signal in the $g1$ frequency group increased by a factor of 5, and in the second $g2$ group by a factor of 40. 

\begin{figure*}
   \centering
   \includegraphics[width=1.0\hsize]{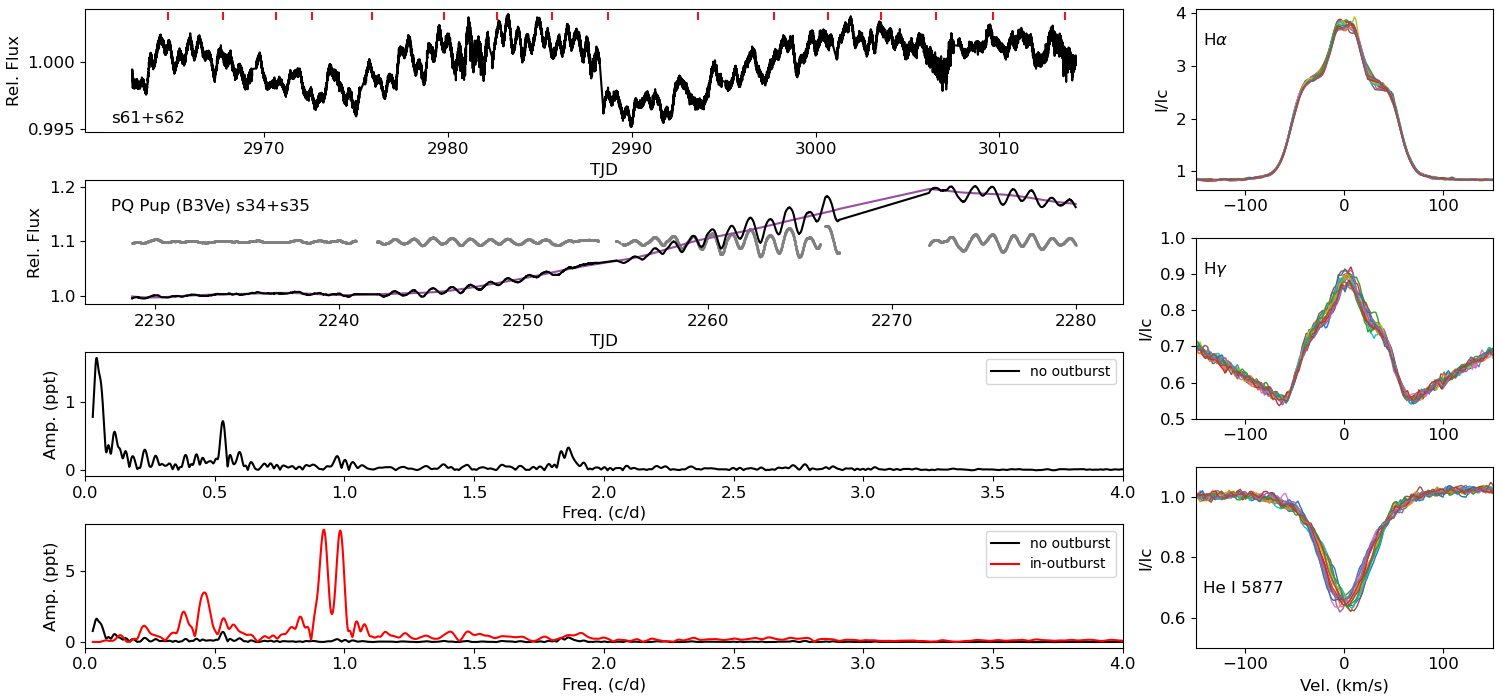}
      \caption{Similar to Fig.~\ref{fig:HD_35345} but for PQ Pup. The top-left panel shows the TESS photometry while the star was apparently not ejecting material. The second-left panel is plotted in the same style as the top panel in Fig.~\ref{fig:HD_35345}. The third-left panel shows the frequency spectrum from the sector 61 and 62 data, and the bottom-left panel adds the frequency spectrum from the sector 34 and 35 data (after TJD 2241). The tick marks in the top panel are the NRES epochs plotted in the right panels (no spectroscopy is available during sectors 34 and 35). 
              }
         \label{fig:PQ_Pup}
   \end{figure*}

Unfortunately, no spectroscopy is available during the outburst in sectors 34 and 35. We acquired 16 NRES spectra during sectors 61 and 62, with observing dates indicated by the tick marks in the top panel of Fig.~\ref{fig:PQ_Pup} and selected corresponding line profiles in the right panels. During this time, the strength of emission lines was slightly decreasing, consistent with a dissipating disk without being replenished by new mass ejection. The line profile variations in e.g., \ion{He}{I}\,$\lambda$5877 are consistent with pulsation.

\clearpage

\section{$\mu$ Cen, $\eta$ Cen, and $\omega$ CMa -- TESS photometry and archival spectra} \label{sec:muCen}

\begin{figure*}[ht]
   \centering
   \includegraphics[width=1.0\hsize]{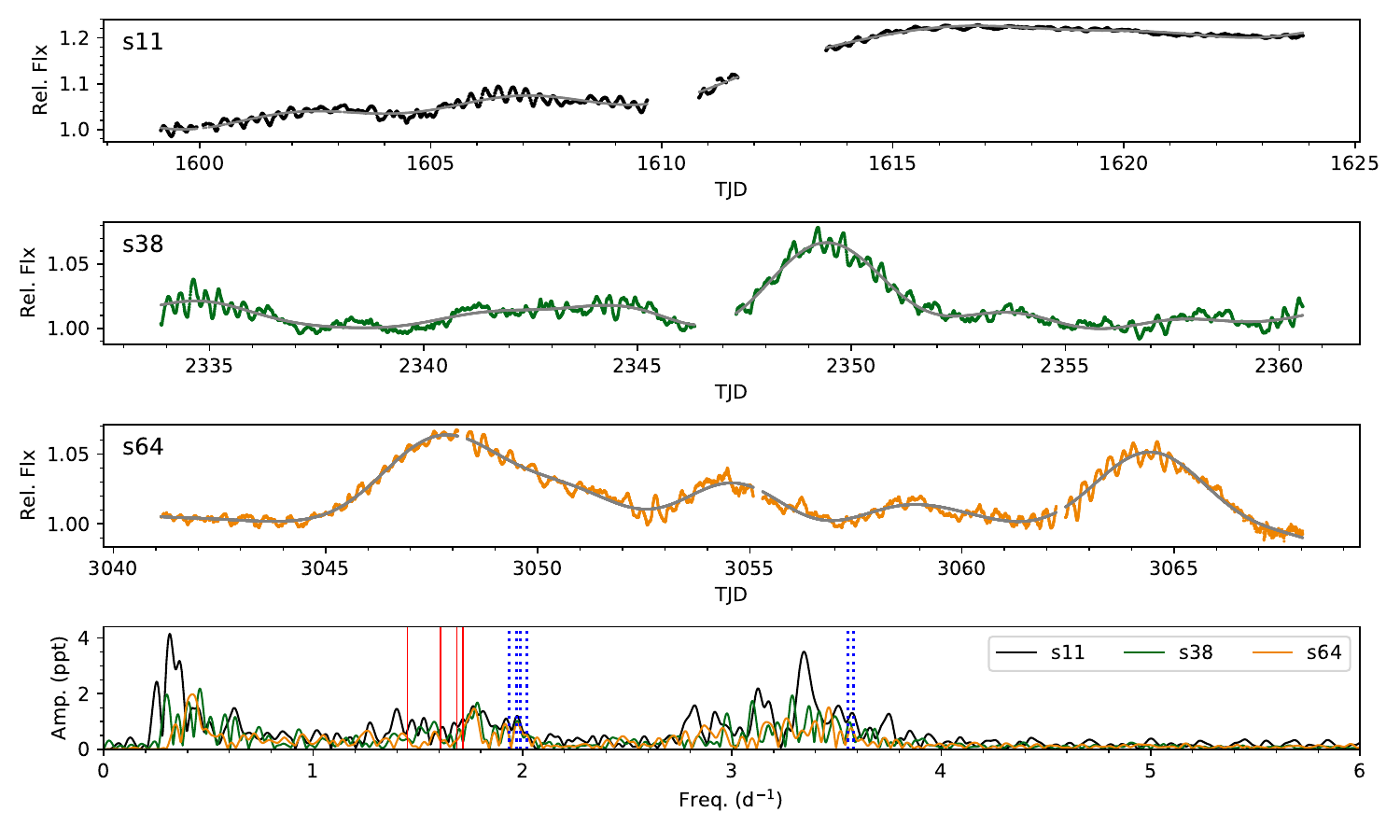}
      \caption{Available TESS data for $\mu$ Cen from sectors 11, 38, and 64. The lighter grey curve describes the low-frequency variability, which was subtracted prior to the calculation of the frequency spectra in the bottom panel. The red and blue vertical lines in the bottom panel are the V/R and spectroscopic pulsation frequencies, respectively. 
              }
         \label{fig:muCen}
   \end{figure*}

TESS has observed $\mu$ Cen in three sectors, as shown in Fig.~\ref{fig:muCen}. As a well known highly active star, it is no surprise that several flickers were recorded by TESS. However, contemporaneous spectroscopy is not available for this system. In the highest amplitude event (sector 11), the brightness increased by about 23\%. The photometric frequency groups are wide and variable. The spectroscopic pulsation frequencies from \citet{1998A&A...336..177R} (vertical blue dotted lines in Fig.~\ref{fig:muCen}) do not obviously stand out in the photometry. In TESS photometry, and considering previous spectroscopic studies \citep{1993A&A...274..356H, 1998A&A...333..125R}, the behavior of $\mu$ Cen on short timescales is similar to the 13 star sample studied in this work. 

$\eta$ Cen and $\omega$ CMa are noteworthy for their long-lasting quasi-permanent {\v{S}}tefl frequencies \citep{2003A&A...411..229R, 2016A&A...588A..56B, 2003A&A...411..167S}. These frequencies (and/or phases) do slightly change over time, and the amplitude of the associated V/R (or EW$_{\rm V}$/EW$_{\rm R}$\xspace) oscillations vary. Nevertheless, the persistence of the {\v{S}}tefl frequencies in $\eta$ Cen and $\omega$ CMa distinguish them from the sample introduced in this work whose {\v{S}}tefl frequencies appear suddenly and typically persist for only several oscillation cycles. 

\begin{figure}[ht]
   \centering
   \includegraphics[width=0.5\hsize]{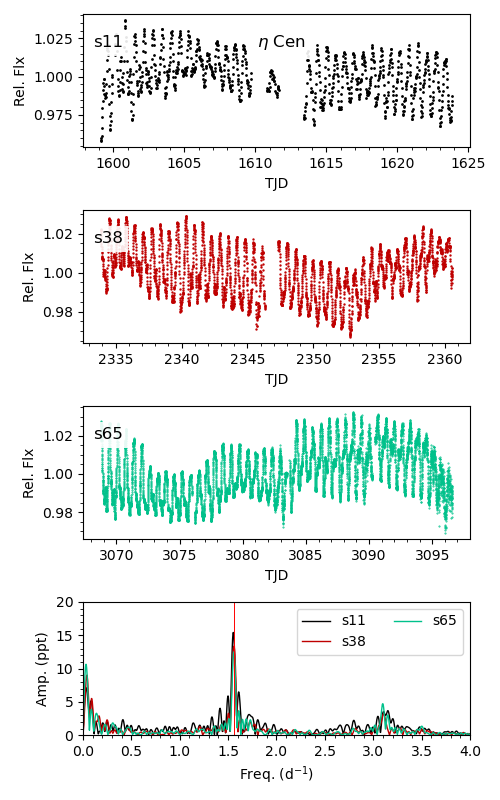}
      \caption{Available TESS photometry for $\eta$ Cen. The bottom panel shows the frequency content of the photometry for each plotted sector.
              }
         \label{fig:etaCen_TESS}
   \end{figure}

\begin{figure}[ht]
   \centering
   \includegraphics[width=0.5\hsize]{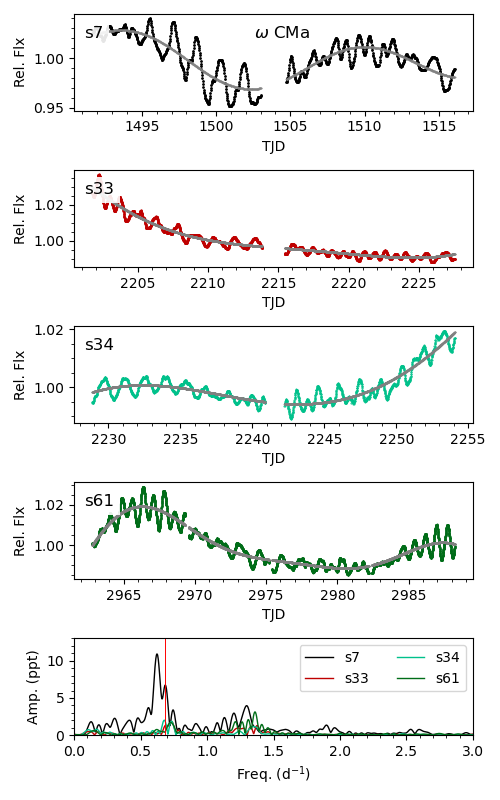}
      \caption{Same as Fig.~\ref{fig:etaCen_TESS} but for $\omega$ CMa.
              }
         \label{fig:omeCMa_TESS}
   \end{figure}

Figures~\ref{fig:etaCen_TESS} and~\ref{fig:omeCMa_TESS} plot the available TESS photometry and its frequency content for $\eta$ Cen and $\omega$ CMa, respectively. For $\eta$~Cen, TESS does not record any short-lived flickers. The variations in the three available TESS sectors are all very similar to each other, and also to the BRITE photometry presented in \citet{2016A&A...588A..56B}. The dominant signal in photometry corresponds to the spectroscopic {\v{S}}tefl frequency, and there is a fairly strong signal at 0.033 d$^{-1}$ in good agreement with BRITE (but with poor frequency resolution). There are additionally several lower-amplitude signals in/near the two main photometric frequency groups. In the TESS photometry for $\omega$ CMa, slow, apparently oscillatory, variations dominant the light curves. Without contemporaneous spectroscopy, it is difficult to determine if these slow variations are associated with disk-build up or are photospheric. Out of the four available TESS sectors, sector 7 exhibits relatively high amplitude frequency groups (possibly hinting at ongoing or enhanced mass loss), whereas the other three sectors are all similar to each other. There does appear to be a photometric signal in TESS associated with the {\v{S}}tefl frequency, but it does not dominate the frequency spectrum as in $\eta$~Cen.

\begin{figure*}[ht]
   \centering
   \includegraphics[width=1.0\hsize]{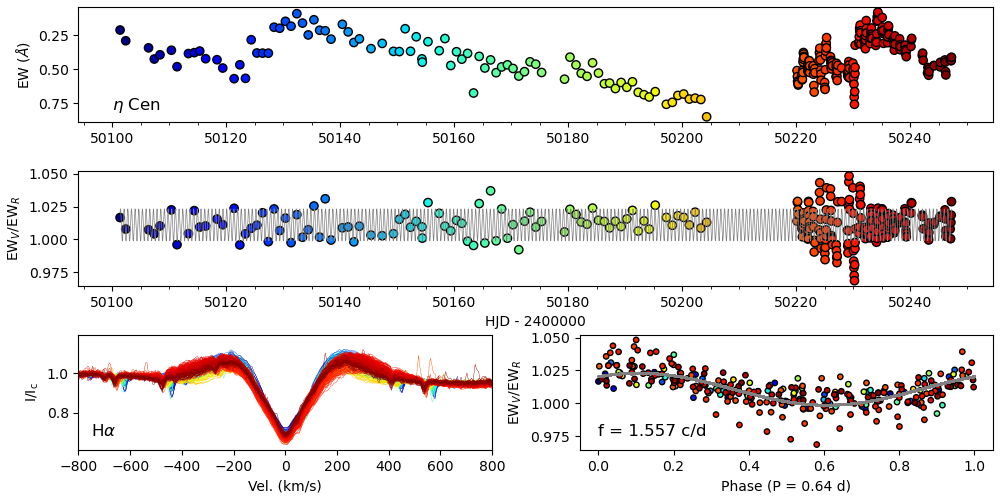}
      \caption{H$\alpha$ data for $\eta$ Cen. \textit{Top:} H$\alpha$ EW. \textit{Middle:}  H$\alpha$ EW$_{\rm V}$/EW$_{\rm R}$\xspace. \textit{Bottom-left:} H$\alpha$ line profiles. \textit{Bottom-right:} H$\alpha$ EW$_{\rm V}$/EW$_{\rm R}$\xspace phased to the circumstellar {\v{S}}tefl frequency. In all panels, the color changes with time from blue to red.
              }
         \label{fig:etaCen_spec}
   \end{figure*}

\begin{figure*}[ht]
   \centering
   \includegraphics[width=1.0\hsize]{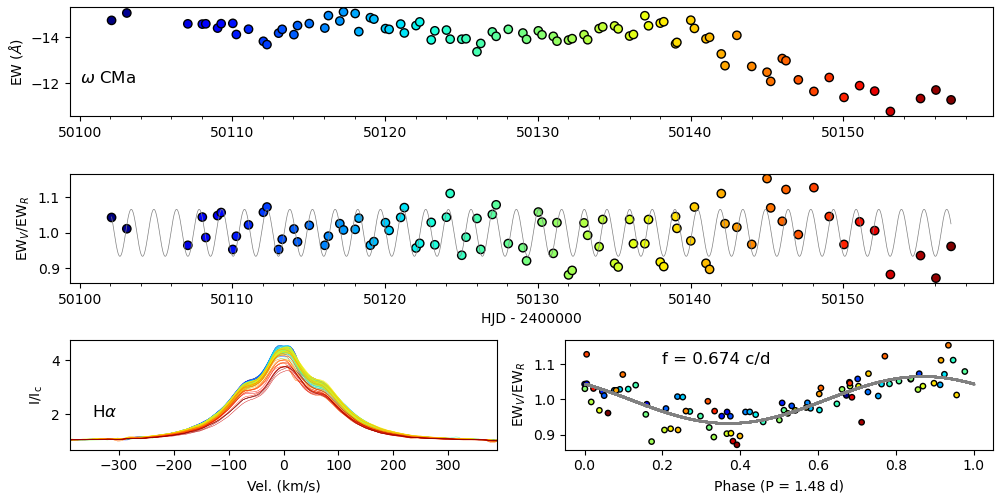}
      \caption{Same as Fig.~\ref{fig:etaCen_spec} but for $\omega$ CMa.
              }
         \label{fig:omeCMa_spec}
   \end{figure*}

Figures~\ref{fig:etaCen_spec} and~\ref{fig:omeCMa_spec} show archival H$\alpha$ data from the HEROS spectrograph for $\eta$ Cen and $\omega$ CMa, respectively. The H$\alpha$ spectra start in January 1996, and extend to June 1996 ($\eta$ Cen) and March 1996 ($\omega$ CMa). Both stars had a fairly strong disk during these timespans. $\eta$ Cen is a shell star, and exhibited variations in its H$\alpha$ EW. Throughout the entire time plotted in Fig.~\ref{fig:etaCen_spec} (about 145 days), the {\v{S}}tefl frequency is evident, with the most obvious change being an increase in amplitude for about 10 days (starting at HJD 2450220) while the EW was numerically decreasing (emission strength was increasing). $\omega$ CMa is viewed at a low inclination angle, and had a roughly constant H$\alpha$ EW for the first 40 days plotted in Fig.~\ref{fig:omeCMa_spec}, and the emission strength (relative to the continuum) decreased in the last $\sim$15 days. The {\v{S}}tefl frequency (here seen in the EW$_{\rm V}$/EW$_{\rm R}$\xspace) is obvious throughout this entire time, appealingly increasing in amplitude towards the end of this dataset. For both stars, the lower-right panels of Figs.~\ref{fig:etaCen_spec} and~\ref{fig:omeCMa_spec} show the EW$_{\rm V}$/EW$_{\rm R}$\xspace measurements phased to their respective {\v{S}}tefl frequencies, demonstrating that these variations are consistent with a (near-)constant frequency and phase. 

It is not entirely clear why $\eta$ Cen and $\omega$ CMa have such long-lasting {\v{S}}tefl frequencies. Perhaps these systems exhibit more continuous mass ejection (but with variable rates of mass flux) compared to the short, punctuated events studied in this work. Given the very different viewing angles, this cannot be explained solely by a line-of-sight inclination angle effect. 

\clearpage 

\section{X-rays in V767\,Cen} \label{sec:Xrays}

V767\,Cen has been observed in X-rays several times notably in July 2021
-- January 2022, when a simultaneous optical and X-ray monitoring was
performed \citep{2022MNRAS.512.1648N}. At the time, the star was undergoing a
several-year decreasing trend of its H$\alpha$ emission but a sudden
emission strengthening appeared, marking the start of an outburst and of
the X-ray campaign. The EW(H$\alpha$) at the time of the X-ray
observations ranged from around --4\AA\ for the exposure taken near peak
emission value and --2\AA\ for exposures taken on the rising branch and
at the end of the outburst. Despite a large change in H$\alpha$, the
recorded X-ray flux appeared remarkably stable considering errors
($\sigma\sim12\%$), as was its hardness. Flux and hardness were also
similar compared to a previous XMM observation taken in January 2007.

Taking advantage of the TESS campaign, an additional observation was
taken with the Swift X-ray telescope on June 1st 2023 (HJD=2460097.016
or TJD=3097.016, ObsID=00014422009). This corresponds to the end of the
TESS observations, and the photometric peak of an outburst - EW(H$\alpha$) was
around --3\AA\ at the time. The EW$_{\rm V}$/EW$_{\rm R}$\xspace cycles for this event were of low amplitude at the Swift epoch, indicating that the ejected material had nearly circularized. 

As the source is rather bright at optical/UV wavelengths, the windowed
timing mode was used, as it was before. The count rate and spectra were
extracted using the on-line Swift
tool\footnote{https://www.swift.ac.uk/user\_objects/ - calibration
available on Nov. 12 2024}. Swift count rates for this new observation
amount to $0.116\pm0.012$\,cts\,s$^{-1}$ in the 0.5--10.\,keV band,
which is again in line with those recorded before
(0.07--0.12\,cts\,s$^{-1}$). The ratio between count rates in
2.--10.\,keV and 0.5--2.0\,keV reaches $0.60\pm0.13$ while it was
0.31--0.67 before.

Spectral fitting was done as in Naz\'e et al. (2022), with an absorbed
two-temperature model ($kT$ fixed to 0.28 and 6.02\,keV). The spectral
parameters are $N_{\rm H}=[0.043+(0.05\pm0.11)]\times
10^{22}$\,cm$^{-2}$ for the interstellar and circumstellar components,
respectively, and $norm(6.02)=0.00222\pm0.00031$\,cm$^{-5}$ (the
normalization of the coolest component being fixed to 8\% of that
value). The observed flux in the 0.5--10.\,keV then amounts to
$(3.72\pm0.51)\times 10^{-12}$\,erg\,cm$^{-2}$\,s$^{-1}$ and the
hardness ratio, defined as the ratio between the fluxes in 2.--10.\,keV
and 0.5--2.0\,keV after correction for the interstellar absorption, is
$1.78\pm0.32$. Again, all those values appear similar to those recorded
before, demonstrating the absence of drastic changes and a rather
constant character for the long-term X-ray emission of V767\,Cen.

Subtle and fast responses could
however not be probed. The previous Swift data indicated possible hints
of a larger luminosity and hardness closer to the start of an event, but
those were only 1 sigma variations: clearly, more sensitive X-ray data
taken very swiftly after the start of an outburst would provide further
indications of the exact X-ray impact of such events.

\end{appendix}

\end{document}